%% file: MASCOT__Metallicity_gradients.tex
\documentclass[a4paper,fleqn,usenatbib]{mnras}

\usepackage{newtxtext,newtxmath}

\usepackage[T1]{fontenc}
\usepackage{ae,aecompl}
\usepackage{fixltx2e}

\usepackage{graphicx}	
\usepackage{amsmath}	
\usepackage{amssymb}	

\usepackage{xcolor}		
\usepackage{mathrsfs} 
\usepackage{breakcites} 
\usepackage{caption} 
\usepackage{multicol}
\usepackage{rotating} 
\usepackage{changepage}  
\usepackage[export]{adjustbox}  
\usepackage{enumitem}   
\usepackage{graphbox} 
\usepackage{pifont}
\usepackage{letltxmacro} 
\usepackage{xparse}

\input{mycommands.tex}

\LetLtxMacro\oldcitep\citep 
\RenewDocumentCommand{\citep}{O{} O{} m}{\oldcitep[#1][#2]{#3}}
\NewDocumentCommand{\citex}{O{} O{} m}{\oldcitep{#3}}

\LetLtxMacro\oldcitet\citet 
\RenewDocumentCommand{\citet}{O{} O{} m}{\oldcitet[#1][#2]{#3}}

\newcommand{\cmark}{\ding{51}}%
\title[Metallicity gradients in MASCOT]{MASCOT: Molecular gas depletion times and metallicity gradients - evidence for feedback in quenching active galaxies}

\author[Bertemes et al.]{C. Bertemes,$^{1}$ \thanks{E-mail: c.bertemes@uni-heidelberg.de}
D. Wylezalek, $^{1}$
M. Albán, $^{1}$
M. Aravena, $^{2}$ 
W. M. Baker,$^{3,4}$ 
S. Cazzoli, $^{5}$
\newauthor
C. Cicone, $^{6}$
S. Mart\'in, $^{7,8}$
A. Schimek, $^{6}$
J. Wagg, $^{9}$
W. Wang $^{1}$
\\
$^{1}$ Zentrum f\"ur Astronomie der Universit\"at Heidelberg, Astronomisches Rechen-Institut M\"onchhofstr, 12-14 69120 Heidelberg, Germany \\
$^{2}$ Núcleo de Astronomía, Facultad de Ingeniería y Ciencias, Universidad Diego Portales, Av. Ejército 441, Santiago, Chile \\
$^{3}$ Kavli Institute for Cosmology, University of Cambridge, Madingley Road, Cambridge, CB3 OHA, UK \\
$^{4}$ Cavendish Laboratory - Astrophysics Group, University of Cambridge, 19 JJ Thompson Avenue, Cambridge, CB3 OHE, UK \\
$^{5}$ IAA - Instituto de Astrofísica de Andalucía (CSIC), Apdo. 3004, E-18008 Granada, Spain \\
$^{6}$ Institute of Theoretical Astrophysics, University of Oslo, PO Box 1029, Blindern 0315, Oslo, Norway \\
$^{7}$ European Southern Observatory, Alonso de Córdova, 3107, Vitacura, Santiago, 763-0355, Chile \\
$^{8}$ Joint ALMA Observatory, Alonso de Córdova, 3107, Vitacura, Santiago, 763-0355, Chile \\
$^{9}$ SKA Observatory, Lower Withington Macclesfield, Cheshire SK11 9FT, UK\\
}

\date{Accepted XXX. Received YYY; in original form ZZZ}

\pubyear{2017}

\begin{document}

\label{firstpage}
\pagerange{\pageref{firstpage}--\pageref{lastpage}}
\maketitle

\begin{abstract}
We present results from the first public data release of the {MaNGA-ARO Survey of CO Targets (MASCOT)}, focussing our study on galaxies whose star-formation rates and stellar masses place them below the ridge of the star-forming Main Sequence. In optically-selected type 2 AGN/LINERs/Composites, we find an empirical relation between gas-phase metallicity gradients $\nabla Z$ and global molecular gas depletion times $\tdep = M_\Hmol/{\rm SFR}$ with ``more quenched'' systems showing flatter/positive gradients. Our results are based on the O3N2 metallicity diagnostic {(applied to star-forming regions within a given galaxy)} which was recently suggested to also be robust against emission by diffuse ionised gas (DIG) and low-ionisation nuclear emission regions (LINERs). We conduct a systematic investigation into possible drivers of the observed  $\nabla Z - \tdep$ relation (ouflows, gas accretion, in-situ star formation, mergers, and morphology). We find a strong relation between $\nabla Z$ or $\tdep$ and centralised outflow strength traced by the \oiii\ velocity broadening. We also find signatures of suppressed star-formation in the outskirts in AGN-like galaxies with long depletion times and an enhancement of metals in the outer regions. We find no evidence of inflows impacting the metallicity gradients, and none of our results are found to be significantly affected by merger activity or morphology. We thus conclude that the observed $\nabla Z - \tdep$ relation may stem from a combination of metal redistribution via weak feedback, and a connection to in-situ star formation via a resolved mass-metallicity-SFR relation. 
\end{abstract}


\section{Introduction}
\label{sec:intro}

One of the key goals in galaxy evolution is to quantify the growth of galaxies, which occurs via the conversion of gas into stars. Specifically, the fuel for star formation is held in the form of a cold, molecular gas phase, which fragments and contracts via cooling until a critical density is reached and gravitational collapse ensues, culminating in the ignition of nuclear fusion and thus the birth of new stars (e.g. \citealt{Schmidt1959, Kennicutt1998a}).  During the actively star-forming phase in the lives of galaxies, their growth is further characterised by an observed tight relation between the rate of star formation SFR and the stellar mass $M_\star$ already in place. This relation is known as the Main Sequence of star-forming galaxies (MS; e.g. \citealt{Brinchmann2004, Noeske2007_MS, Renzini2015}). It has been suggested to arise as a combination of the Kennicutt-Schmidt relation between molecular gas mass surface density and SFR surface density \citep{Schmidt1959, Kennicutt1998a}, and the ``molecular gas Main Sequence'' between between the stellar and molecular gas mass surface densities \citep{Lin2019, Baker2022a}. The star-forming branch around the MS can be contrasted to the locus of passive galaxies, and their ensemble forms a bimodal distribution in the SFR-$M_\star$ plane. The regime in between both populations, dubbed the Green Valley, is sparsely populated, which has been suggested to point to rapid quenching as galaxies transition from the star-forming into the passive regime towards the end of their life (see \citealt{Salim2014} for a review).

Studies of the cold neutral gas phase hold important clues for characterising and understanding SFR fluctuations and the quenching of galaxies, and thus represent a prime element in compiling a theory of galaxy evolution. For instance, it has been shown that galaxies below the Main Sequence do not only have a lower quantity of molecular gas, but are also less efficient at converting their remaining gas into stars (e.g. \citealt{Saintonge2017}; see also \citealt{Saintonge2022} for a detailed review). The recent rise of state-of-the-art integral field unit (IFU) and multi-object spectroscopy (MOS) facilities enables studies to go beyond investigating how molecular gas properties shape galaxies in a global way by resolving galaxies as extended dynamic systems. For instance, the EDGE-CALIFA (Extragalactic Database for Galaxy Evolution - Calar Alto Integral Field Area survey; \citealt{Bolatto2017}) and ALMaQUEST surveys (ALMA-MaNGA QUEnching
and STar formation; \citealt{Lin2020}) have conducted resolved CO emission line radio observations on samples of galaxies with optical-IFU data. The EDGE-CALIFA collaboration performed CO follow-up observations of 177 galaxies with existing CALIFA data with the CARMA (Combined Array for Millimeter-wave Astronomy) interferometer, which will further be expanded by ongoing observations with the APEX telescope (295 targets to date; \citealt{Colombo2020}). The goal of this survey is to study the resolved relationships between molecular gas and e.g. stellar mass and SFR, ionised gas kinematics, dust extinction, as well as physical and chemical conditions within galaxies. For instance, \citet{Colombo2020} found evidence for a scenario in which the quenching of galaxies starts with an initial central shortage of molecular gas, but then in a second phase a decrease in star forming efficiency becomes the most important driver for permanently pushing galaxies into quiescence. Within the ALMaQUEST survey (described in more detail in Section \ref{subsec:AQ}), starbursts selected to feature at least a $50 \%$ elevation of the SFR in the central region were shown to be primarily driven by an increased star-forming efficiency \citep{Ellison2020}, which is the ratio of the SFR over the molecular gas mass $M_\Hmol$. On the other hand, Green Valley galaxies show both suppressed star-forming efficiencies and molecular gas fractions ($M_\Hmol /M_\star$) with respect to the MS, both in star-forming as well as `retired' spaxels \citep{Lin2022}. {We also refer the reader to \citet{Sanchez2021a,Sanchez2021b} for a summary of the resolved and integrated relations between molecular gas mass, SFR and stellar mass, and the connection between the different scales.}

Complementing the ALMaQUEST and EDGE-CALIFA survey, \citet{Wylezalek2022} recently published the first data release of our MaNGA-ARO Survey of CO Targets (MASCOT). Using 12m single-dish $^{\rm 12}$CO(1-0) observations with the Arizona Radio Observatory, we have collected data for 187 galaxies from the MaNGA survey (Mapping Nearby Galaxies at Apache Point Observatory; \citealt{Bundy2015}) to date, selected to span a broad range in stellar masses ($\sim 1.5$ dex) and $\rm sSFR = SFR/M_\star$ ($>3$ dex). Our observations thus enhance the number statistics of galaxies with simultaneously available optical-IFU data and (galaxy-integrated) CO observations. Further observations are still ongoing. We note that MASCOT is highly complementary to ALMaQUEST in terms of the field of view probed by the CO observations (while drawing from the identical underlying optical-IFU survey), and therefore we include the public global ALMaQUEST data in \citet{Wylezalek2022} together with the MASCOT DR1 for convenience.

In this paper, we present results based on the first data release. In particular, we combine the study of molecular gas properties with resolved metallicities and kinematics, as the latter are, amongst others, sensitive to gas accretion, feedback mechanisms and in-situ star formation. We also investigate to what extent inflows/outflows have any measurable impact on star formation via influencing the availability of fuel, or its consumption.

In recent years, resolved chemical abundance studies have brought to light the diversity of gas-phase metallicity gradients present within galaxies, reflecting the variety of physical processes that shape them. The majority of galaxies in the nearby Universe are found to exhibit negative gas-phase metallicity gradients, i.e. they are more metal-rich in their central regions than in the outskirts ({see e.g. \citealt{Sanchez2014} and} the review by \citealt{Maiolino2019}). This is qualitatively consistent with an inside-out growth scenario in which stellar populations in the centre have more time to chemically pollute their surroundings. Further, \citet{Belfiore2017} and \citet{Poetrodjojo2018} found evidence for a mass - metallicity gradient relation complementing the established mass - metallicity relation. The latter has been suggested to emerge from a relation on resolved scales between local metallicities and stellar mass surface densities tracing morphology \citep{BarreraBallesteros2016, Lutz2021}. {The dependence of the radial metallicity distribution on both mass and morphology has also been shown in \citet{Sanchez2020, Sanchez2021a} (Figs 15 and 16, respectively). \citet{Boardman2021, Boardman2022} further found metallicity gradients to steepen with galaxy size at fixed mass, which they traced back to variations of the local stellar mass surface density relative to the total stellar mass. }Resolved metallicities have also been found to show a weaker dependence on global \HI\ mass fraction \citep{Franchetto2021} or absolute \HI\ mass \citep{Lutz2021}, which is consistent with a `local gas regulator model' balancing gas accretion, star formation and outflows. Using a numerical chemical evolution model, \citet{Lian2018, Lian2018b} argue that reproducing both the observed global mass-metallicity relation and metallicity gradients requires including the impact of chemically enriched outflows or radial variations in the intial mass function.  Similarly, \citet{Hemler2021} show that feedback processes by AGN and stars in Illustris TNG50 flatten metallicity gradients over time to reproduce the observed population at low redshift, while FIRE simulations predict that bursty star formation and feedback can significantly change metallicity gradients on sub-Gyr timescales at high redshift \citep{Ma2017}. 
\citet{Baker2022b} further demonstrate that local metallicities in MaNGA galaxies depend more strongly on the global SFR than on the local one, which may be interpreted in terms of stellar feedback redistributing chemically enriched material.

Studying the connection between the chemical distribution, gas flows, star formation and molecular gas properties is of particular interest in the regime below the Main Sequence, including the Green Valley, to uncover information about the drivers of quenching. However, (resolved) metallicity studies have traditionally been sparse in the below-MS regime given the prevalence of diffuse ionised gas (DIG) contamination. One approach is to limit the analysis to star-forming regions within those galaxies with DIG emission (e.g. \citealt{Sanchez2018}). Recent works have further investigated to what extent DIG emission as well as low-ionization (nuclear) emission-line regions (LINERs/LIERs) may or may not bias strong line metallicity calibrators \citep{Zheng2017, Kumari2019, ValeAsari2019}. For the O3N2 metallicity diagnostic (based on \Halpha, \NII, \Hbeta, and \OIII) in particular, the bias on recovered metallicity gradients was found to be limited to $\lesssim 0.05$ dex and, importantly, symmetric (such that it is unlikely to give rise to artificial correlations in any statistical analysis). Motivated by these findings (summarised in more detail in Section \ref{subsec:choice_O3N2}), we therefore focus on the below/on-MS regime in this work, using the O3N2 diagnostic to study the interplay between metallicity gradients and molecular gas properties, and furthermore include LI(N)ERs and Composites in our analysis. {To exercise caution, we will still use metallicity gradients derived via star formation-dominated sub-fields within a given galaxy.}

The paper is organised as follows: In Section \ref{sec:Data}, we describe the data products we use, including a more detailed description of the MASCOT observations. We also motivate our choice of metallicity diagnostics, and present the sample selection. In Section \ref{sec:Analysis}, we present the main focus of this paper, namely a relation found between metallicity gradients $\nabla Z$ and molecular gas depletion time $\tdep = M_\Hmol / {\rm SFR}$ within AGN, LINERs and Composites below/on the MS, while also performing sanity checks about the robustness of this relation. In Section \ref{sec:Discussion}, we expand our data analysis by investigating the viability of several possible drivers behind the $\nabla Z - \tdep$ relation: gas flows, in-situ star formation, mergers and morphology. Finally, we proceed to summarise our findings in Section \ref{sec:Conclusion}. Throughout this paper, we assume a standard cosmology with $H_0 = 70 \ \kms \ {\rm Mpc}^{-1}$, $\Omega_m = 0.3$ and  $\Omega_{\Lambda} = 0.7$.

\section{Data and methods}
\label{sec:Data}

\subsection{MASCOT}
\label{subsec:MASCOT}

\begin{figure*}
\centering
\includegraphics[width=0.85\textwidth]{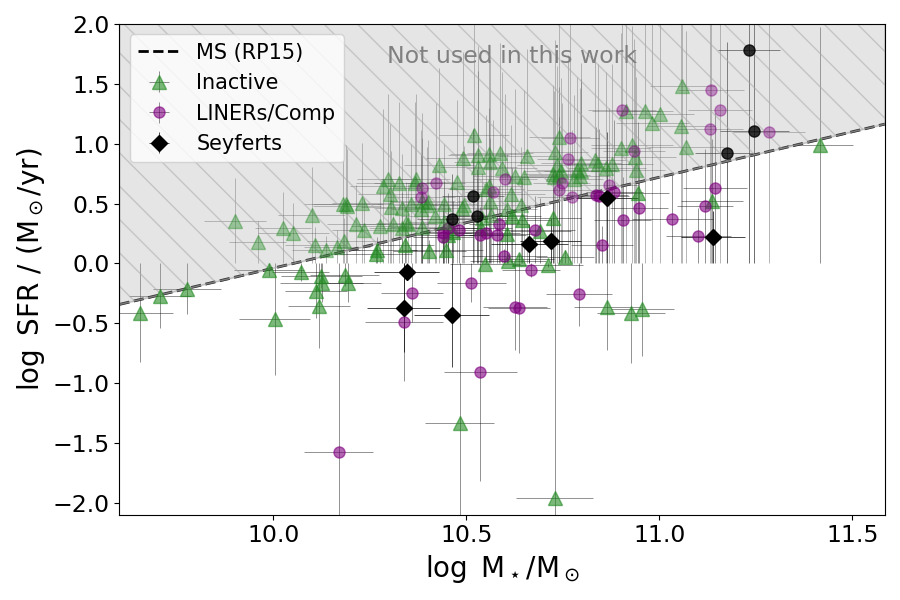}
\caption{Location of the MASCOT sample within the plane of star formation rate SFR versus stellar mass $M_\star$. The \citet{Renzini2015} Main Sequence (MS) relation is shown as a dashed line. In this work, we focus on the below/on-MS regime and thus discard the targets above the MS ridge as illustrated by the grey hatched region. The green datapoints correspond to inactive galaxies, the purple ones to LINERs and Composites, and the black ones to Seyfert AGN, according to a \citet{BPT1981} classification based on their central kpc. }
\label{fig:MS_plane}
\end{figure*}

Our primary dataset is the MaNGA-ARO Survey of CO Targets (MASCOT; first data release presented in \citealt{Wylezalek2022}). MASCOT is an ESO Public Spectroscopic Survey using the Arizona Radio Observatory (ARO) to conduct single-dish CO(1-0) follow-up observations of galaxies with existing optical-IFU data from the Mapping Nearby Galaxies at Apache Point Observatory (MaNGA) survey, which will be summarised below in Section \ref{subsec:MaNGA}. The main objective of MASCOT is to investigate the relation between molecular gas properties and spatially resolved optically-derived quantities. In particular, the size of the sample (consisting of $\sim 20\% $ of AGN/LINERs/Composites) enables us to investigate how such relations vary for galaxies with different levels of star-forming and AGN activity, and examine the role of morphology and environment. 

Observations were carried out using the 3mm receiver on the 12m ARO antenna (equivalent to ALMA Band 3, 84 – 116 GHz). {The ARO beam size is $\sim 55$".} To inform our observing strategy, we performed on-the-fly data reduction and stayed on-target until we either obtained a detection at the level of $\gtrsim 5 \sigma$, reached a sensitivity limit of $\mathrm{rms} = 0.25$ mK (using velocity bins of $\sim 50 {\rm km/s}$), reached the end of a given observing block, or the telescope required a point-focus. Moreover, we prioritised targets according to their visibility (i.e. low air masses), whilst avoiding pointing the antenna directly into the Sun or the wind. Besides visibility considerations, our sample was further designed to span a broad range in stellar masses with $\log M_\star > 9.5 M_\odot$ and specific SFRs focused around the Main Sequence, as well as including a number of Green Valley galaxies and starbursts. In more detail, the first data release consists of 187 galaxies spanning over $1.5$ dex in stellar mass and over 3 orders of magnitude in sSFR. Fig \ref{fig:MS_plane} shows the position of our targets in the MS plane, supplemented by the ALMaQUEST survey (see Section \ref{subsec:AQ} below). The dashed line shows the local MS relation obtained by \citet{Renzini2015}. In this work, we exclude all galaxies above the dashed line and select those falling below it to focus on the regime below the MS, as well as on it (given the uncertainties in the datapoints and the intrinsic scatter around the MS of $\sim 0.3$ dex). Our sample contains inactive galaxies, as well as LINERs/Composites and Seyfert AGN, which are coloured in green, purple and black, respectively, according to a \citet{BPT1981} classification within the central $1$ kpc (discussed in more detail in Section \ref{subsec:sample}).

We provide the binned and unbinned CO profiles, along with other raw data products (e.g. CO fluxes and luminosities) and derived quantities (e.g. \Hmol\ masses and kinematics of the fitted CO lines), on the MASCOT website \footnote{\url{https://wwwstaff.ari.uni-heidelberg.de/dwylezalek/mascot.html}}. The \Hmol\ masses are derived using the CO-to-\Hmol~conversion factor $\alpha_{\rm CO}$ from \citet{Accurso2017} which depends on metallicity, as well as offset in SFR from the star-forming Main Sequence relation (which we take from \citealt{Renzini2015}). The first data release of MASCOT makes use of the stellar masses and SFRs from the Pipe3D catalog (\citealt{Sanchez2016b, Sanchez2018}; described in more detail in Section \ref{subsec:Pipe3D}) to calculate molecular gas fractions and depletion times. While the \citet{Accurso2017} $\alpha_{\rm CO}$ conversion was established based on the \citet{PP04_metal} calibrator, metallicities based upon it are not included in the Pipe3D catalog. \citet{Wylezalek2022} therefore uses the Pipe3D oxygen abundances based on the \citet{Maiolino2008} prescription, and corrects for a constant offset of $0.05$ with respect to \citet{PP04_metal} metallicities within our mass range \citep{Sanchez2017}. 
{In this work, we derive $\Hmol$ masses in two ways: firstly, following \citet{Wylezalek2022} with the metallicity-dependent CO-to-\Hmol~conversion outlined above, and secondly, using a constant $\alpha_{\rm CO}$ factor instead.}

\subsection{ALMaQUEST}
\label{subsec:AQ}
We are supplementing our ARO CO(1-0) observations with data from the ALMA-MaNGA QUEnching and STar formation survey(ALMaQUEST; \citealt{Lin2020}). The ALMaQUEST project used the Atacama Large Millimeter Array (ALMA) to perform spatially resolved CO(1-0) follow-up observations of 46 galaxies included in the MaNGA survey (described in Section \ref{subsec:MaNGA} below) on kpc scales. Their targets were selected to span a broad range of star-forming activity ranging from starbursts down to CO-fainter Green Valley galaxies. The field-of-view at the CO(1-0) frequency is $\sim 50$", broadly similar to the ARO beam size of $\sim 55$" from our MASCOT observations. Galaxy-integrated measurements from ALMaQUEST are included in \citet{Wylezalek2022} along with the first MASCOT data release.

\subsection{MaNGA}
\label{subsec:MaNGA}

As mentioned above, a primary science goal of the MASCOT survey (Section \ref{subsec:MASCOT}) is to study the connection between resolved optically-derived galaxy parameters and molecular gas properties via CO(1-0) observations. The required optical-IFU data is taken from the public Mapping Nearby Galaxies at Apache Point Observatory (MaNGA) survey, which constitutes the underlying population from which the MASCOT sample is drawn. 
We now proceed to describe the MaNGA survey and relevant data products in more detail.

\subsubsection{MaNGA survey overview}

The Mapping Nearby Galaxies at Apache Point Observatory (MaNGA) survey \citep{Bundy2015} is part of the fourth generation of the Sloan Digital Sky Survey (SDSS; \citealt{Blanton2017_SDSS-IV}). Launched in July 2014, MaNGA has been collecting optical-IFU spectra for $\sim 10$k galaxies at low redshift ($\langle z \rangle \sim 0.03$) using the two BOSS (Baryonic Oscillation Spectroscopic Survey)
spectrographs mounted on the 2.5 metre Sloan telescope at the Apache Point Observatory
(APO). The red and blue arm of the BOSS spectrographs cover the wavelength range from  $3600$ \AA\ to $10300$ \AA\ \citep{Smee2013}, with a median spectral resolution $\langle R \rangle \sim 2000$. The MaNGA sample is designed to follow a roughly flat distribution in $\log M_\star$ above $9 M_\odot$ \citep{Wake2017}, with about two thirds of the targets being observed out to $1.5 \ R_e$ (``primary'' sample) and the rest being covered out to $2.5 \ R_e$ (``secondary'' sample). We refer the reader to \citet{Law2015} and \citet{Yan2016a, Yan2016b} for details on the observational strategy and Data Reduction Pipeline (DRP). The final DRP datacube products are interpolated on a grid of $0.5$" x $0.5$" spaxels, corresponding to roughly $1-2$ kpc  (while the effective angular resolution amounts to $\sim2.5$"). {All of the MASCOT targets are included in SDSS-IV DR15, and thus we use version 2.4.3 of the MaNGA dataproducts.}

\subsubsection{Stellar masses, SFRs and metallicity (gradients) from Pipe3D}
\label{subsec:Pipe3D}

{In this work, we make use of version 2.4.3} of the public Pipe3D Value Added Catalog \citep{Sanchez2016b, Sanchez2018}. This dataset contains estimates of key galaxy properties - including stellar population properties, star formation histories, emission line models and stellar absorption indices - as recovered by running the Pipe3D full spectral fitting pipeline using the spectra of MaNGA galaxies on a spaxel-by-spaxel basis. {Pipe3D is an iterative fitting procedure following a similar process to an MCMC inference \citep{Lacerda2022},} which derives the best-fit stellar velocity and velocity dispersion fields in a first step, then proceeds to constrain the amount of dust attenuation with fixed stellar kinematics in a second iteration. The results from these first two steps are then used to fit the continuum as a linear combination of simple stellar population (SSP) templates (with a Salpeter initial mass function) to recover galaxy quantities, while independently modelling the emission lines with multiple Gaussians. The adopted stellar library combines GRANADA theoretical stellar templates \citep{Martins2005} with the MILES stellar library \citep{SanchezBlazquez2006, Vazdekis2010, FalconBarroso2011}.

In this work, we use the Pipes3D integrated stellar mass, SFR as well as metallicities and gradients thereof. For our choice of metallicity calibrator, we maintain consistency with \citet{Wylezalek2022} in which the first MASCOT data release was presented and use the Pipe3D oxygen abundances at $1 R_e$ based on the \citet{Maiolino2008} prescription ("oh\_re\_fit\_m08"; M08). As we derive \Hmol\ masses using the metallicity-dependent conversion from \citet{Accurso2017} and as those authors used the O3N2 prescription from \citet{PP04_metal}, we apply a correction of $+0.05$ dex to the Pipe3D M08 metallicities to account for the offset. However, for the metallicity \textit{gradients} $\nabla Z$, we follow a slightly different approach. {Specifically, we use metallicity gradients derived in two different ways: firstly, given that we are including in our analysis the locus below the Main Sequence where DIG levels are elevated, we choose to use an O3N2-based metallicity prescription since the latter was suggested to be reasonably robust in DIG and LI(N)ER regions \citep{Zheng2017,Kumari2019, ValeAsari2019} with no systematic bias (but a symmetric scatter), as we will discuss in more detail in Section \ref{subsec:choice_O3N2}. } 
We therefore choose to use directly the Pipe3D oxygen abundance gradient based on the O3N2 calibrator from \citet{Marino2013} ("alpha\_oh\_re\_fit\_o3n2") evaluated within $0.5-2 \ R_e$ {(which are based on those regions within galaxies ionised by star formation)}. Secondly, we also use Pipe3D gradients based on the \citet{Tremonti2004} calibrator inferred from the R23 ratio, as the latter has been shown to be less sensitive to ionisation (\citealt{Kewley2019}; Fig 9), at least within the metallicity range of our dataset. We will rely on the O3N2-based metallicity gradients for the main presentation of our results, and conduct a sanity check based on the R23 diagnostic instead in Section \ref{subsec:Sanity_checks}.

\subsubsection{Resolved SFRs and stellar masses recovered by a radially resolved full spectral fitting procedure}
\label{subsec:OwnFitting}

We carry out a radially resolved full spectral fitting procedure to incorporate the possible impact of the star formation distribution and resolved stellar masses in our analysis. Our approach follows the stellar population synthesis (SPS) method, in which a given spectrum is fitted as a superposition of the emissions from stellar populations of different ages and metallicities. We run our fitting procedure on radial spectra, obtained  by dividing each galaxy into annuli of $0.2 \ R_e$ width tracing the elliptical Petrosian apertures, and apply aperture corrections to recover the total g-band flux from the NASA-Sloan Atlas catalog (NSA; \citealt{Blanton2011}).
\label{lastpage}

While we defer a detailed description of our fitting procedure to an upcoming paper \citep{Bertemes2022c}, it can be briefly summarised as follows: We use the Bagpipes code \citep{Carnall2018} to model the observed emission in galactic annuli via a superposition of stellar populations following a 2-component star formation history (SFH) consisting of a lognormal and a decoupled recent burst of constant SFR operating over the past 30 Myr. Each of the two SFH components assumes a single stellar metallicity for all of its stars (which for the recent burst corresponds to the gas-phase metallicity). Bagpipes is based on the 2016 version of the \citet{Bruzual2003} models, and assumes a \citet{Kroupa2002} IMF. We simultaneuously fit the MaNGA spectra and associated NSA photometry in the optical wavelength range of $3700 - 7400$ \AA, assuming a \citet{Calzetti2000} extinction law. We further assume a fixed ratio of nebular to stellar attenuation $\eta = 1/0.44$ due to birthcloud dust, as also retrieved by \citet{Calzetti2000}. Ionised gas emission lines are modelled self-consistently in Bagpipes with CLOUDY \citep{Ferland1998}. However, we mask the \OII\ and \SII\ lines, as it was found empirically that they were challenging to reproduce simultaneously with the Balmer lines, \OIII, and \NII, which we prioritise as they hold significant information about the SFR as well as physical and chemical conditions in the ISM. We attribute this challenge to the relatively simplified CLOUDY photoionisation model grid used in terms of assumptions about the birthcloud age, metallicity, and ionisation parameter. Further, we introduce more freedom to the $\Halpha$ line fitting via a normalisation parameter allowed to vary between $1-1.5$. The rationale behind this parameter is that at fixed SFR, the \Halpha\ luminosity predicted by Bagpipes (based on its underlying CLOUDY grid) is known to feature a systematic offset with respect to the values predicted by \citet{Kennicutt2012} of up to $0.2$ dex depending on metallicity and ionisation. Moreover, we set a logarithmic prior on the time of peak star formation, which was found to be required for consistency with the cosmic SFR density in the photometric fitting runs in \citet{Carnall2019b}. Finally, we impose a prior following a student's t-distribution on the jump in SFR between the recent burst and the early SFH, which will favour a smooth transition in the absence of constraining information \citep{Leja2019a}.

We find that when running our fitting procedure on our sample defined in Section \ref{subsec:sample} below, our recovered global SFRs agree well with the Pipe3D measurements, with only a minor offset of $0.12$ dex and a scatter of $0.19$ dex. Similarly, our recovered stellar masses are offset from the Pipe3D reference values by $<0.001$ dex, with a scatter of $\sim 0.13$ dex. Motivated by this agreement, we will use our SPS fitting procedure to derive resolved stellar masses and SFRs within annuli inside galaxies.

\subsection{Choice of metallicity calibrator}
\label{subsec:choice_O3N2}

In this work, we are are focusing on the interplay between resolved metallicities and molecular gas properties in galaxies that are located \textit{below/on the Main Sequence (MS)}. Below the MS, metallicity studies have traditionally been sparse due to the ubiquity of diffuse ionised gas (DIG) regions, and concerns about the latter potentially contaminating common metallicity diagnostics. It has been suggested that DIG emission may bias strong line diagnostics by contributing to the line fluxes entering the calculation \citep{Zhang2017}.

{We first note that the gas-phase metallicity gradients which we take from the Pipe3D catalog are based on regions consistent with being ionised predominantly by star formation \citep{Sanchez2018}. To address concerns about potential secondary contamination by DIG, we proceed to discuss relevant results from the literature.} Recently, in this context, \citet{Kumari2019} established a metallicity prescription for DIG- and LI(N)ER-dominated regions. Exploiting Multi Unit Spectroscopic Explorer (MUSE) observations at $50-100$ pc scale resolution in 24 nearby star-forming galaxies, the authors selected a sample of individual DIG/LI(N)ER - HII region pairs located sufficiently close to each other to be assumed to share a single metallicity. In particular, the study encouragingly found the O3N2 diagnostic (based on \Halpha, \NII, \Hbeta, and \OIII) to be remarkably robust in DIG/LI(N)ER regions, showing only a minor offset of $<0.04$ dex from the ground truth derived in the HII regions, with $0.05$ dex scatter in either direction. Seyfert-classified spaxels located in the outskirts were also included in the analysis. 

On the topic of possible DIG contamination, a similar conclusion was reached by \citet{ValeAsari2019} for a set of $\sim 1400$ MaNGA galaxies, with the additional challenge that the $\sim$kpc scale resolution does not allow resolving individual HII regions. {For a careful exploration of how a loss of spatial resolution can impact and smoothen recovered metallicity gradients, we refer the reader to \citet{Mast2014}.} \citet{ValeAsari2019} used high-S/N star-forming spaxels and characterised the contribution of DIG to relevant emission lines via the equivalent width of the \Halpha\ line. The latter has been favoured as a tracer of DIG over the \Halpha\ surface brightness, since it assesses how faint the line emission is relative to the underlying stellar continuum (rather than in absolute terms) and can therefore identify multiple DIG regions stacked along the line of sight \citep{Lacerda2018}. The work by \citet{ValeAsari2019} concluded that the O3N2 metric was not significantly affected by DIG contamination (while a moderate impact was recovered for N2 based on \nii / \Halpha ). 
On the other hand, an earlier study by \citet{Zhang2017} analysed $365$ face-on star-forming MaNGA galaxies, and found the O3N2 index to be biased in DIG-dominated regions identified by their low \Halpha\ surface brightness density (with the above-mentioned caveat). However, they also found the associated bias in O3N2-based metallicity gradients to be limited to $\pm 0.05 \ {\rm dex}/R_e$ in either direction. Due to the symmetric character of the bias, one can argue that given sufficient number statistics, any DIG contamination could thus not cause any artificial correlations of the gas-phase metallicity gradient with other galaxy properties, but merely weaken any existing trends at worst.

Going back to LINER emission, we note that in LINERs with low \Halpha\ equivalent width thought to be photoionised by post-AGB stars rather than AGN {\citep{Binette1994, Singh2013}}, \citet{OliveiraJr2022} favoured the N2 (\nii / \Halpha -based) metallicity calibrator over O3N2, motivated by photoionisation models of post-AGB populations. However in this work, we will select only LINERs whose \Halpha\ equivalent width suggests them to be ionised by an AGN as discussed in the next Section \ref{subsec:sample}, and the considerations on potential DIG contamination outlined above may further disfavour N2. 

All things considered, the literature reinforces our confidence in applying the O3N2 metallicity diagnostic taken from the Pipe3D catalog as discussed in Section \ref{subsec:Pipe3D} to galaxies showing DIG emission{, especially given that the Pipe3D metallicity gradients are calculated using only star-forming regions within these sources (and excluding the central $0.5 \ R_e$).} Further, while observational studies investigating the impact of LI(N)ER emission on different metallicity calibrators are still sparse to date, the above-mentioned results of \citet{Kumari2019} motivate us to also explore O3N2-derived metallicity gradients within the LINERs, Seyferts and Composites in the MASCOT sample, which we dub ``AGN-like'' objects. We note that, while the majority of our AGN-like objects are not classified as Seyferts, it has been  cautioned against using the O3N2 metallicity prescription in Seyfert AGN due to its sensitivity to ionisation - the theoretical predictions in \citet{Kewley2019} suggest it to vary by over $2$ dex e.g. at solar metallicity when evaluated over an ionisation grid spanning $2$ dex from $\log \ U = -3.98$ to $-1.98$ (Fig 9). In Section \ref{sec:Analysis}, we therefore also use the R23 metallicity diagnostic, which \citet{Kewley2019} predict to be robust against ionisation at the metallicities spanned by our sample ($\log \ {\rm O/H}>8.6$).

\subsection{Sample selection}
\label{subsec:sample}

Motivated by the findings in the previous Section \ref{subsec:choice_O3N2} on the applicability of the O3N2 metallicity diagnostic in the presence of DIG and LINER emission, we focus on below/on-MS galaxies and include LINERs/Composites/AGN in our sample. In detail, we select our sample as follows:
\begin{itemize}[leftmargin=*,labelindent=15pt]
    \item We start by using the \citet{Renzini2015} Main Sequence prescription to select all targets from the MASCOT (+ALMaQUEST) sample that fall below the ridge corresponding to the dashed line in Fig \ref{fig:MS_plane}. {We note that given the instrinsic scatter of the MS of $\sim 0.3$ dex and the uncertainties in SFR and $M_\star$, a significant part of our sample is formally consistent with being on-MS. While we do not define a pure below-MS sample, the average properties of our sample capture those of a less active and transitioning population and can be contrasted against the average properties of those galaxies falling above the MS ridge. Henceforth, we refer to the selected sources as ``below-MS sample'' for the sake of simplicity.}
    \item Among these, we exclude edge-on galaxies with $b/a<0.2$ (for which the limited view on their centres makes metallicity gradients challenging to derive).
    \item We exclude type I AGN using the \citet{Comerford2020} catalog as host galaxy properties are challenging to derive in cases where the central QSO outshines its host.
    \item We further exclude major mergers identified by visual inspection (meaning constellations with 2 galaxies falling into the ARO beam). In galaxies undergoing a major merger with disturbed morphology, radial profile analyses are complicated by the fact that spaxels at the same distance from the centre may not be directly comparable to each other if the galaxy is disrupted in a non-axisymmetric way. Mergers may moreover affect the metallicity distribution within galaxies via mixing processes and inflows, as we will discuss in more detail in Section \ref{subsec:mergers}. 
    \item Finally, we split our sample to consider separately purely star-forming galaxies and ``AGN-like'' objects which include LINERs, Composites and Seyferts. The latter are identified via the NII and SII BPT diagram \citep{Kewley2001, Kauffmann2003c} using the line ratios extracted from a circular aperture of $1$ kpc around the centre. For spaxels that fall only partially within the circle, we use as weights the fraction of their area which belongs to the circle. We further exclude spaxels that are flagged as unusable by the MaNGA DAP (``DONOTUSE'' flag\footnote{\url{https://sdss-mangadap.readthedocs.io/en/latest/metadatamodel.html}}), as well as those with S/N$<3$ for any of the BPT lines or with negative line fluxes. Finally, we follow \citet{Sanchez2018} by requiring AGN-like objects to have an \Halpha\ equivalent width above $1.5$ \AA~(to remove LINERs photoionised by post-AGB stars rather than AGN). The methodology will be further detailed in an upcoming paper \citep{Alban2022}.
\end{itemize}

We discard galaxies that were undetected in our CO observations.
Our final sample thus selected contains 79 galaxies. Among these, 38 are LINERs/Composites/Seyferts and 41 are inactive galaxies.

\section{Results}
\label{sec:Analysis}

\subsection{A link between gas-phase metallicity gradients and depletion times in AGN-like galaxies below/on the MS}
\label{subsec:main_relation}

\begin{figure*}
\centering

\begin{minipage}{\textwidth}
\begin{tabular}{c c}
\begin{minipage}{0.25\textwidth}
\centering
{\LARGE Optical AGN \\ + LINERs \\[1.1ex] + Composites}
\end{minipage}  &
\raisebox{-.65\height}{\includegraphics[width=0.6\textwidth, trim={0cm 0cm 0cm 0.35cm},clip]{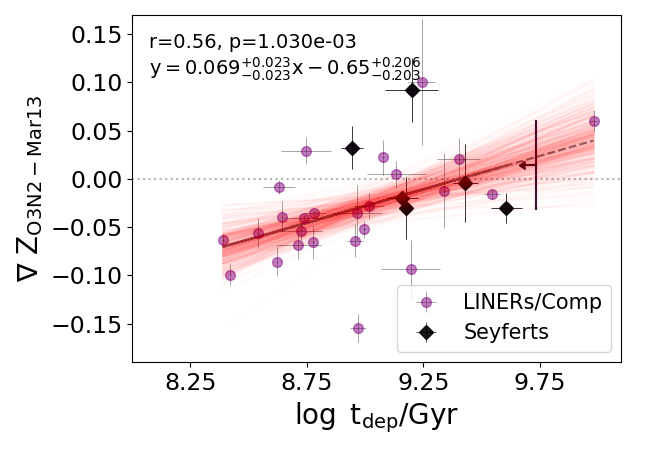}}
\end{tabular}
\end{minipage}

\begin{minipage}{\textwidth}
\begin{tabular}{c c}
\begin{minipage}{0.25\textwidth}
\centering
{\LARGE Inactive \\[1.1ex] galaxies}
\end{minipage}  &
\raisebox{-.65\height}{\includegraphics[width=0.6\textwidth, trim={0cm 0cm 0cm 0.35cm},clip]{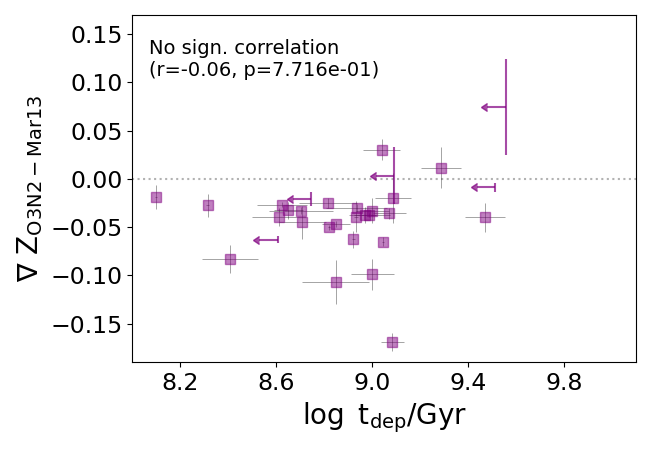}}
\end{tabular}
\end{minipage}

\begin{minipage}{\textwidth}
\begin{tabular}{c c}
\begin{minipage}{0.25\textwidth}
\centering
{\LARGE LINERs \\[1.1ex] + Composites}
\end{minipage}  &
\raisebox{-.65\height}{\includegraphics[width=0.6\textwidth, trim={0cm 0cm 0cm 0.35cm},clip]{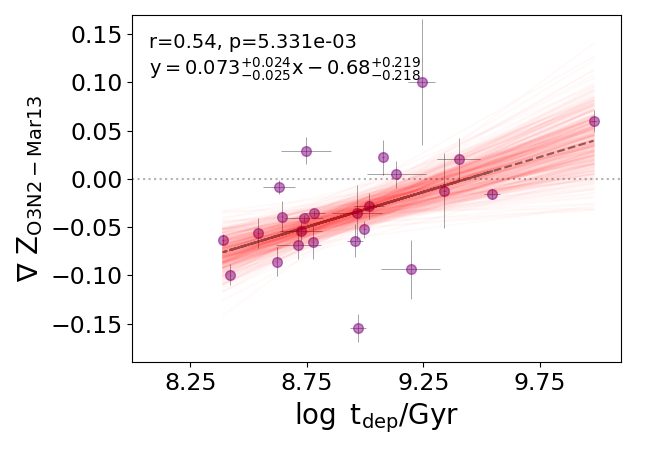}}
\end{tabular}
\end{minipage}

\caption{Metallicity gradients $\nabla Z$ as a function of depletion time $\tdep = {\MHmol /\rm SFR}$ below the MS. \textit{Top panel}: Relation for AGN, LINERs and Composites. Seyfert AGN are highlighted in black, whereas LINERs and Composites are shown in purple. \textit{Middle panel}: For the inactive galaxies in our sample, we find no correlation. \textit{Bottom panel}: Relation for LINERs and Composites - discarding AGN due to concerns about the applicability of the O3N2 metallicity prescription, but including LINERs given the robustness of the O3N2 calibration. In the top and bottom panels, the best-fit relation is overplotted as a black line, with uncertainties illustrated by the red transparent lines. Upper limits are shown by arrows and excluded from the analysis. The correlation is significant in all panels except for the inactive galaxies (see annotated p-values and r-values). {It further holds when excluding the datapoint with longest \tdep in the top and bottom panel with $p=0.003/0.017$ and $r=0.52/0.48$, respectively.} We here use gas-phase metallicity gradients based on the O3N2 diagnostic and calculate \tdep\ based on a metallicity-dependent CO-to-\Hmol\ conversion $\alpha_{\rm CO}$. However, we revisit the relations in this Figure in Figs \ref{fig:check_ConstAlphaCO} and \ref{fig:check_T04}, where we use a constant $\alpha_{\rm CO}$ and a different metallicity calibrator as a sanity check. }
\label{fig:main_plot}
\end{figure*}

For the optically-selected type 2 AGN, LINERs and Composites \textit{below/on the MS} from our sample described above in Section \ref{subsec:sample}, we find a relation between gas-phase metallicity gradients $\nabla Z$ and molecular gas depletion times $\tdep = M_\Hmol/{\rm SFR}$, which is shown in Fig~\ref{fig:main_plot} in the top panel. The correlation is strong with an associated p-value of $\sim 0.001$ and r-value of $0.56$, as annotated in the top of the figure together with the best linear fit (slope: $0.07 \pm 0.02$). When excluding the datapoint with longest \tdep, the statistical coefficients are merely changed to $p=0.003$ and $r=0.52$. We report that such a relation is not present \textit{above the MS} (Fig \ref{fig:Zgrad_vs_tdep_AboveMS} in the appendix). In Fig~\ref{fig:main_plot}, the Seyfert AGN are coloured in black, while the LINERs and Composites are shown in purple, and for CO non-detections, upper limits are indicated via an arrow (but not included in the analysis). The best-fit relation is overplotted as a black line and the red lines represent uncertainties on the fit. The latter are obtained via MCMC-sampled fits from the \texttt{linmix} module\footnote{\url{https://linmix.readthedocs.io/en/latest/}}, which repeatedly perturbs the datapoints within their uncertainties and reruns the fitting. Each such fit yields slightly different coefficients, and the median coefficients over all iterations are used to define the best-fit relation.

{As a further statistical check, we validate the robustness of the correlation by constructing 1000 bootstrapped samples which are drawn from the observed sample with replacement and using them to calculate a distribution of r-values. The latter peaks at $r=0.55$ and we confirm that it does not contain $r=0$ in its associated $95 \%$ confidence interval (Appendix \ref{app:bootstrap}). On a different note, we point out that {Fig~\ref{fig:main_plot} displays depletion times based on the metallicity-dependent CO-to-\Hmol~conversion outlined in Section \ref{subsec:MASCOT}, but we will address and test that assumption in Section \ref{subsec:Sanity_checks}}. }

Moving to inactive galaxies, the middle panel of Fig~\ref{fig:main_plot} displays the $\nabla Z - \tdep$ plane for those system that show no optical signs of LINER or AGN character in their central spaxels according to the BPT selection. In this case, we find no evidence for a correlation (although the picture could conceivably change when increasing the number statistics after a future MASCOT data release). 

As mentioned in Section \ref{subsec:choice_O3N2}, the applicability of the O3N2 metallicity prescription is not guaranteed in Seyfert AGN given the calibrator's sensitivity to ionisation. On the other hand, \citet{Kumari2019} found the diagnostic to be very robust in LINERs and DIG regions based on studying HII–DIG/LI(N)ER pairs in nearby spirals (as well as including non-central Seyfert-classified regions). In Section \ref{subsec:Sanity_checks}, we will conduct a sanity check using a different metallicity diagnostic. For now, in the bottom panel of Fig \ref{fig:main_plot}, we show the relation excluding Seyfert AGN and focusing on LINERs/Composites only. The correlation coefficient $r = 0.54$ and p-value $p=0.005$ still support a strong correlation.  We also checked that the correlation is still significant with $p=0.017$ when excluding both the Seyferts and the single datapoint with the longest depletion time. The $\nabla Z - \tdep$ relation therefore does not depend on the inclusion of type 2 Seyfert AGN.

The depletion time $\tdep = M_\Hmol/{\rm SFR}$ on the x-axis is inversely related to how efficiently a galaxy is converting its remaining \Hmol\ gas into stars. It is thus an important quantity in galaxy evolution, particularly in the below-MS regime, where one may simplistically interpret it as a measure of how quenched a system is, though the picture becomes more complicated when considering that not all systems below the MS are indeed in the process of being quenched - some may conceivably move back to the MS in the future following SFR fluctuations (see e.g. \citealt{Chauke2019} for rejuvenation events in LEGA-C galaxies).  Within this context, \citet{Colombo2020} recently used resolved molecular gas properties and optical-IFU data from the EDGE-CALIFA survey {to compare centrally-star forming and centrally-quenched galaxies. While this type of analysis is complicated by CO-non detections, their findings suggest that once galaxies are moved below the MS due to a central shortage of molecular gas, the depletion time then becomes the main driver that pushes them further towards quiescence when they are already gas-poor and centrally quenched}. The relation we find in Fig~\ref{fig:main_plot} thus supports a picture in which any mechanism(s) that quench AGN-like galaxies also flatten or invert metallicity gradients in the process.

{We note that the existence of a $\nabla Z - \tdep$ relation is in qualitative agreement with recent findings from \citet{Sanchez2021a}. These authors used optical dust attenuation as a tracer to derive molecular gas mass estimates (see \citealt{Concas2019}) for a compilation of publicly accessible IFU datasets, and explored the behaviour of both the star-forming efficiency SFE and metallicity gradients as a function of mass for different morphological types (their Fig 16). While both quantities show a negative dependence on mass for the general population (which would a priori suggest a positive correlation between $\nabla Z$ and SFE), the results strongly depend on morphology. In particular, for intermediate-type galaxies that are located preferentially below the MS (such as S0), the trend between mass and $\nabla Z$ inverts, which suggests $\nabla Z$ to be inversely correlated with SFE in these objects, and thus positive correlated with the depletion time $\tdep$. Our findings are thus qualitatively consistent with the results from \citet{Sanchez2021a} based on indirect molecular gas estimates. We will revisit the role of morphology in Section \ref{subsec:morphology}.}

\begin{figure*}
\centering
\includegraphics[width=0.4\textwidth]{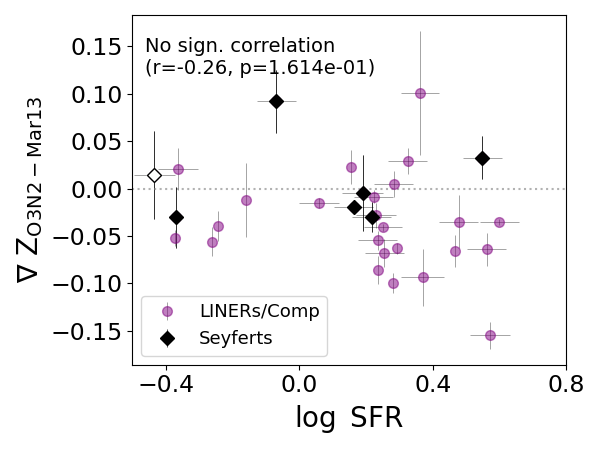}
\includegraphics[width=0.4\textwidth]{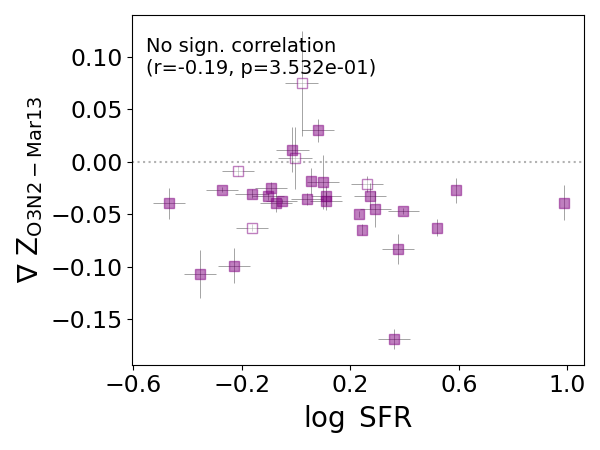}
\includegraphics[width=0.4\textwidth, trim={0cm 0cm 0cm 0.3cm},clip]{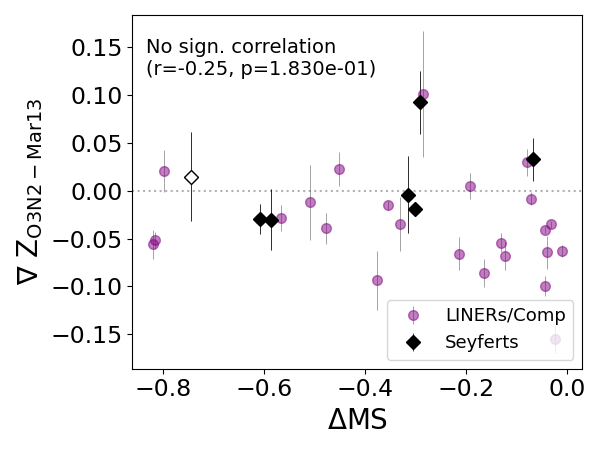}
\includegraphics[width=0.4\textwidth, trim={0cm 0cm 0cm 0.3cm},clip]{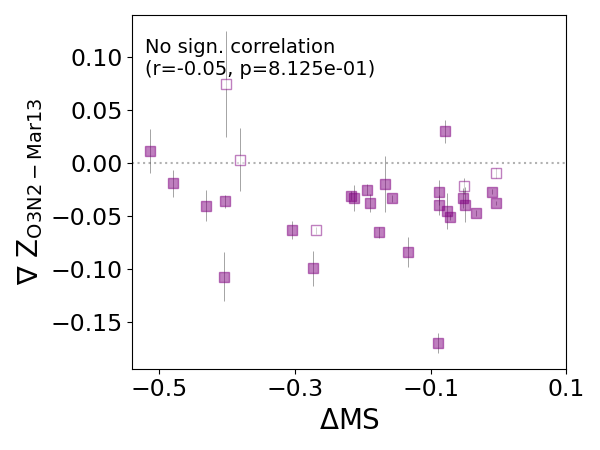}

\caption{Metallicity gradients plotted against SFR values in the top row, and against MS offset $\rm \Delta  MS = \log \ SFR - \log \ SFR_{RP15}$ in the bottom row. In the left column corresponding to AGN, LINERs and Composites, the optically-selected Seyfert AGN are highlighted in black, while the purple datapoints correspond to LINERs and Composites. The right panels focus on inactive galaxies instead. In active and inactive panels alike, we report no correlation between $\nabla Z$ and SFR, nor between $\nabla Z$ and $\rm \Delta  MS $. We thus conclude that we cannot trace back the $\nabla Z - \tdep$ relation from Fig \ref{fig:main_plot} to a pure relation with SFR or MS offset. Instead, variations in $\nabla Z$ are primarily driven by \tdep (relation in Fig~\ref{fig:main_plot}). }
\label{fig:Zgrad_SFR}
\end{figure*}

Given the observed link between longer depletion times and declining star-forming activity below the MS (e.g. \citealt{Saintonge2017}, \citealt{Wylezalek2022}), it is instructive to contrast the metallicity gradients also against SFRs and MS offsets instead of depletion time. In the top panels of Fig~\ref{fig:Zgrad_SFR}, we look for a relation between $\nabla Z$ and SFR, with AGN/LINERs/Composites shown in the left panel and inactive galaxies in the right panel. In the left panel, the black symbols denote BPT-selected Seyfert AGN, while the purple dots show LINERs and Composites. Galaxies that are CO-undetected are shown via open symbols. We note that while we use the Pipe3D SSP-derived SFR values in this plot, the associated SSP-derived errors are likely to be underestimated since they merely describe how easily a best-fit SFR value can be found among all templates. We therefore choose to use \Halpha -derived uncertainties. Both for active and inactive galaxies alike, there is no evidence for a correlation between $\nabla Z$ and SFR. We will however revisit the impact of star formation in a resolved fashion in Section \ref{subsec:SF}. 

{In the bottom panels of Figure \ref{fig:Zgrad_SFR}, we also contrast the $\nabla Z$ values to the MS offset $\rm \Delta  MS = \log \ SFR - \log \ SFR_{RP15} $ with respect to the \citet{Renzini2015} prescription for the MS, again finding no correlation for active (\textit{left panel}) and inactive (\textit{right panel}) galaxies alike. In Section \ref{subsec:morphology}, we further show that the metallicity gradients are not correlated with the amount of molecular gas within our sample. These findings indicate that the metallicity gradients are primarily driven by the depletion time rather than SFR or MS offset. Thus, the crucial factor is not the absolute star formation, but rather the level of star formation in relation to how much building material is still left in the form of cold molecular gas.}

\begin{figure*}
\centering
\includegraphics[width=0.4\textwidth, trim={0cm 0cm 0cm 0.cm},clip]{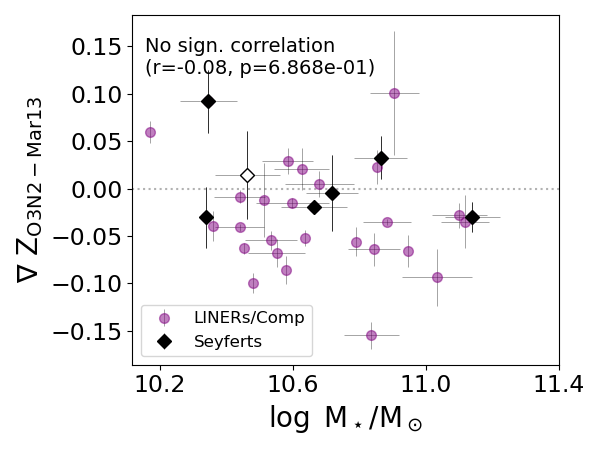}
\includegraphics[width=0.4\textwidth, trim={0cm 0cm 0cm 0.cm},clip]{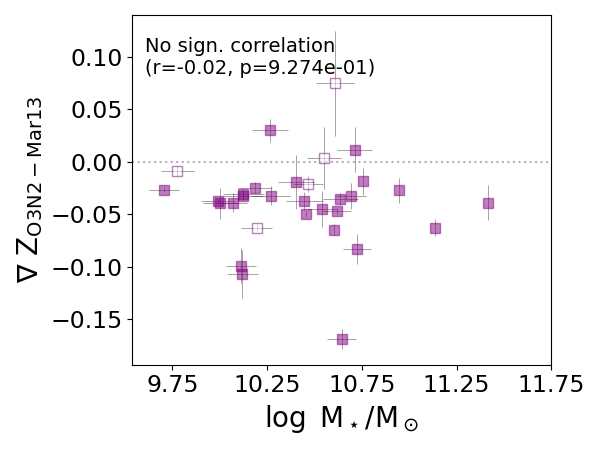}
\caption{Metallicity gradients $\nabla Z$ plotted against stellar mass $M_\star$. We do not find any correlation in either AGN/LINERs/Composites (left panel) or inactive galaxies (right panel), so we conclude that mass effects cannot explain the $\nabla Z - \tdep$ relation found in Fig~\ref{fig:main_plot}.}
\label{fig:Zgrad_lMstar}
\end{figure*}

Finally, it is necessary to check for the presence of mass effects among subsamples which may impact our results. In particular, we note that as can be seen in Fig \ref{fig:MS_plane}, the AGN/LINERs/Composites in our sample have somewhat higher stellar masses than the inactive galaxies on average, which could a priori complicate a direct comparison between them. For those galaxies that pass our selection criteria, we find that inactive galaxies are distributed around a median mass of $\log \ M_\star=10.45$ with a spread of $0.37$ dex, while the AGN/LINERs/Composites show a median mass of $\log \ M_\star=10.63$ with $0.25$ dex scatter. Thus, the masses of selected active and inactive are comparable within $1 \sigma$. Importantly, in Figure \ref{fig:Zgrad_lMstar}, we show that within our sample there is no correlation between metallicity gradients and stellar mass either for AGN/LINERs/Composites (\textit{left panel}) or inactive galaxies (\textit{right panel}). Thus, we do not suspect that mass effects play a role in the $\nabla Z - \tdep$ relation in the AGN-like objects in our study.

\subsection{Sanity checks investigating potential unphysical drivers of the relation}
\label{subsec:Sanity_checks}

\begin{figure*}
\centering
\includegraphics[width=0.4\textwidth]{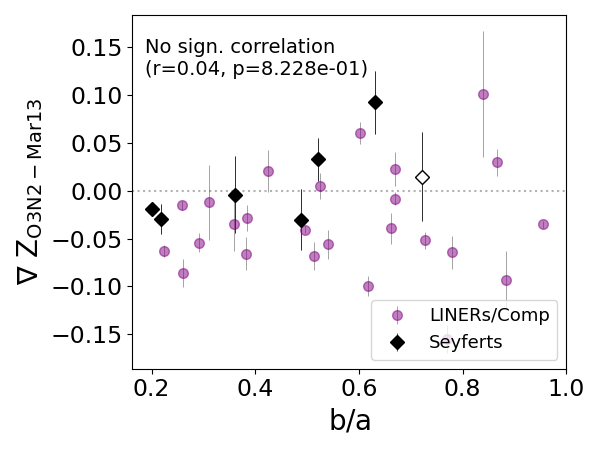}
\includegraphics[width=0.4\textwidth]{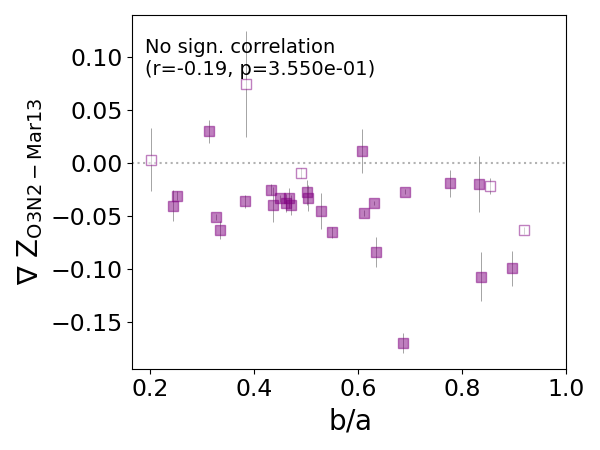}
\caption{Gas-phase metallicity gradients $\nabla Z$ plotted against the minor-to-major axis ratio $b/a$. In the left panel focussing on AGN/LINERs/Composites, Seyferts are highlighted in black. The right panel corresponds to inactive galaxies. The open symbols denote CO-undetected sources. In either case, we do not find any correlation between $\nabla Z$ and $b/a$, which argues against inclination effects affecting our analysis.}
\label{fig:Zgrad_vs_baRatio}
\end{figure*}

\begin{figure*}
\centering
\noindent
\begin{adjustwidth}{-0.7cm}{-0.5cm}
\begin{minipage}{1.12\textwidth}
    \noindent
    \begin{minipage}{0.32\textwidth}
    \centering
    \large \phantom{aaaa} AGN/LINERs/Composites \\[1.5ex]
    \includegraphics[width=\textwidth, trim={0cm 0cm 0cm 0.35cm},clip]{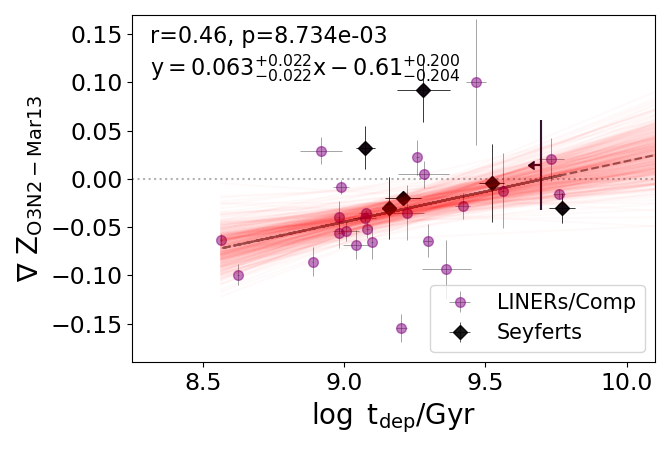}
    \end{minipage}
    \begin{minipage}{0.32\textwidth}
    \centering
    \large \phantom{aaaa} Inactive galaxies \\[1.5ex]
    \includegraphics[width=\textwidth, trim={0cm 0cm 0cm 0.35cm},clip]{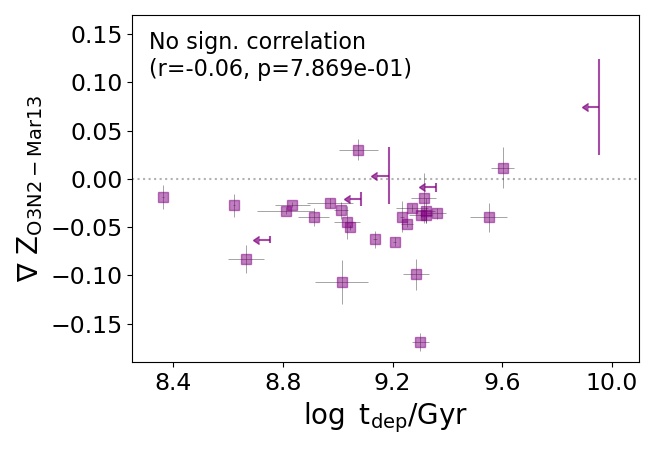}
    \end{minipage}
    \begin{minipage}{0.32\textwidth}
    \centering
    \large \phantom{aaaa} LINERs/Composites \\[1.5ex]
    \includegraphics[width=\textwidth, trim={0cm 0cm 0cm 0.35cm},clip]{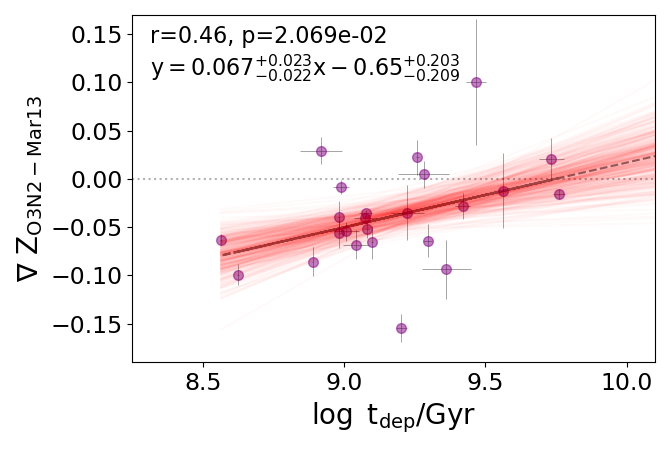}
    \end{minipage}
\end{minipage}
\end{adjustwidth}
\caption{Same as Fig~\ref{fig:main_plot}, but recalculating the depletion time with a constant CO-to-\Hmol\ conversion factor $\alpha_{\rm CO}=4.36 \ {\rm \ M_\odot / (K \ km \ s^{-1} \ pc^{-2}) }$ corresponding to the canonical Milky Way value.}
\label{fig:check_ConstAlphaCO}
\end{figure*}

\begin{figure*}
\centering
\noindent
\begin{adjustwidth}{-0.7cm}{-0.5cm}
\begin{minipage}{1.12\textwidth}
    \noindent
    \begin{minipage}{0.32\textwidth}
    \centering
    \large \phantom{aaaa} AGN/LINERs/Composites \\[1.5ex]
    \includegraphics[width=\textwidth, trim={0cm 0cm 0cm 0.35cm},clip]{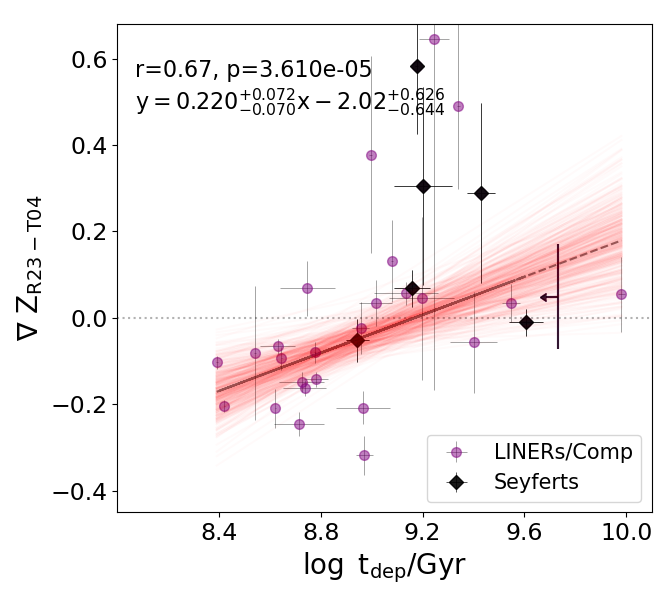}
    \end{minipage}
    \begin{minipage}{0.32\textwidth}
    \centering
    \large \phantom{aaaa} Inactive galaxies \\[1.5ex]
    \includegraphics[width=\textwidth, trim={0cm 0cm 0cm 0.35cm},clip]{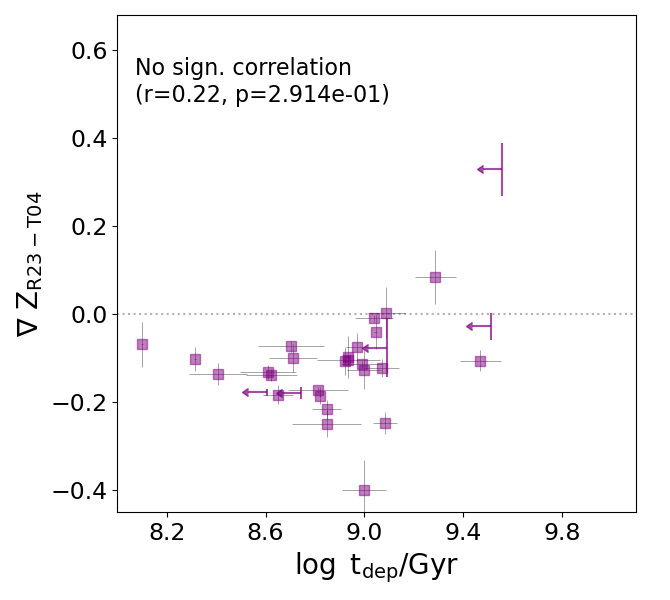}
    \end{minipage}
    \begin{minipage}{0.32\textwidth}
    \centering
    \large \phantom{aaaa} LINERs/Composites \\[1.5ex]
    \includegraphics[width=\textwidth, trim={0cm 0cm 0cm 0.35cm},clip]{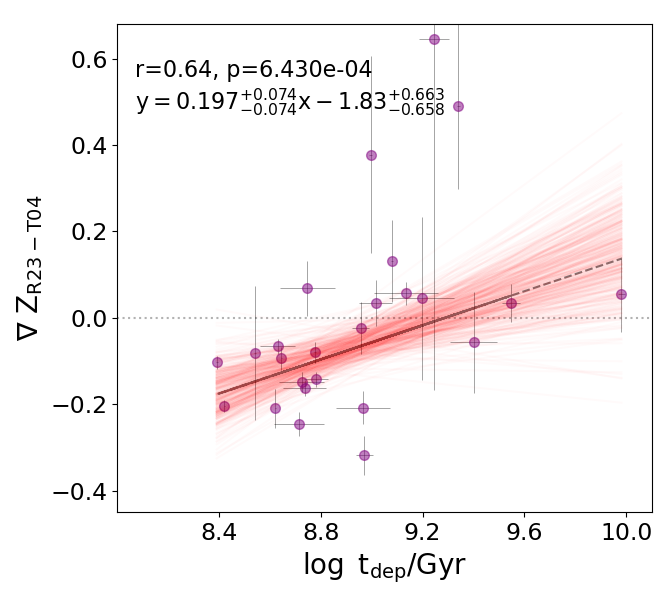}
    \end{minipage}
\end{minipage}
\end{adjustwidth}
\caption{Same as Fig~\ref{fig:main_plot}, but using the Pipe3D gas-phase metallicity gradients based on the \citet{Tremonti2004} R23 calibrator instead of the \citet{Marino2013} O3N2 prescription. }
\label{fig:check_T04}
\end{figure*}

Before we go into a discussion on possible physical interpretations for the $\nabla Z$ - \tdep\ relation presented in Section \ref{sec:Analysis}, we will first test the robustness of these results with respect to unphysical drivers including key aspects of our adopted methodology.

{First, we search for potential inclination effects affecting our analysis in Fig \ref{fig:Zgrad_vs_baRatio} by contrasting the metallicity gradients to the minor-to-major axis ratio $b/a$ for AGN-like objects (left panel; with Seyferts highlighted in black) and inactive galaxies (right panel). CO-undetected galaxies are shown as open symbols. Encouragingly, we clearly do not find any correlation between $\nabla Z$ and $b/a$. We therefore do not expect inclination effects to play a significant role in our analysis.}

Further, it is important to note that in Fig~\ref{fig:main_plot} the x-and y-axes - \tdep and $\nabla Z$ - are not fully independent of each other. The molecular gas masses entering \tdep are derived from our observed CO fluxes using the $\alpha_{\rm CO}$ conversion factor from \citet{Accurso2017} (see Section \ref{subsec:MASCOT}), which depends on galaxy-integrated metallicity (as well as offset from the star-forming Main Sequence). 
It is therefore necessary to check how our results may vary when using a constant $\alpha_{\rm CO}$ factor to derive \Hmol\ masses and associated depletion times. In Fig~\ref{fig:check_ConstAlphaCO}, we use a constant value of $4.36 \ {\rm \ M_\odot / (K \ km \ s^{-1} \ pc^{-2}) }$ for $\alpha_{\rm CO}$ corresponding to the canonical Milky Way value, and plot the resulting $\nabla Z$ - \tdep\ relation in an analogous way to Fig \ref{fig:main_plot}. For the AGN/LINERs/Composites (left panel) and LINERs+Composites (right panel), we still obtain a significant $\nabla Z - \tdep $ relation with similar slopes than in Fig~\ref{fig:main_plot} (as well as no relation for inactive galaxies in the middle panel). {Throughout the rest of the paper, we will continue using the metallicity-dependent $\alpha_{\rm CO}$ factor to derive depletion times and other molecular gas properties.}

As a second check, we verify that our results do not depend entirely on our choice of metallicity prescription. For this, we explore using gas-phase metallicity gradients from the Pipe3D catalog based on the \citet{Tremonti2004} R23 calibration instead of the \citet{Marino2013} O3N2 diagnostic, since the former has been shown to be practically insensitive to the ionisation parameter $U$ in the regime with $\log ({\rm O/H}) + 12 > 8.5$ when evaluated over  $-3.98<U<-1.98$ (\citealt{Kewley2019}). We retrieve a metallicity range of $8.62 < \log ({\rm O/H}) + 12< 8.89$ for all galaxies passing our sample selection. We show the resulting $\nabla Z$ - \tdep\ relation based on the R23 calibrator in Fig~\ref{fig:check_T04}, where the left panel again corresponds to AGN+LINERs+Composites, the middle panel to inactive galaxies, and the right panel to LINERs+Composites. In all cases except for the inactive galaxies, we recover a significant correlation. Interestingly, compared to Fig~\ref{fig:main_plot}, we obtain a significantly steeper slope ($0.22$ versus $0.07 \ {\rm Gyr^{-1}}$). However, we choose to proceed with the metallicity gradients obtained via the \citet{Marino2013} O3N2 method, given that many of our galaxies are affected by DIG and/or LINER emission, in which the robustness of the O3N2 calibrator has been tested as summarised in Section \ref{subsec:choice_O3N2} (while such a test has, to our knowledge, not been conducted for R23).

\section{Discussion}
\label{sec:Discussion}

In this Section, we will discuss various interpretations of the origin of the $\nabla Z - \tdep $ relation presented in Section \ref{subsec:main_relation}.

\subsection{Outflow scenario}
\label{subsec:outflows}

\begin{figure*}
\begin{adjustwidth}{-0.9cm}{-0.7cm}
\centering
\includegraphics[height=0.43\textwidth]{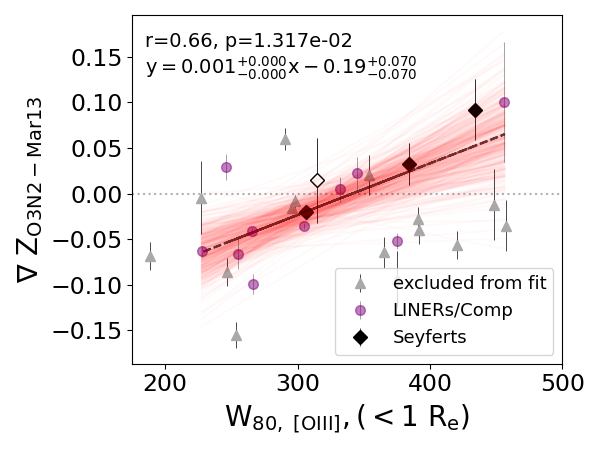}
\includegraphics[height=0.43\textwidth]{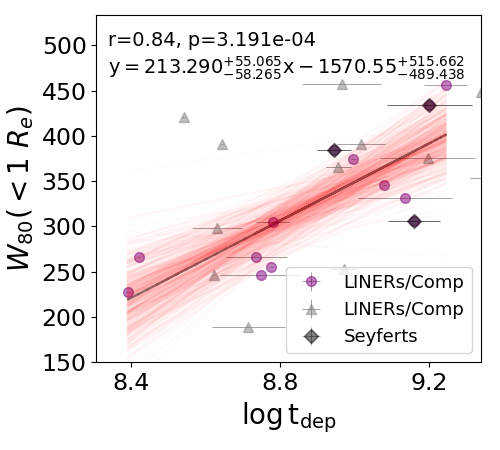}
\end{adjustwidth}
\caption{\textit{Left panel}: Gas-phase metallicity gradients $\nabla Z$ plotted against the mean \oiii\ $W_{80} (< R_e)$ parameter (averaged over all spaxels out to $1 \ R_e$). We find a strong correlation between $\nabla Z$ and average \oiii\ $W_{80}$ in Seyfert AGN (black datapoints) and LINERs/Composites (purple dots). We suggest that this relation emerges via weak chemically-enriched outflows. The targets plotted in grey are excluded from the fit due to exhibiting too many spaxels in which we deem the recovered $W_{80}$ values to be unreliable ($S/N < 10$, `DONOTUSE' flag, or negative flux). Only targets with $>10$ \% of reliable spaxels are included in the fit. The best linear fit is shown as a black solid line, while the red lines show fits recovered when perturbing the datapoints within their uncertainties. The r-value, p-value and coefficients of the best fit are further annotated in the top of the panel. \textit{Right panel}: There is also a strong correlation between \oiii\ $W_{80} (< R_e)$ and depletion times \tdep , such that the AGN-like galaxies with the strongest \oiii\ line broadening are the least efficient at forming stars out of the available molecular gas.
}
\label{fig:Zgrad_vs_W80_0ptRe}
\end{figure*}

{The metallicity gradients within galaxies are responsive to internal galaxy processes as well as gas accretion and feedback.} In this Section we consider the scenario of metal-enriched outflows redistributing metals and facilitating chemical mixing within galaxies. 
In this picture, we hypothesise that the below-MS galaxies with longer depletion times in our Figure \ref{fig:main_plot} show flatter metallicity gradients as they are 
(simplistically) at a later evolutionary stage such that feedback has been operating for a longer time than in those galaxies with short \tdep that are efficiently forming stars. 

\subsubsection{Background}

Observational outflow studies have traditionally had a strong focus on the most luminous AGN and starbursts; in recent years however, ionised outflows have been shown to be present in LINERs (e.g. \citealt{Cazzoli2018, HermosaMunoz2022}), low-luminosity AGN as well as ``typical'' star-forming galaxies {(e.g. \citealt{LopezCoba2019, LopezCoba2020})} though with limited impact as assessed via coupling efficiency or mass outflow rates \citep{Wylezalek2020, Avery2021}. {In particular, in the context of AGN feedback, it has been suggested that it is the integrated effect over time that makes galaxies quench since black hole mass estimates have been found to hold the most predictive power over galaxies' quenching status \citep{Bluck2020,Piotrowska2021}. In this sense, outflows below the MS - despite being likely weak - could be the ``final straw'' that pushes galaxies towards quiescence after having been subjected to gas removal/heating on long timescales via repeated energy injection.}
In the context of stellar feedback, from a simulations perspective, \citet{Gibson2013} find that even weak SN feedback leads to efficient gas mixing on cosmological timescales as manifested by a flattening of metallicity gradients towards low redshifts in the MUGS suite \citep{Stinson2010} of cosmological zoom-in simulations, while stronger feedback produced flat metallicity gradients throughout cosmic time. The adiabatic blastwave model feedback used in these results operates via thermal energy injection into the gas surrounding stellar particles (with the ``strong'' model given twice the heating power of the ``weak'' model, plus radiation feedback). It therefore does not necessarily manifest in terms of strong outflows and may thus be challenging to constrain observationally, but we will nevertheless explore a quick kinematic analysis to see if any outflow signatures can be found. The importance of AGN feedback has further been highlighted in Illustris TNG50 simulations, where \citet{Hemler2021} argue that it likely contributes significantly to the shallow metallicity gradients recovered at low redshift (next to stellar feedback), especially in high-mass galaxies where kinetic mode AGN feedback kicks in. {In EAGLE galaxies, \citet{Trayford2019} also demonstrate that in mock spaxels with high stellar mass surface density $\Sigma_\star$, switching on AGN feedback leads to substantially lower local metallicities (their Fig 7). The strength of the effect was found to depend on $\Sigma_\star$, such that within a given galaxy, one would expect the denser central regions to be most affected.}

Metal-enriched outflows on galactic scales have been invoked across redshifts to reproduce the observed chemistry of the circum-galactic medium (see the review by \citealt{Tumlinson2017}). 
At low redshift, there have also been observational hints that feedback mechanisms may play a strong role in shaping the metallicity distribution within galaxies. For instance, \citet{Baker2022b} investigate the predictors of resolved metallicities in star-forming MaNGA galaxies, and find intringuingly that there is a stronger anti-correlation with global SFR than with the local SFR (though both local and global quantities matter), which is broadly consistent with large-scale stellar feedback-driven winds ejecting or redistributing metals. \citet{Avery2021} further find evidence for an enhanced dust attenuation within the outflows of MaNGA galaxies, which could potentially be traced back to metal enrichment, while \citet{Chisholm2018} confirmed the high level of metal enrichment in the outflows of seven local star-forming galaxies using HST/COS spectra.

\subsubsection{Kinematic analysis}

In light of the considerations above, as a first step, we are searching for kinematic signatures of ouflows in the \oiii\ line tracing ionised gas. First, we use the \oiii\ maps from the MaNGA DAP and impose a signal-to-noise threshold ${\rm S/N > 3}$ \footnote{while discarding spaxels with a ``DONOTUSE'' flag, as well as negative fluxes}. We further discard galaxies in which less than $10$ \% of spaxels fulfill the aforementioned criteria. In each selected spaxel, we then calculate $W_{80}$, defined as the spectral width that encompasses $80$ percent of the total line flux. For this purpose, the line profiles are fit with a single or double Gaussian, where a second component is only permitted if it has a signal-to-noise ${\rm S/N > 3}$ to prevent noise fitting. Skylines are masked. We then calculate the weighted average $W_{80}$ value over all spaxels falling within a circular aperture of radius $0.1  R_e$, where each spaxel is weighted by the fraction of its area that falls into the annulus. 
The methodology will be described in more detail in an upcoming paper investigating kinematic signatures in different types of AGN \citep{Alban2022}.

In the left panel of Figure \ref{fig:Zgrad_vs_W80_0ptRe}, we first attempt to identify indications of kinematic feedback by AGN, stars or a combination thereof in the more central regions by contrasting the metallicity gradients to the mean \oiii\ $W_{80}$ within $1 \ R_e$ for our sample of AGN-like galaxies. Grey dots show the location of galaxies that were discarded due to less than $10$ \% of their spaxels fulfilling our selection criteria outlined above, and are excluded from the correlation analysis. Black datapoints denote Seyfert AGN, while LINERs/Composites are shown in purple. There is a significant correlation of central \oiii\ line width with $r=0.66$ and $p \sim 0.013${, such that the AGN-like systems with broader lines have flatter or positive metallicity gradients. Given the link we found between $\nabla$ Z and depletion time, we expect $W_{80} (< 1\ R_e)$ to correlate with \tdep\ as well, which we confirm in the right panel of Fig \ref{fig:Zgrad_vs_W80_0ptRe}} ($r=0.84$, $p<0.001$). {For both relations in Fig \ref{fig:Zgrad_vs_W80_0ptRe}, we draw 1000 bootstrapped samples from the original datapoints with replacement and calculate 1000 associated r-values, and verify that the $95 \% $ confidence intervals on $r$ do not include zero (Appendix \ref{app:bootstrap})}. We also check that the $\nabla Z - W_{80} (< 1 \ R_e)$ relation persists when using a different metallicity diagnostic or a constant CO-to-$\Hmol$ conversion factor (Appendix \ref{app:W80_check_T04}).

We interpret these findings as a strong indicator of a causal link between outflows {(which could be stellar or AGN-driven)} and the chemical distribution in our sample of AGN-like objects, as well as star-formation efficiency. We argue again that although one may naïvely expect more massive galaxies to have broader $W_{80}$ values, the trend between $\nabla Z$ and $W_{80} (< 1 R_e)$ cannot be explained by obvious mass effects, given that we found no relation between $\nabla Z$ and stellar mass in Fig \ref{fig:Zgrad_lMstar} (Section \ref{subsec:Sanity_checks}) {, although we will discuss the influence of the gravitational potential further in the paragraph below}. For now, we highlight that we checked that the same trend is not seen in inactive galaxies (which should in principle show potential well effects similar to the AGN-like objects). We note that this finding does not necessarily imply that the outflows are traced back to {AGN feedback with no contribution from stars}. We also note that the $W_{80}$ values fall below $<500 \ {\rm \kms}$ and would thus not be regarded as outflow signatures in studies of galaxies on the Main Sequence. {However, below the Main Sequence we expect signatures to weaken for both stellar and AGN feedback, as AGN fuelling decreases given a shortage of gas, which is also reflected in the fact that the majority of our below-MS AGN-like objects are LINERs/Composites rather than Seyferts.} As outlined above, weak feedback may still play a significant role in this regime. We thus favour an interpretation in which weak outflows may not be trivial to detect when using line widths as a tracer, but still exist. 

{Before expanding the kinematic analysis, we first address the point that emission line widths are not a pure tracer of outflows, but are expected to also respond to the potential well. The MaNGA spaxels in which we derive \oiii\ $W_{80}$ values sample a finite range in radii and thus a spread of orbital velocities, and beam smearing may further complicate the picture, with both of these effects being a function of inclination. The correlation seen in Fig \ref{fig:Zgrad_vs_W80_0ptRe} above is therefore likely influenced to some extent by potential well effects. 
All other factors being equal, galaxies with a higher dynamical mass within the central 1 $R_e$ would experience a stronger line broadening (assuming a fixed spaxel size). According to the resolved mass-metallicity relation {\citep{RosalesOrtega2012, Sanchez2013, BarreraBallesteros2016}}, more concentrated galaxies would broadly be expected to have higher central metallicities (though with a large scatter), which may simplistically translate into more negative metallicity gradients. In this case one may naïvely expect an anti-correlation between $\nabla Z$ and $W_{80}$ in active and inactive galaxies alike, unlike the positive correlation we find in Fig \ref{fig:Zgrad_vs_W80_0ptRe} in AGN-like objects only. Further, we found no mass dependence nor inclination dependence of the metallicity gradients (Figs \ref{fig:Zgrad_lMstar} and \ref{fig:Zgrad_vs_baRatio}). In Section \ref{subsec:morphology} {as well as the appendix \ref{app:morph_app}} focussing on morphological tracers, we will see that we do not find evidence for any correlation between $\nabla Z$ and the stellar mass surface density (Fig \ref{fig:Zgrad_vs_mustar}){, nor with the central stellar or \Halpha\ velocity dispersions (top row of Fig \ref{fig:Zgrad_vs_BTratio})}. In the Appendix (Section \ref{app:sanity_checks_w80}), we search more directly for potential dependencies of the $W_{80}$ parameter, but find no evidence for any trend with stellar mass, stellar mass surface density or inclination (Fig \ref{fig:W80_vs_baratio}). We therefore interpret \oiii\ $W_{80}$ as being primarily a tracer of outflows.
}

\begin{figure*}
\centering
\begin{adjustwidth}{-0.5cm}{-0.5cm}
\includegraphics[width=0.33\linewidth]{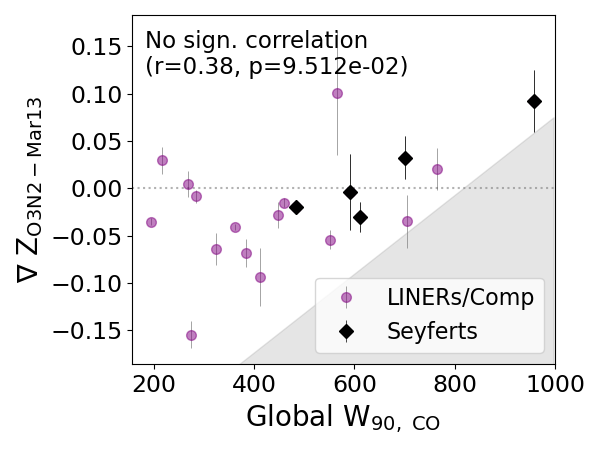}
\includegraphics[width=0.33\linewidth]{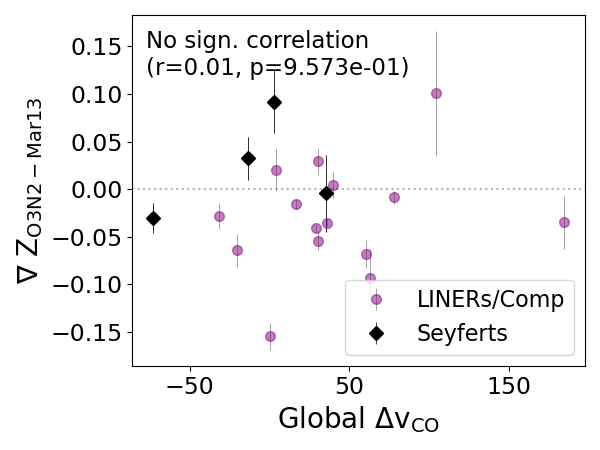}
\includegraphics[width=0.33\linewidth]{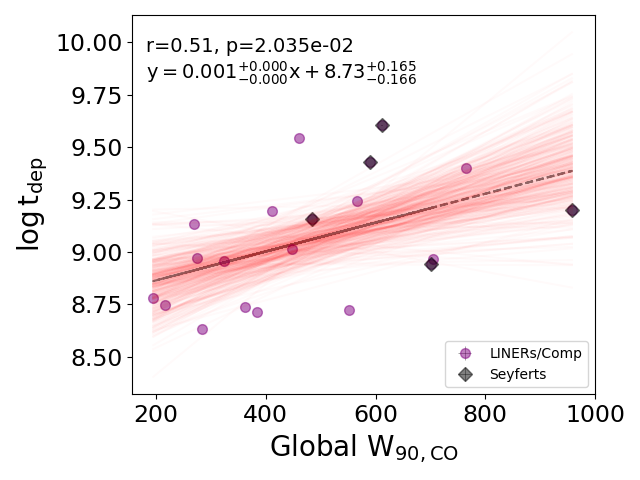}
\end{adjustwidth}
\caption{\textit{Left panel}: Metallicity gradients plotted against the $W_{90}$ measurement of the CO line width from the galaxy-integrated CO profiles of our AGN, LINERs and Composites. \textit{Middle panel}: Metallicity gradients plotted against the median velocity offset between the peak of our galaxy-integrated CO profiles and the frequency predicted for a CO emitter at the exact systemic velocity of a given galaxy (i.e. $115.27$ GHz at rest-frame). \textit{Right panel}: Depletion times contrasted against the global CO $W_{90}$ values. While we find a possible correlation which could hint at molecular outflows reducing the star-formation efficiency, we caution that given the unresolved character of the CO measurements, the datapoints with CO $W_{90} < 500 \ {\rm km s^{-1}}$ are likely driven to a significant extent by the potential well. If one were to remove that contribution, all datapoints would be shifted to lower $W_{90}$, especially the low-$W_{90}$ ones. }
\label{fig:Zgrad_vs_COW90}
\end{figure*}

To explore a possible impact of feedback (AGN or stellar) on global scales, we contrast the metallicity gradients against the galaxy-integrated \oiii\ $W_{80}$ (median over all selected spaxels) for our AGN/LINERs/Composites. In this case we report that we do not find any evidence for a correlation ($p=0.06$; omitted from Fig \ref{fig:Zgrad_vs_W80_0ptRe}), though we note that that the possibility of a delayed impact cannot be excluded.

We then proceed to study the cold gas kinematics of the AGN-like objects in Fig~\ref{fig:Zgrad_vs_COW90}. In the left panel, we plot the metallicity gradients as a function of the $W_{90}$ value obtained from our galaxy-integrated velocity profiles, i.e. the width containing $90$ \% of the CO line flux {(which is more sensitive than $W_{80}$ for narrow CO line profiles with coarse spectral bins spanning $\sim 50 \rm km \ s^{-1}$)}. The CO line is fitted by a single or double Gaussian as required, as outlined in \citet{Wylezalek2022}. We do not find any significant relation, although we note an ``empty triangle'' region shaded in grey such that no objects fall into the regime with large $W_{90}$ and substantially negative $\nabla Z$. {We note that the galaxy-integrated character of our CO measurements implies that the CO $W_{90}$ is expected to be dominated by the potential well for those objects showing low line widths. Those datapoints with $W_{90} \gtrsim 500 \ {\rm km \ s^{-1} }$ however are unlikely to be solely driven by the rotation of the galaxy, and may exhibit potential indications for a $\nabla Z - W_{90}$ correlation, which is however not possible to verify given the low number statistics in this regime. 
We further find no relation in the middle panel of Fig~\ref{fig:Zgrad_vs_COW90} when contrasting the metallicity gradients to the {velocity offset between our observed CO line peak and the line predicted for a CO emitter at the exact systemic velocity of a given galaxy (i.e. $115.27$ GHz at rest-frame)}. Finally, we note that when contrasting the CO $W_{90}$ measurements against depletion times (instead of metallicity gradients as in the first panel)}, a potential relation arises, but we caution again that especially the systems with low $W_{90}$ are expected to be strongly driven by the spread in orbital velocities contained within the entire galaxy (with gas moving both towards and away from an observer). If the relation was shown to persist when removing the contribution of the rotation curve to $W_{90}$ e.g. by modelling galaxies as inclined rotating disks (which should strongly decrease the low-$W_{90}$ values in particular), this would be evidence for molecular gas outflows reducing the star formation within AGN-like galaxies. However, we defer a more in-depth study of the CO kinematic profiles to a future analysis.
We further checked that in inactive galaxies, there is no relation between global cold gas kinematics and metallicity gradients or $t_{\rm dep}$. We note that our single-dish observations do not allow us to explore resolved kinematics of the neutral gas phase, and molecular outflow signatures are more challenging to detect on a global scale. 

To conclude this section, considering the \oiii\ ionised gas kinematics in AGN-like galaxies in Fig~\ref{fig:Zgrad_vs_W80_0ptRe} we suggest that low-velocity outflows {(of stellar or AGN origin)} below the Main Sequence could plausibly drive chemical mixing in AGN/LINERs/Composites while prolonging depletion times (i.e. reducing star-forming efficiency) and possibly driving those galaxies towards quiescence in a ``last straw'' scenario. 
{Our findings raise the question of how metal-mixing timescales compare to AGN lifetimes. AGN have been suggested to ``flicker'' on and off in many consecutive short cycles of $\sim 10^5 \ {\rm yr}$ amounting to a total lifetime of $10 \ {\rm Myr}- 1 {\rm Gyr}$ as estimated from e.g. the Soltan argument \citep{Schawinski2015}. However, it is not clear a priori how the outflow strength in AGN varies across their lifetime. Our naïve assumption is that the galaxies with stronger outflow signatures in Fig~\ref{fig:Zgrad_vs_W80_0ptRe} (and flatter/positive metallicity gradients) are ``older'' to allow more time for metal redistribution, though the exact time required for that is also challenging to constrain. For instance, using a toy model describing metal production based on the observed SFR density profiles of $z \sim 1$ galaxies, \citet{Simons2021} argue that in the absence of any redistribution mechanism, galaxies would develop rapidly declining metallicity gradients incompatible with observations on timescales of $\sim 10 - 100 \ \rm Myr$, which may thus be regarded as an upper limit to metal mixing timescales. }

\subsection{Connection to fluctuations in local star-forming activity} 
\label{subsec:SF}

It is well established that the global metallicities of galaxies are linked to both global stellar mass and SFR (the ``fundamental metallicity relation''; \citealt{Ellison2008}, \citealt{Mannucci2010}). It has further been proposed that this global relation arises from local scalings between the resolved stellar mass, SFR and oxygen abundance. {In particular, \citet{Sanchez2021a} show mathematically that once a local relation is in place, a global one naturally follows.} \citet{BarreraBallesteros2016} confirm the existence of a local mass-metallicity relation {(see also \citealt{RosalesOrtega2012, Sanchez2013}), while they find no evidence for a secondary relation with SFR surface density $\Sigma_{\rm SFR}$, similarly to \citet{AlvarezHurtado2022}}. On the other hand, \citet{SanchezAlmeida2019} and \citet{Baker2022b} report a local anti-correlation between $\Sigma_{\rm SFR}$ and metallicity at fixed mass in MaNGA galaxies (see also the review by \citealt{Maiolino2019}). This has been attributed to localised inflows fuelling star formation whilst diluting the enriched gas, though \citet{Baker2022b} highlight the even stronger importance of the global SFR in setting local metallicities. The significance of the global SFR could possibly be attributed to the transport of metals by galactic-scale winds driven by stellar feedback.

\begin{figure*}
\centering
\begin{minipage}{\textwidth}
\begin{adjustwidth}{-0.7cm}{-0.5cm}
\includegraphics[height=0.22\textwidth]{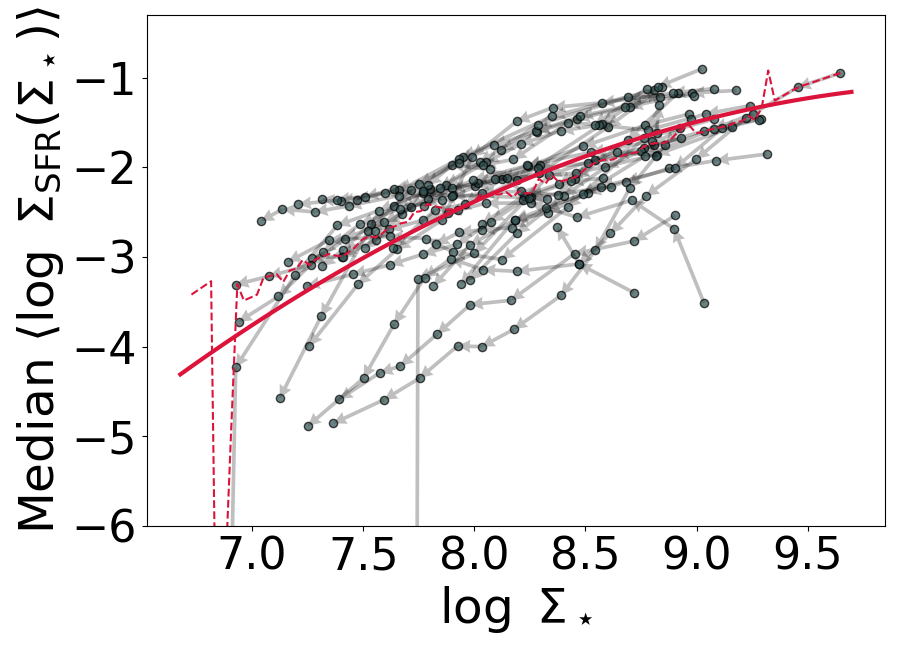}
\hspace*{0.3cm}
\includegraphics[height=0.235\textwidth, trim={0cm 10cm 19cm 0cm},clip]{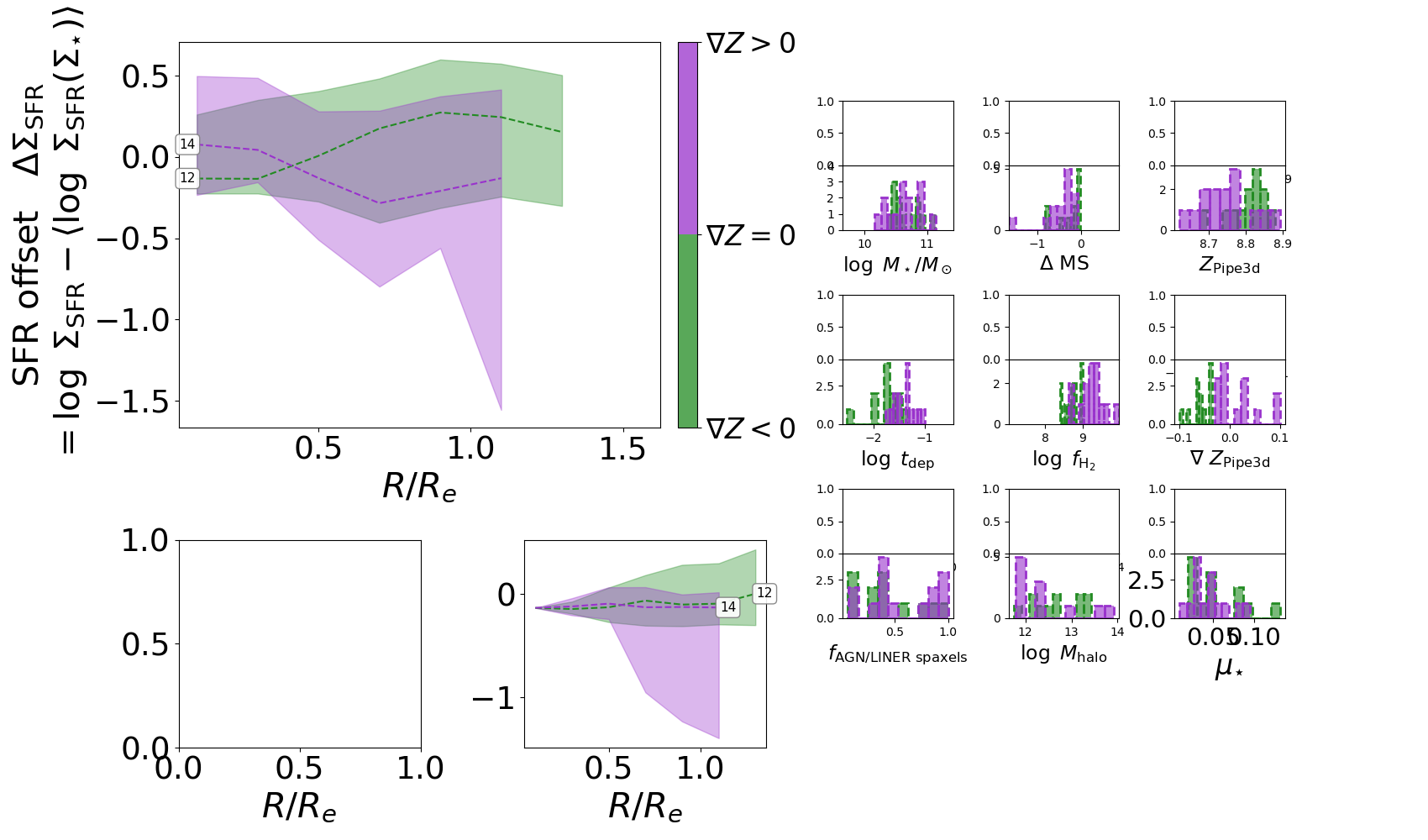}
\hspace*{0.3cm}
\includegraphics[height=0.235\textwidth, trim={0cm 10cm 19cm 0cm},clip]{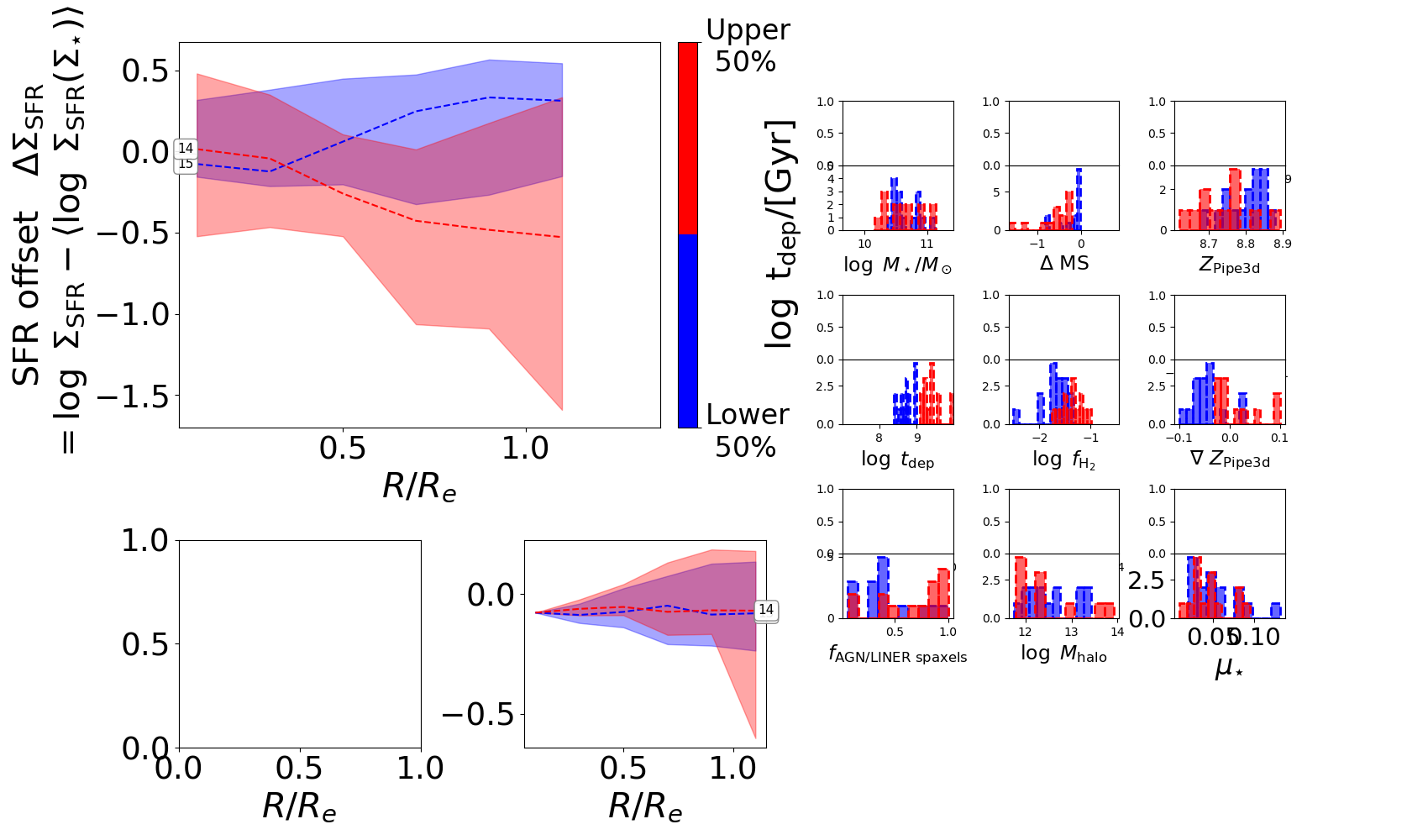}
\end{adjustwidth}
\end{minipage}
\caption{\textit{Left panel}: The individual dots show the $\Sigma_{\rm SFR}$-$\Sigma_\star$ distribution of our AGN/LINERs/Composites sample using annuli of $0.2 \ R_e$ width, with the red solid line showing the median SFR surface density $\langle \Sigma_{\rm SFR} (\Sigma_\star)\rangle$ for a given mass. Each grey background curve corresponds to a single galaxy, with each data point on the curve corresponding to an individual annulus, such that the arrows show the path taken in the $\Sigma_{\rm SFR} - \Sigma_{\star}$ plane when going from the centre-most annulus to the outermost one. \textit{Right panel}: Radial profiles of the offset $\Delta \Sigma_{\rm SFR}$ from the median of the $\Sigma_{\rm SFR}$-$\Sigma_\star$ distribution spanned by our sample (solid line in top panels), i.e. offset between SFR surface density $\Sigma_{\rm SFR}$ within a given annulus and the median SFR surface density $\langle \Sigma_{\rm SFR} (\Sigma_\star)\rangle$ evaluated at the stellar mass contained within the annulus. In each panel, the sample is sub-divided into galaxies with longer/shorter depletion times (upper/lower 50th percentile of \tdep ), as indicated by red/blue colouring. The dashed lines show the median profile in a given \tdep\ bin, with associated spread illustrated by the shaded regions. The AGN-like objects with long depletion times show a similar central star-forming activity at fixed mass than those with short depletion times, but are comparatively more passive in the outskirts. }
\label{fig:SigSFR_Offset_profiles}
\end{figure*}

In this section, we explore the possibility that the $\nabla Z - \tdep$ relation may arise due to variations in local in-situ star formation which simultaneously link to the resolved metallicities and affect the local depletion time $\tdep = M_\Hmol / {\rm SFR}$. To evaluate variations in SFR whilst controlling for stellar mass, we derive the radially resolved $\Sigma_{\rm SFR}$-$\Sigma_\star$ distribution for our below-MS sample as follows: We subdivide each galaxy in our sample into annuli of $0.2 \ R_e$ width, and run our full spectral fitting procedure (described in section \ref{subsec:OwnFitting} along with a comparison between our recovered global $\rm SFR$ \& $M_\star$ and the Pipe3D reference values) on the annular spectra. We thus recover the SFR surface density and stellar mass surface density in individual annuli, which we show as a scatter plot in the left panel of Fig~\ref{fig:SigSFR_Offset_profiles} for our AGN-like objects, with the median baseline shown as a red solid line. {Both quantities have been deprojected.}

We then proceed to calculate, for each AGN/LINER/Composite, the radial profile of the offset $\Delta \Sigma_{\rm SFR}$ between $\Sigma_{\rm SFR}$ and the median value at fixed mass ($\langle \Sigma_{\rm SFR} (\Sigma_\star)\rangle$) in our sample, and analyse how they vary as a function of depletion time. In the middle panel of Fig~\ref{fig:SigSFR_Offset_profiles}, we classify the resulting profiles in two bins according to their metallicity gradient, as indicated by the colour (green/purple = negative/positive $\nabla Z$). The dashed lines indicate the median profile in each $\nabla Z$/\tdep\ bin, and the shaded regions indicate the spread between profiles. While the spread in profiles is large, there may be a potential indication of elevated star formation (while controlling for mass) in the outskirts of AGN-like galaxies with negative metallicity gradients compared to those with $\nabla Z > 0 $, while their centre is already quenched. Given the relation we found between $\nabla Z$ and \tdep\, we also inspect the role of the depletion time in setting the $\Delta \Sigma_{\rm SFR}$ profiles in the right panel of Fig ~\ref{fig:SigSFR_Offset_profiles}. The colour-coding reflects a binning into the upper and lower 50th percentile of \tdep values (blue/red = short/long $t_{\rm dep}$). We find a trend with \tdep such that objects with longer depletion time (corresponding to positive metallicity gradients) show a suppression of star formation in their outskirts compared to objects with short \tdep\ whilst exhibiting similar levels of star formation in the centre. Within the context of the resolved fundamental metallicity relation, the drop in SFR in the outskirts may drive higher local metallicities at large radii (e.g. via a lack of diluting \Hmol\ gas), thus flattening or inverting gradients compared to AGN/LINERs/Composites with short $t_{\rm dep}$. 

It is interesting to note that despite our results on outflows in the previous Section \ref{subsec:outflows}, the more quenched galaxies with long depletion times and positive $\nabla Z$ are  more suppressed in star formation in the outskirts than in the centre. The weak character of the outflows may indicate that while they can mix gas efficiently within galaxies, they do not permanently remove it from their centre. In fact, the deficit of star formation in the outskirts is more broadly in line with e.g. radiative feedback scenario heating up the surrounding medium to prevent new inflows (see \citealt{Brennan2018} for a prediction of preventative feedback restricting the accretion of diluting gas in cosmological zoom-in simulations). For inactive galaxies (omitted from Fig~\ref{fig:SigSFR_Offset_profiles}), we could not easily identify any trend with $\tdep$ in the shape of $\Delta \ \Sigma {\rm SFR}$ profiles which exhibited large spreads.

To summarise, we find a tentative link between positive metallicity gradients and suppressed star formation in the outskirts in AGN-like galaxies, which further coincides with long depletion times. Thus the metallicity gradients of our sample of AGN-like objects may indeed be partially explained by a connection to in-situ star formation, though our results on the importance of outflows in the previous Section \ref{subsec:outflows} suggest that this is not the whole picture. We conclude for now that in AGN-like objects, the observed $\nabla Z - \tdep$ relation may arise as a combination of at least two drivers. We proceed to consider other factors (inflows, mergers and morphological factors) in the following sections.

\subsection{Inflow scenario}
\label{subsec:inflows}

\begin{figure*}
\begin{adjustwidth}{-0.7cm}{-0.5cm}
\centering
\noindent
\begin{minipage}{1.12\textwidth}
\includegraphics[height=0.25\textwidth]{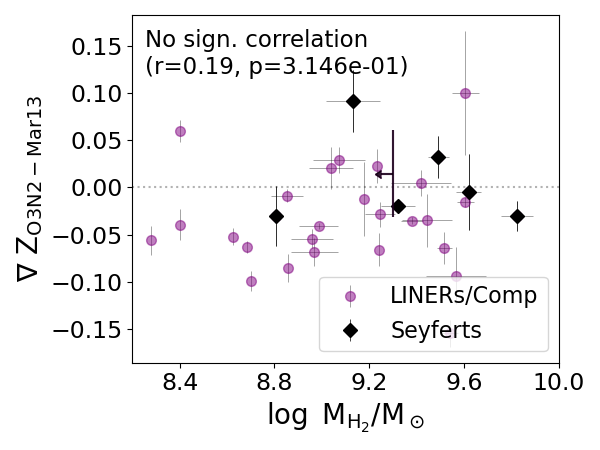}
\includegraphics[height=0.25\textwidth]{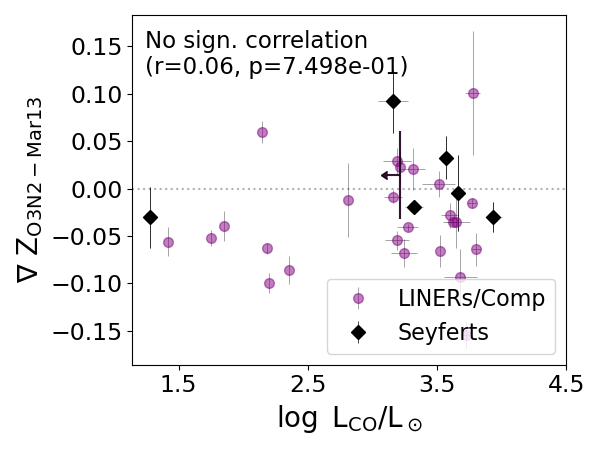}
 \includegraphics[height=0.245\textwidth, trim={0cm 0cm 0cm 0.3cm},clip]{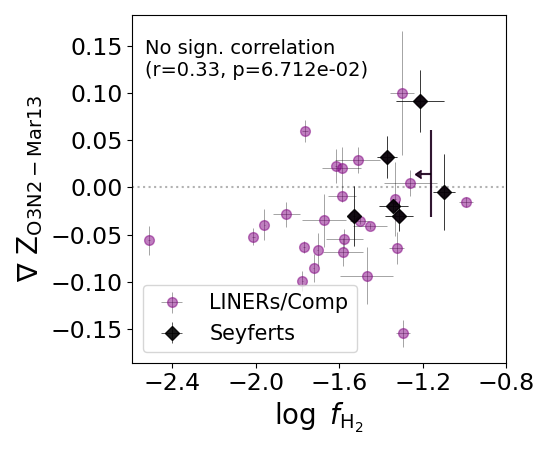}
\end{minipage}
\end{adjustwidth}
\caption{Metallicity gradients plotted against molecular gas mass $M_\Hmol$ (left) and CO luminosity (middle) and molecular gas fraction $M_\Hmol / M_\star$ (right). We do not see any correlation in any of these planes.}
\label{fig:Zgrad_vs_lMH2_LCO}
\end{figure*}

In this Section we consider the possibility of radial inflows of pristine/metal-poor gas funneled directly to the central regions of galaxies causing flatter/positive metallicity gradients by diluting the central regions in those galaxies that have long depletion times. For nearby galaxies on the MS, \citet{Lutz2021} found local O3N2-based metallicities to be correlated with the galaxy-integrated \textit{atomic} gas fraction. We note that the inflow scenario may not be as straight-forward for the galaxies in our sample, given that these all fall below the Main Sequence by selection and one may therefore in general expect them to be more gas-deficient (e.g. \citealt{Saintonge2017}). Nevertheless, it is not impossible for galaxies below the MS to contain a significant amount of gas without star formation being fuelled (though this appears to be more often the case for atomic hydrogen, see e.g. \citealt{Janowiecki2020}). For instance the gas may be stabilised against fragmentation and collapse via morphological transformation \citep{Martig2009}. In Section \ref{subsec:morphology}, we will discuss potential impacts of morphology in more detail. 
Here we explore instead how the metallicity gradients behave as a function of CO luminosity, \Hmol\ mass and \Hmol\ fraction for our sample to check if there is any direct evidence of recent gas inflows manifesting via an enhanced amount of gas. 

In Figure \ref{fig:Zgrad_vs_lMH2_LCO}, we contrast $\nabla Z$ to the molecular gas mass $M_\Hmol$ in the left panel, to the CO luminosity $\rm L_{CO}$ in the middle panel and the molecular gas fraction $\fHmol = M_{\rm H_2} / M_\star$ in the right panel. In either case, we do not find any evidence for a correlation. 
For inactive galaxies omitted from Fig~\ref{fig:Zgrad_vs_lMH2_LCO}, we similarly do not find any evidence for inflows driving the observed metallicity gradients. We note however that localised inflows that do not propagate to galaxy-wide scales would be challenging to detect given our unresolved CO observations. Within this context, interestingly, a recent statistical study of the resolved fundamental metallicity relation in MaNGA galaxies \citep{Baker2022b} found that the strongest predictors of local metallicity were not the local SFR surface density (and local stellar mass surface density), but instead the global SFR and global stellar mass. The fact that the local SFR is only of secondary importance in setting local metallicities {(see also \citealt{AlvarezHurtado2022})} argues also against localised inflows being a primary driver (as inferred via the resolved KS relation; \citealt{Kennicutt1998a, Schmidt1959}).

\subsection{Merger scenario}
\label{subsec:mergers}

Aside from in-situ chemical evolution and metal transport due to inflows or feedback processes, the metallicity distribution within galaxies may be affected by merging events during their lifetime. 
Galaxy interactions are thought to flatten metallicity gradients as a consequence of chemical mixing as well as metal-poor gas inflows reaching the centre. This behaviour has indeed been observed e.g. in LIRGs at various merger stages \citep{Rich2012}, as well as optically- and IR-selected early-stage spiral–spiral interactions \citep{Rupke2010}.
On the other hand, in a sample of 36 post-merger MaNGA galaxies, the local metallicity has been found to be suppressed in the outskirts while not being significantly affected in the central regions \citep{Thorp2019}.

While we have already attempted to exclude ongoing mergers from our analysis by identifying them visually from the SDSS multi-band images as noted in Section \ref{sec:Analysis}, it is still possible for our sample to include post-merger galaxies which are no longer identifiable via morphological signatures. 
Therefore, while a full kinematic axis analysis is beyond the scope of this work, we perform a simple visual inspection of the stellar and gas-phase velocity maps taken from the {MaNGA DAP with hybrid binning (\citealt{Belfiore2019})\footnote{In the hybrid binning scheme, the emission line maps are providing line measurements in individual spaxels (while the stellar continuum is analysed in Voronoi bins)}} to identify galaxies that show potential signs of a misalignement between their stellar and gas-phase dynamics manifesting in different kinematic major axes of rotation. We point out that even this simple exercise can become challenging in the below-MS regime due to the weakness of the lines required to derive velocity maps. In general, we conclude that for the majority of our sample, the kinematics of the stellar and gas-phase components are consistent with rotating in an aligned manner. However, we flag a number of objects that show \textit{potential} signs of misalignement. In detail, we verified that our results are not strongly affected when removing the AGN-like galaxies with the following IDs: 8588-6101, 8595-3703, 8084-6103. Updated versions of Figs~\ref{fig:main_plot} and \ref{fig:Zgrad_vs_W80_0ptRe} excluding the targets listed above are presented in the appendix (Fig~\ref{fig:GradZ_vs_ltdep_Exclude_someVel}).

Finally, it is still an open question for how long the optical and kinematic signature of mergers remain detectable. {For instance, using \texttt{GADGET-3/SUNRISE} major merger simulations to construct mock velocity and velocity dispersion fields at the resolution of MaNGA, \citet{Nevin2021} broadly retrieve kinematic observability timescales of $0.9-6 \ {\rm Gyr}$. On the other hand, \citet{Lotz2008} find that post-merger galaxies may become undetectable after just $200$ Myr depending on merger parameters.} We cannot exclude that long-term effects from past mergers affect our sample, but one may argue that any such effects on the gas-phase metallicities would be ``washed out'' again by the other factors of influence that impact the chemical reservoir more imminently: in-situ chemical evolution, inflows, and outflows. {As mentioned in Section \ref{subsec:outflows}, while studies constraining the timescale for metal redistribution are yet sparse, \citet{Simons2021} argue that it should be limited to $\lesssim 100$ Myr, as otherwise galaxies may rapidly develop metallicity gradients more negative than observed (as predicted by modelling metal production based on observed SFR surface density profiles in $z \sim 1$ 3D-HST galaxies).}

\subsection{Morphological scenario}
\label{subsec:morphology}

\begin{figure*}
\centering
\begin{adjustwidth}{-0.7cm}{-0.5cm}
\includegraphics[height=0.35\textwidth, trim={0cm 0cm 0cm 0.3cm},clip]{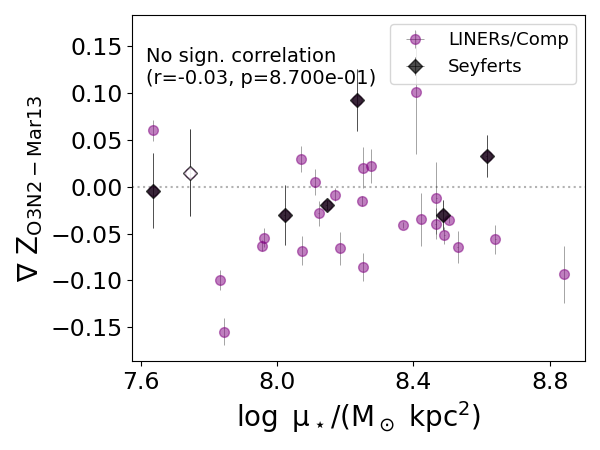}
\includegraphics[height=0.37\textwidth, trim={1cm 10cm 20cm 0},clip]{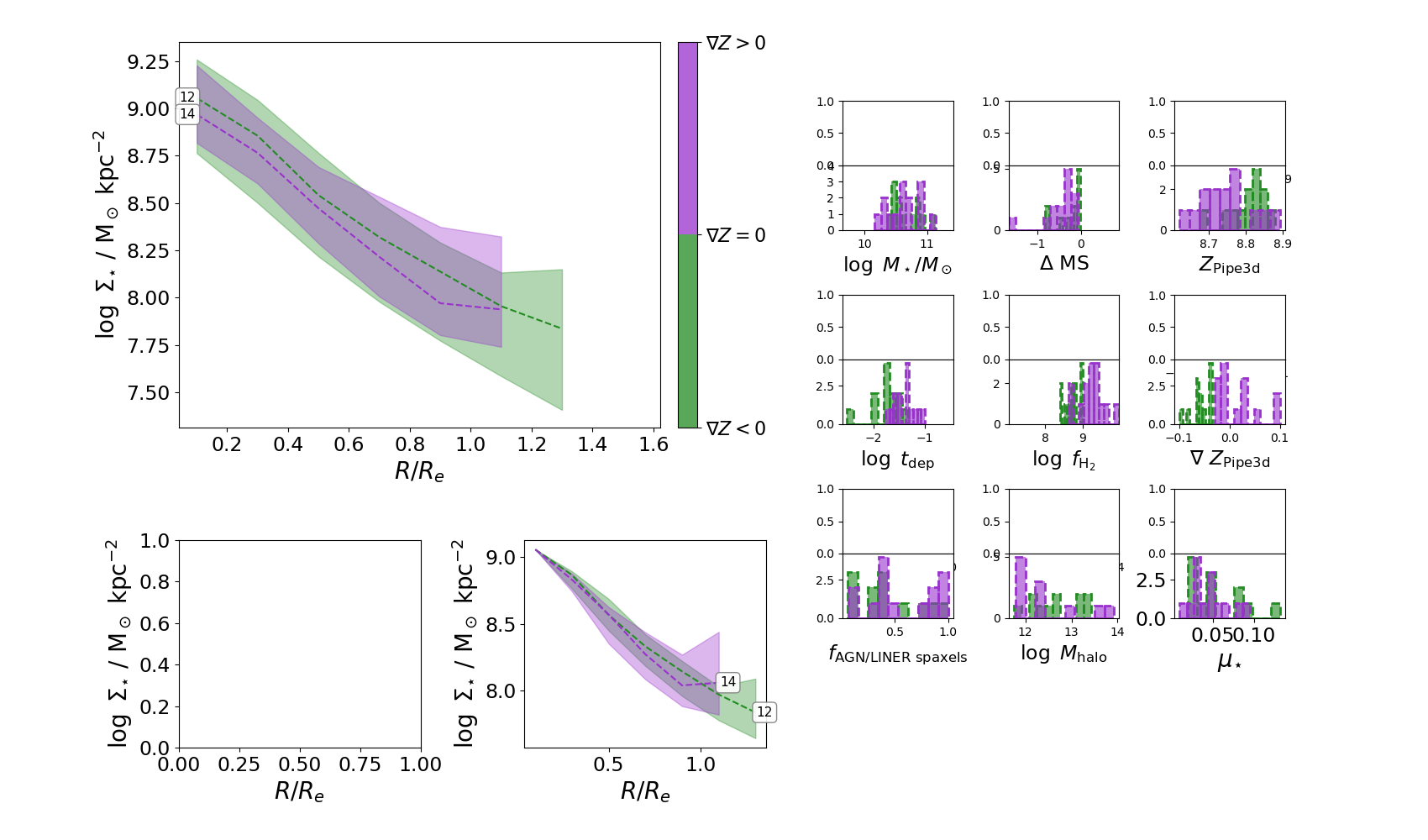}
\end{adjustwidth}
\caption{\textit{Left panel}: Metallicity gradients plotted against global stellar mass surface density $\mu_\star = 0.5 \cdot M_\star/(\pi {R_e}^2)$ as a tracer of morphology for AGN, LINERs and Composites. We find no correlation. \textit{Right panel}: Median profiles of stellar mass surface density $\Sigma_\star$ as a function of galacto-centric radius in units of the half-light radius $R_e$. The colour-coding refers to bins of $\nabla Z$ where objects were divided into a bin with $\nabla Z>0$ (purple) and one with $\nabla Z<0$. The median profiles in each $\nabla Z$ bin are shown as dotted lines with associated spread illustrated by the shaded regions. Again, we find no link between $\nabla Z$ and stellar mass surface density profiles.}
\label{fig:Zgrad_vs_mustar}
\end{figure*}

As mentioned already in Section \ref{subsec:inflows}, morphological effects may influence the gas properties and metal distribution within galaxies. For example, in a morphological quenching scenario \citep{Martig2009}, the gas disk would be stabilised against fragmentation and galaxies could thus retain a significant central gas reservoir diluting local metallicities e.g. from a recent inflow even at low star-forming activity. {Using a compilation of different public optical-IFU datasets, \citet{Sanchez2020} recently showed galaxies of different morphological types exhibited different shapes in their radial metallicity profiles (whilst also segregating into mass bins).} In actively star-forming galaxies, \citet{Lutz2021} recently found evidence that the metallicity gradients are primarily set by variations in stellar mass surface density. {From the resolved mass-metallicity relation {\citep{RosalesOrtega2012, Sanchez2013, BarreraBallesteros2016}}, one may also expect metallicity gradients to be influenced by the stellar mass surface density profiles in the low-$\Sigma_\star$ regime, although the scatter in the relation is large.} We first use the stellar mass surface density within $1 \ R_e$ ($\mu_\star = 0.5 \cdot M_\star/(\pi {R_e}^2)$) as a morphological tracer to assess broadly how morphological factors may contribute to our observed $\nabla Z - \tdep$ relation. In the left panel of Fig~\ref{fig:Zgrad_vs_mustar}, we contrast the metallicity gradients to $\mu_\star$ for AGN, LINERs and Composites, finding no correlation. 

We then move to exploring the resolved profiles of stellar mass surface density as a function of $\nabla Z$, using the results from our full spectral fitting procedure described in Section \ref{subsec:OwnFitting}. In the right panel of Fig~\ref{fig:Zgrad_vs_mustar}, we show our recovered profiles of stellar mass surface density $\Sigma_\star = M_{\star, \rm ann} / A_{\rm ann} $, denoting the mass of stars per surface area $A_{\rm ann}$ within a given annulus of width $0.2 \  R_e$. The colour-coding reflects a binning according to $\nabla Z$ (green/purple = objects with negative/positive metallicity gradients). The dotted lines denote the median profiles, and the spread is indicated by the shaded regions. Clearly the two curves corresponding to different $\nabla Z$ bins are fully consistent with each other, such that we do not find any evidence for the stellar mass gradients being linked to metallicity gradients. 

Finally, we also report that when using the Pipe3D $v/ \sigma$ ratio of velocity over velocity dispersion within $1.5 \ R_e$ as a tracer of how rotation-dominated a given galaxy is, we find no trend with metallicity gradients, contrary to recent results at higher redshift (out to $z = 2.5$) by \citet{Sharda2021}. Likewise, we report no correlation between $\nabla Z$ and the photometrically-inferred bulge-to-total ratio $B/T$ from the PyMorph catalog as a morphological tracer. Further, we find no relation with the central velocity dispersion within $2.5$ arcsec of either the stellar kinematics $\sigma_\star$ (when excluding the single datapoint with lowest $\sigma_\star$ ) or the gas dynamics as traced by \Halpha . We thus conclude that, at least by using the simple morphological tracers in this Section, we do not find any evidence for the metallicity gradients in our sample being influenced by morphological effects, {contrary to the literature results summarised above. We note however that our sample size does not allow us to investigate the impact of morphology in separate mass bins. As a result, morphological trends may still be present, but could be diluted by the diversity of masses within our sample.}

\section{Conclusions}
\label{sec:Conclusion}

In this paper, we presented a selection of early results from the MASCOT survey \citep{Wylezalek2022}, focusing on an empirically found relation between gas-phase metallicity gradients $\nabla Z$ and molecular gas depletion times \tdep\  in LINERs, Seyfert AGN and Composites below/on the Main Sequence (MS). We used oxygen abundance gradients derived via the \citet{Marino2013} O3N2 calibrator, and specifically decided to explore the below/on-MS regime affected by DIG emission given recent results by e.g. \citet{Kumari2019} supporting the robustness of the O3N2 metallicity calibrator in DIG/LI(N)ER-dominated regions (as discussed in more detail in Section \ref{subsec:choice_O3N2}). {We still exercised caution by using the metallicity gradients from the Pipe3D catalog, which were derived based on regions within galaxies that are predominantly ionised by star formation.} We excluded edge-ons, type 1 AGN, and visually identified major mergers from the analysis. 
Our primary results can be summarised as follows:

\begin{itemize}
    \item The $\nabla Z - \tdep$ relation we report is strongly significant for our sample of ``AGN-like objects'' combining AGN, LINERs and Composites ($p \sim 0.001, r=0.56$). When excluding Seyfert AGN (e.g. due to concerns about the accuracy of the metallicity gradients), the relation still holds for LINERs+Composites with $p \sim 0.005, r=0.54$ (Fig~\ref{fig:main_plot}), while no relation is found in inactive galaxies. \\
    {The relation is intriguing given that variations in depletion time (or inversely, star forming efficiency) have been suggested to be the driving factor that pushes galaxies into the passive branch once they experienced an initial central lack of molecular gas \citep{Colombo2020}. In that sense, the discovery of a $\nabla Z - \tdep$ relation \textit{below the MS} may suggest that whichever mechanism drives quenching in AGN-like objects also flattens or inverts metallicity gradients in the process.} \\
    \item The relation cannot be traced back to a link between $\nabla Z$ and MS offset, nor a trend with global SFR (Fig~\ref{fig:Zgrad_SFR}). Further, we cannot explain the relation via the presence of mass effects, as we report no trend between $\nabla Z$ and stellar mass (Fig~\ref{fig:Zgrad_lMstar}), and none with inclination (Fig \ref{fig:Zgrad_vs_baRatio}).  \\
    \item We also checked that the relation persists when using a constant CO-to-\Hmol\ conversion factor instead of a metallicity-dependent one (Fig~\ref{fig:check_ConstAlphaCO}), and when using metallicity gradients based on the \citet{Tremonti2004} R23 calibrator instead of the \citet{Marino2013} O3N2 prescription (Fig \ref{fig:check_T04}). \\
\end{itemize}

We then explored different possible physical drivers of the $\nabla Z - \tdep$ relation in AGN/LINERs/Composites in Section \ref{sec:Discussion}:

\begin{itemize}
    \item We contemplated the scenario of metal redistribution via chemically enriched outflows, i.e. assuming $\nabla Z$ is flatter in galaxies with long depletion times because they are simplistically at a later evolutionary stage where feedback already had more time to operate. 
    In AGN/LINERs/Composites, we indeed found a strong link between $\nabla \ Z$ and the mean width ("$W_{80}$") of the ionised \OIII\ line within $1 \ R_e$ (Fig~\ref{fig:Zgrad_vs_W80_0ptRe}), as well as between $W_{80}$ and $t_{\rm dep}$. We interpret this trend as a strong indication of feedback-induced chemical mixing taking place as galaxies transition to the red cloud. 
    \\ 
    Interestingly, the observed velocity broadening is very modest at $<500 \ {\rm km/s}$. 
    {However, in a picture where galaxies quench due to the integrated effect of feedback over prolonged time, weak feedback below the MS may represent a ``final push'' that permanently decreases star-forming efficiencies. For a detailed discussion, including results from simulations, 
    we refer the reader to Section \ref{subsec:outflows}.} 
    \\
    \item We then analysed the impact of local star formation (at fixed stellar mass). Potentially, metallicity gradients may be traced back to local variations in star-formation due to a resolved fundamental metallicity relation, which in turn may impact the global depletion time $\tdep = M_\Hmol / {\rm SFR}$. Indeed our sample of AGN/LINERs/Composites shows a connection between positive $\nabla Z$ / longer depletion times and a relative suppression of SFR surface density in the outskirts (Fig \ref{fig:SigSFR_Offset_profiles}), unlike inactive galaxies. Next to the impact of feedback, in-situ star formation may thus contribute to shaping the $\nabla Z- \tdep$ relation in our AGN-like objects. 
    \\
    \item Similarly to outflows, inflows may influence metallicity gradients (though our galaxies reside below the MS). However, we find no trend when contrasting metallicity gradients to $\Hmol$ mass, CO luminosity, or molecular gas fraction $f_\Hmol = M_\Hmol / M_\star$ (Fig~\ref{fig:Zgrad_vs_lMH2_LCO}), and thus no direct indication of gas accretion affecting the metallicity distribution. However, we note that our unresolved single-dish CO observations do not allow us to exclude the possibility of localised inflows influencing our results. 
    \\
    \item We further considered morphological effects. For instance, galaxies could decrease in star-forming efficiency whilst preserving a gas reservoir that is dynamically stabilised against fragmentation \citep{Martig2009}. 
     
    However, we find no evidence for metallicity gradients depending on stellar mass surface density within $1 \ R_e$ (Fig~\ref{fig:Zgrad_vs_mustar}), and no difference between the stellar mass surface density profiles of AGN-like objects with positive and negative $\nabla Z$. 
    Using different morphological tracers - the $v/\sigma$ ratio of velocity over velocity dispersion within $1.5 \ R_e$, a photometrically-inferred bulge-to-total mass ratio, or central velocity dispersion within $2.5$ arcsec  (stellar or \Halpha) - offered no further clues.
    \\
    \item Given that merger activity would naturally impact the metallicity distribution within galaxies, we briefly search for merger signatures within our sample by visually comparing the \Halpha\ and \oiii\ gas-phase velocity maps to the stellar velocity maps. We confirmed that our results were not strongly affected when conservatively discarding any galaxies which showed potential signs of misalignement in stellar and gas-phase kinematics. 
\end{itemize}

In conclusion, we propose that in our AGN-like targets, the observed $\nabla Z - \tdep$ relation arises partially as a consequence of chemical mixing due to centrally-driven {(stellar or AGN) outflows} that increase in impact as time goes on and depletion times become longer, and is partially connected to fluctuations in local star formation. The latter is consistent with a resolved fundamental metallicity relation which may arise from a lack of diluting gas leading to both lower SFRs and higher local metallicities in the outskirts of quenching AGN-like objects. 

{We finally note that, given the sample size of our below-MS objects, we have not considered the impact of environment on metallicity in this work (f.ex. enriched accretion onto satellites, see \citealt{Schaefer2019}). We defer an investigation of environmental effects on the molecular gas properties and other galaxy parameters using our full sample (below- and above-MS) to a later paper.}

\section*{Acknowledgements}

The entire MASCOT team would like to warmly thank the staff
at the Arizona Radio Observatory, in particular the operators of the 12m Telescope, Clayton, Kevin, Mike and Robert, for their continued support and help with the observations. 
We further thank A. Bluck, B. Easeman, T. Davis, R. Maiolino, J. Piotrowska, P. Schady, J. Trayford and S. Wuyts for fruitful discussions. We thank the referee S. F. Sánchez for helpful and constructive comments.

DW and CB are supported through the Emmy Noether Programme of the German Research Foundation. MA acknowledges support from FONDECYT grant 1211951, CONICYT + PCI + INSTITUTO MAX PLANCK DE ASTRONOMIA MPG190030, CONICYT+PCI+REDES 190194 and ANID BASAL project FB210003. W.B. acknowledges support from the ERC Advanced Grant 695671, “QUENCH” and from the Science and Technology Facilities Council (STFC).

This research made use of Marvin, a core Python package and web framework for MaNGA data, developed by Brian Cherinka, José Sánchez-Gallego, Brett Andrews, and Joel Brownstein \citep{Cherinka2019}\footnote{\url{http://sdss-marvin.readthedocs.io/en/stable/}}. 

This project makes use of the MaNGA-Pipe3D dataproducts. We thank the IA-UNAM MaNGA team for creating this catalogue, and the ConaCyt-180125 project for supporting them.

Funding for the Sloan Digital Sky Survey IV has been provided by the Alfred P. Sloan Foundation, the U.S. Department of Energy Office of Science, and the Participating Institutions. SDSS-IV acknowledges support and resources from the Center for High Performance Computing at the University of Utah. The SDSS
website is www.sdss.org. SDSS-IV is managed by the Astrophysical Research Consortium for the Participating Institutions of the SDSS Collaboration including
the Brazilian Participation Group, the Carnegie Institution for Science, Carnegie Mellon University, Center for Astrophysics | Harvard
\& Smithsonian, the Chilean Participation Group, the French Participation Group, Instituto de Astrofísica de Canarias, The Johns Hopkins University, Kavli Institute for the Physics and Mathematics of the Universe (IPMU) / University of Tokyo, the Korean Participation
Group, Lawrence Berkeley National Laboratory, Leibniz
Institut f\"ur Astrophysik Potsdam (AIP), Max-Planck-Institut f\"ur Astronomie
(MPIA Heidelberg), Max-Planck-Institut f\"ur Astrophysik
(MPA Garching), Max-Planck-Institut f\"ur Extraterrestrische Physik (MPE), National Astronomical Observatories of China, New Mexico State University, New York University, University of Notre Dame,
Observatário Nacional / MCTI, The Ohio State University, Pennsylvania State University, Shanghai Astronomical Observatory, United Kingdom Participation Group, Universidad Nacional Autónoma de México, University of Arizona,University of Colorado Boulder, University
of Oxford, University of Portsmouth, University of Utah,
University of Virginia, University of Washington, University of Wisconsin, Vanderbilt University, and Yale University.

\section*{Data availability}
The data underlying this article are available in \citet{Wylezalek2022} and its supplementary material, as well as on the MASCOT website: \url{https://wwwstaff.ari.uniheidelberg.de/dwylezalek/mascot.html}

\bibliographystyle{mnras}
\bibliography{CB_bib_short} 

\clearpage
\appendix

\section{Galaxies above the MS do not show a $\nabla  Z - \tdep$ relation}
\label{app:AboveMS}

\begin{figure*}
\centering
\includegraphics[width=0.49\textwidth]{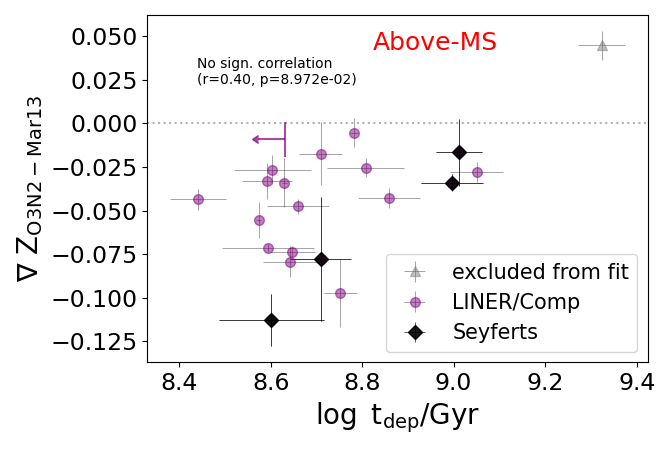}
\includegraphics[width=0.49\textwidth]{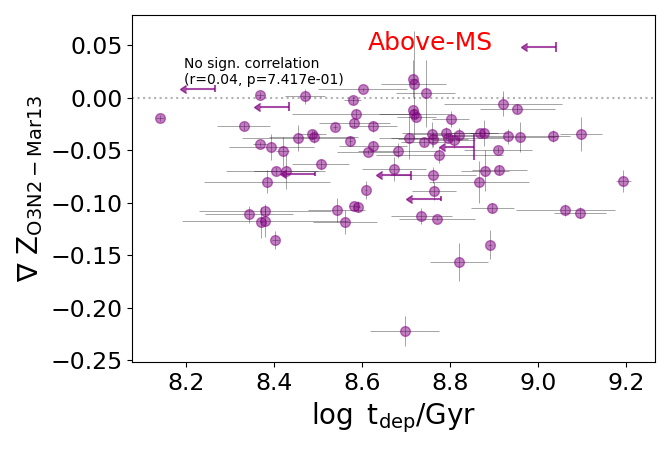}
\caption{Location of \textit{above-MS galaxies} from the MASCOT sample in the plane of metallicity gradients $\nabla Z$ against depletion time $\tdep = {\MHmol /\rm SFR}$. In the left panel focussing on AGN/LINERs/Composites, Seyferts are highlighted in black, and the datapoint in grey is excluded to avoid it being the sole driver of our correlation analysis. The right panel corresponds to inactive galaxies. The arrows denote upper limits for CO-undetected sources. Above the Main Sequence, there is no correlation in either case. Thus, the $\nabla Z - \tdep$ relation shown in Fig \ref{fig:main_plot} is unique to AGN-like objects located below the MS.}
\label{fig:Zgrad_vs_tdep_AboveMS}
\end{figure*}

{In the work presented in this paper, we focused on galaxies located below the Main Sequence. For the purpose of a quick comparison, in Fig \ref{fig:Zgrad_vs_tdep_AboveMS}, we check whether the MASCOT targets lying above the Main Sequence also show any correlation between their metallicity gradients $\nabla Z$ and their depletion time $\tdep$, such as the $\nabla Z - \tdep$ relation seen in AGN-like objects below the MS (Fig \ref{fig:main_plot}). In active and inactive galaxies alike (left and right panel, respectively), we do not find any correlation in the above-MS regime. For AGN-like objects, we further note that with the exception of the single datapoint with longest \tdep\, all metallicity gradients are negative in the above-MS systems in Fig \ref{fig:Zgrad_vs_tdep_AboveMS}, while below the MS there are several (nine) AGN-like objets with positive gradients in Fig \ref{fig:main_plot}. One complicating factor for systems above the Main Sequence is the higher contribution from stellar feedback which may not be limited to the central regions within galaxies, and may affect both the metallicity distribution and the depletion time. }

\section{Statistical bootstrapping analysis}
\label{app:bootstrap}

\begin{table*} 
\vspace*{-0.5cm}
\captionsetup{width=1.\textwidth}
\caption{
Median r-values (Spearman's rank coefficient) and associated confidence intervals derived by a bootstrapping analysis for several of the correlations discussed in this work using 1000 bootstrapped samples.
}
\begin{adjustwidth}{0cm}{0cm}
{
{\begin{tabular}{l l l l l l}
\hline
Parameter A & Parameter B & Median $r$-value between A \& B & $95$\% confidence interval on $r$ &  Confirming $r \neq 0$? & Figure \\
\hline
\hline
$\nabla Z$ & \tdep & 0.55 & [0.25, 0.76] & \cmark & Fig \ref{fig:main_plot} \\
$\nabla Z$ & \oiii\ $W_{80} (< 1 R_e )$ & 0.7 & [0.05, 0.98] & \cmark & Fig \ref{fig:Zgrad_vs_W80_0ptRe} \\
\oiii\ $W_{80} (< 1 R_e )$ & \tdep & 0.82 & [0.48, 0.97] & \cmark & Fig \ref{fig:Zgrad_vs_W80_0ptRe} \\
\hline
 \end{tabular} 
 }\\
 \vspace*{0.05cm}
 } 
\end{adjustwidth}
 \label{tab:bootstrap} 
 \end{table*}  
\captionsetup{width=0.8\textwidth}

{In this Section, we use a bootstrapping approach to further test the statistical robustness of several key relations we presented herein. 
For each correlation, we use the associated sample to draw 1000 bootstrapped samples with replacement - these have the same number of datapoints as the original sample, but may feature repeated entries while omitting other points. For each such bootstrapped sample, we calculate an associated Spearman's rank coefficient to assess how the correlation changes when random datapoints drop out or are repeated. From the recovered distribution of r-values, we then proceed to calculate the $95$\% confidence interval.}

{The results from this exercise are summarised in Table \ref{tab:bootstrap}. The first two columns name the parameters that are subject of our correlation analysis in AGN-like objects. Specifically, we focus on the relations between metallicity gradients $\nabla Z$, depletion time \tdep, and mean \oiii\ $W_{80}$ within 1 $R_e$ (Figs \ref{fig:main_plot} and \ref{fig:Zgrad_vs_W80_0ptRe}). The median and $95$\% confidence interval from our recovered distribution of $r$-values are listed in the third and fourth columns, while the fifth column specifies whether the statistical soundness of the correlation is confirmed (which holds if the $95$\% confidence interval is not covering $r=0$). The last column links to the Figure illustrating each correlation.}

The bootstrapping analysis suggest that all three correlations investigated in Table \ref{tab:bootstrap} are sound. $\nabla Z$ and \tdep are clearly correlated, with $r \in [0.25/0.76]$ and $<r> = 0.55$. Similarly, the correlations between resolved \oiii\ $W_{80} (< 1 \ R_e)$ and $\nabla Z$ or \tdep are robust with $<r> = 0.7/0.82$ and $r \in [0.05/0.98]/[0.48, 0.97]$, respectively.

\section{Discussion on the interpretation of the \oiii~line width as a tracer of outflows}
\label{app:sanity_checks_w80}

\begin{figure*}
\centering
\begin{adjustwidth}{-0.7cm}{-1cm}
\includegraphics[width=0.32\linewidth]{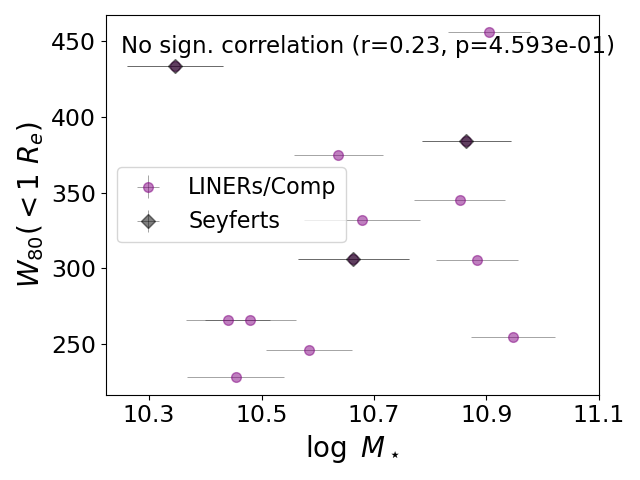}
\includegraphics[width=0.32\linewidth]{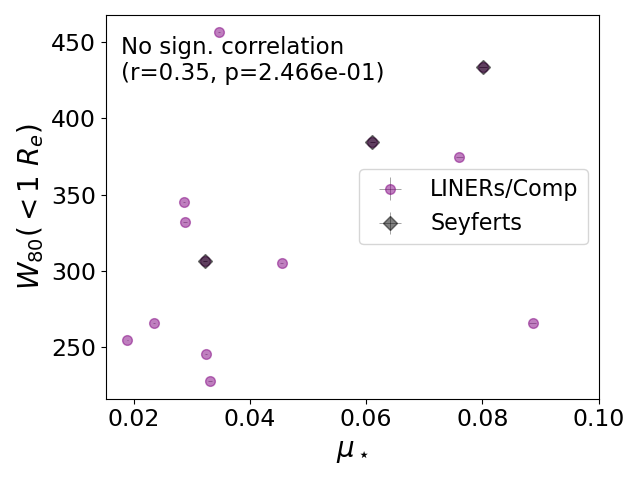}
\includegraphics[width=0.32\linewidth]{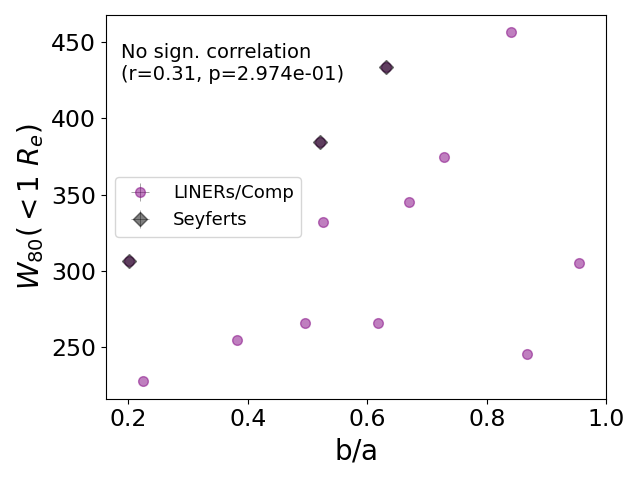}
\end{adjustwidth}
\caption{\textit{Left panel:} $W_{80}$ values contrasted against stellar mass, yielding no significant correlation. \textit{Middle panel:} $W_{80}$ values plotted against stellar mass surface density $\mu_\star = 0.5*M_\star / (\pi * R_e^2)$ , again showing no evidence for a correlation. \textit{Right panel}: $W_{80}$ values contrasted against the minor-to-major axis ratio $b/a$. There is no correlation and specifically no evidence for artificially heightened $W_{80}$ values in edge-on systems, where single spaxels may probe a larger diversity of radii.}
\label{fig:W80_vs_baratio}
\end{figure*}

{In Section \ref{subsec:outflows}, we explored how the gas-phase metallicity gradients in our samples of AGN-like objects were impacted by outflows, using as a tracer the \oiii\ $W_{80}$, defined as the linewidth encompassing $80 \%$ of the \oiii\ line flux. However, as discussed in Section \ref{subsec:outflows}, outflows are not the only drivers of emission line broadening, such that $W_{80}$ is not expected to be a pure tracer of outflows. While the $W_{80}$ values are derived in single spaxels (e.g. without any stacking), individual spaxels still probe a finite range of radii and associated velocities which  broadens emission lines especially in central spaxels (while at large radii the rotation curve is flattening). Beam smearing could exacerbate this issue. 
The line broadenings we observe below the MS are weak, with $W_{80}$ values staying below $\sim 500 \ \rm km \ s^{-1}$, and could thus in principle be contaminated substantially potential well effects. We also note that there could be a complicating hidden dependency, since galaxies with a larger potential well may also harbour a more massive black hole, which may therefore drive stronger outflows, as well as increase $W_{80}$. In this Section, we nevertheless conduct a number of sanity checks to reinforce our confidence in the use of the \oiii\ $W_{80}$ parameter as a tracer of outflows.}

{In the left and middle panels of Fig \ref{fig:W80_vs_baratio}, we first contrast the \oiii\ $W_{80}$ values of the AGN-like objects against the stellar mass and stellar mass surface density within one half-light radius $\mu_\star = 0.5*M_\star / (\pi * R_e^2)$ as a direct tracer of the potential well. Seyfert AGN are highlighted in black. In either case, there is no correlation. In the right panel, we proceed to plot $W_{80}$ against the minor-to-major axis ratio $b/a$ tracing the inclination. If the $W_{80}$ values were driven primarily by potential well effects, then edge-on systems would be most affected, since indidvidual spaxels would probe a broader range of radii. However, we recover no evidence for $\nabla Z$ being inclination dependent.}

\section{Use of other morphological tracers}
\label{app:morph_app}

\begin{figure*}
\centering
\includegraphics[width=0.49\textwidth, trim={0cm 0cm 0cm 0.3cm},clip]{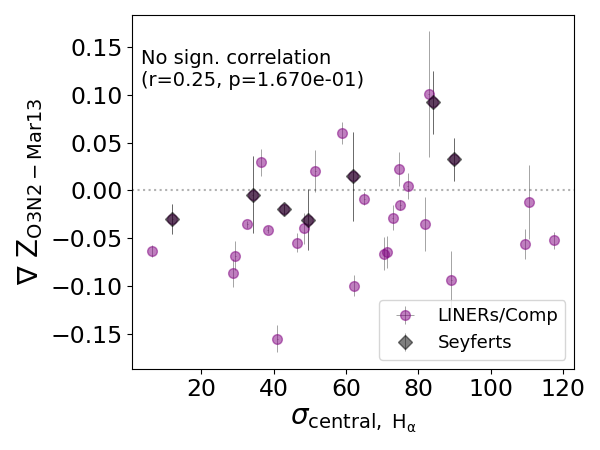}
\includegraphics[width=0.49\textwidth, trim={0cm 0cm 0cm 0.3cm},clip]{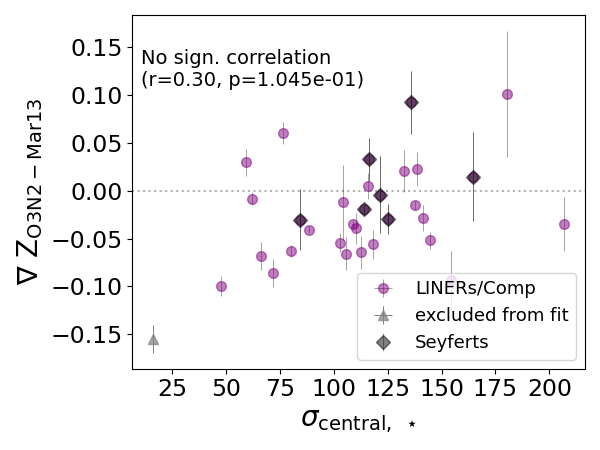}
\includegraphics[width=0.49\textwidth, trim={0cm 0cm 0cm 0.3cm},clip]{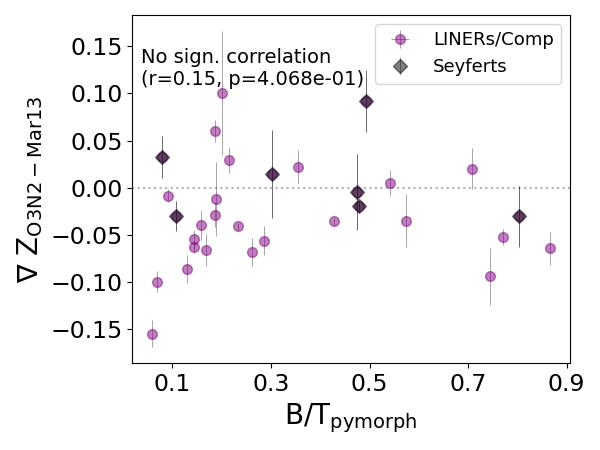}
\includegraphics[width=0.49\textwidth, trim={0cm 0cm 0cm 0.3cm},clip]{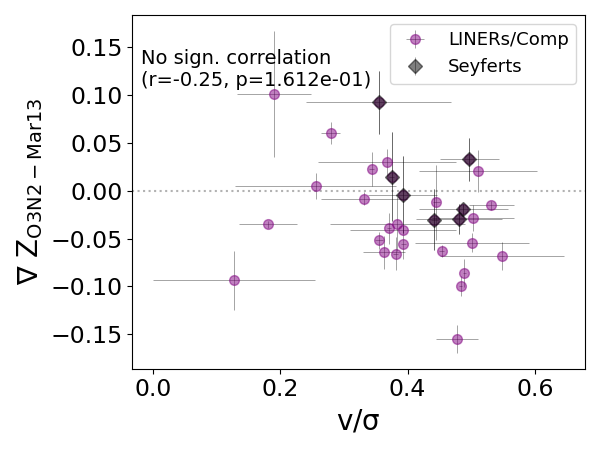}
\caption{{Metallicity gradients plotted against various morphological tracers: the velocity dispersion $\sigma_{\rm central, \ \Halpha }$ (\textit{top left panel}), the central stellar velocity dispersion $\sigma_{\rm central, \ \star }$ (\textit{top right panel}) the bulge-to-total ratio $B/T$ (\textit{bottom left panel}) and the $v/\sigma$ ratio of velocity over velocity dispersion (\textit{bottom right panel}). We find no evidence for a correlation in either case.}
}
\label{fig:Zgrad_vs_BTratio}
\end{figure*}

{In this section, we return to the question of whether the metallicity gradients in our AGN-like objects are connected to their morphological states. While we found no obvious trends with stellar mass surface density in Section \ref{subsec:morphology}, we here explore a variety of alternative morphological tracers.}

{In Fig~\ref{fig:Zgrad_vs_BTratio}, we contrast $\nabla Z$ to the central \Halpha\ velocity dispersion $\sigma_{\rm central, \ \Halpha }$ (\textit{top left panel}), the central stellar velocity dispersion $\sigma_{\rm central, \ \star }$ (\textit{top right panel}), the photometrically-inferred bulge-to-total ratio $B/T$ (\textit{bottom left panel}) and the $v/\sigma$ ratio of velocity over velocity dispersion (\textit{bottom right panel}) as a measure of how rotation-dominated a galaxy is.
and $v/\sigma$ was taken from the Pipe3D catalog \citep{Sanchez2016b, Sanchez2018}, while the $B/T$ ratio was taken from the PyMorph catalog \citep{Fischer2019, DominguezSanchez2022}. In any case, we do not find any evidence for a link between morphology and metallicity gradients in AGN-like objects. This finding may be counterintuitive, as one would expect the metallicity distribution to be affected by variations in local stellar mass surface density according to the local mass-metallicity relation \citep{BarreraBallesteros2016}. Our results suggest that in AGN-like galaxies, the other drivers of metallicity gradients we identified - outflows and the distribution of star formation - play a more important role than the mass distribution. 
}

\section{The $\nabla Z - W_{80}$ relation revisited with R23-based metallicities or constant $\alpha_{\rm CO}$}
\label{app:W80_check_T04}

\begin{figure*}
\centering
\includegraphics[width=0.49\textwidth]{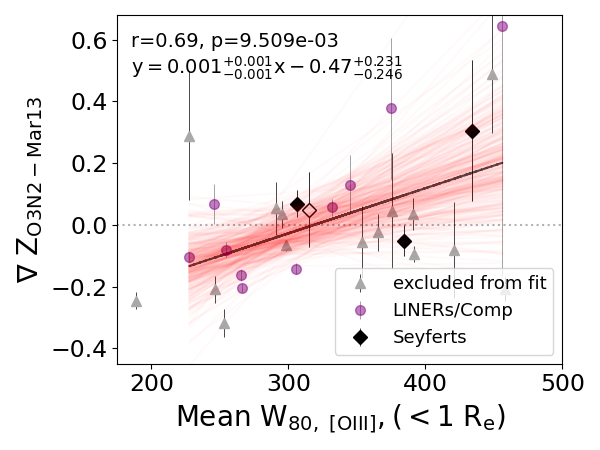}
\includegraphics[width=0.49\textwidth]{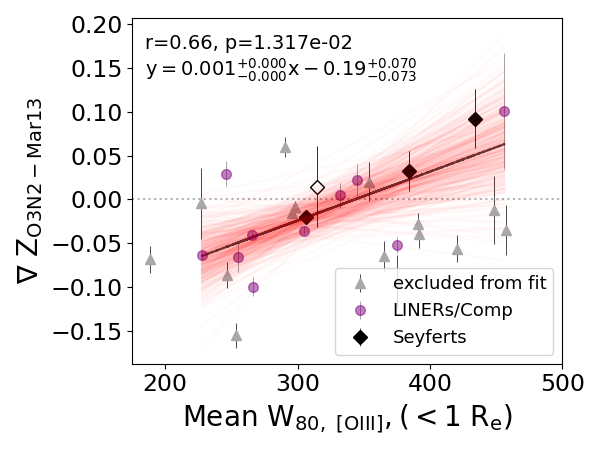}
\caption{\textit{Left panel}: Same as Fig \ref{fig:Zgrad_vs_W80_0ptRe}, but based on the \citet{Tremonti2004} R23 metallicity diagnostic instead of the \citet{Marino2013} O3N2 prescription. \textit{Right panel}: Same as Fig \ref{fig:Zgrad_vs_W80_0ptRe}, but recalculating the depletion time with a constant CO-to-\Hmol\ conversion factor $\alpha_{\rm CO}=4.36 \ {\rm \ M_\odot / (K \ km \ s^{-1} \ pc^{-2}) }$ corresponding to the canonical Milky Way value.}
\label{fig:Sanity_check_GradZ_W80}
\end{figure*}

We now briefly explore how the $\nabla Z - W_{80} (< 1 \ R_e)$ relation presented in Section \ref{subsec:outflows} for AGN/LINERs/Composites below/on the MS may depend on our choice of metallicity calibrator or on assumptions about the metallicity dependence of the CO-to-$\Hmol$ conversion factor $\alpha_{\rm CO}$. In Section \ref{subsec:Sanity_checks}, we showed that the $\nabla Z - \tdep$ relation is still recovered when using the \citet{Tremonti2004} R23 diagnostic instead of the O3N2 calibrator, and also when using a constant CO-to-$\Hmol$ conversion factor based on the canonical Milky Way value rather than the metallicity-dependent conversion based on \citet{Accurso2017} used throughout the rest of the paper. We repeat this exercise for the $\nabla Z - W_{80} (< 1 \ R_e)$ relation in Fig \ref{fig:Sanity_check_GradZ_W80}, showing that the relation persists both when switching to the R23 metallicity calibrator (left panel) or when setting $\alpha_{\rm CO} \equiv 4.36 \ {\rm \ M_\odot / (K \ km \ s^{-1} \ pc^{-2}) }$ in the calculation of the depletion time (right panel).

\section{Discarding potential post-merger systems}
\label{app:merger_app}

\begin{figure*}
\centering
\noindent
\begin{adjustwidth}{-0.7cm}{-0.5cm}
\begin{minipage}{1.12\textwidth}
    \noindent
    \begin{minipage}{0.32\textwidth}
    \centering
    \large \phantom{aaaa} AGN/LINERs/Composites \\[1.5ex]
    \includegraphics[width=\textwidth, trim={0cm 0cm 0cm 0.35cm},clip]{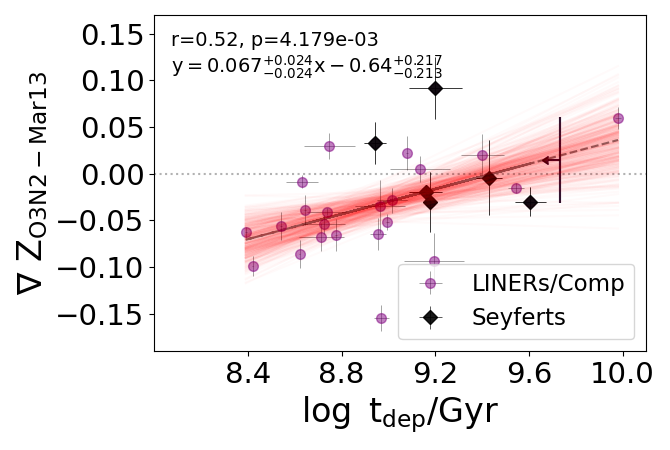}
    \end{minipage}
    \begin{minipage}{0.32\textwidth}
    \centering
    \large \phantom{aaaa} Inactive galaxies \\[1.5ex]
    \includegraphics[width=\textwidth, trim={0cm 0cm 0cm 0.35cm},clip]{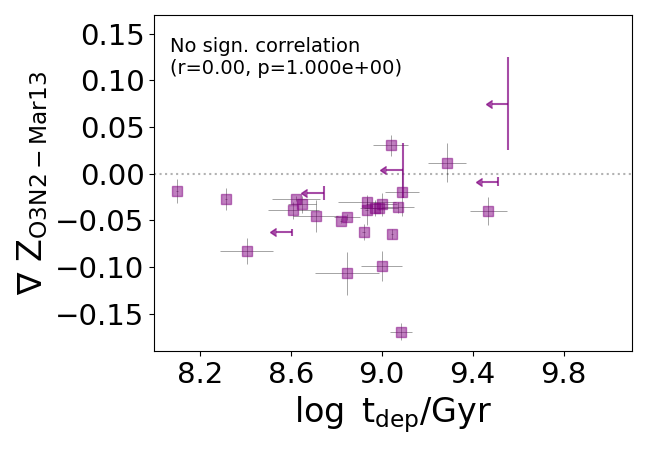}
    \end{minipage}
    \begin{minipage}{0.32\textwidth}
    \centering
    \large \phantom{aaaa} LINERs/Composites \\[1.5ex]
    \includegraphics[width=\textwidth, trim={0cm 0cm 0cm 0.35cm},clip]{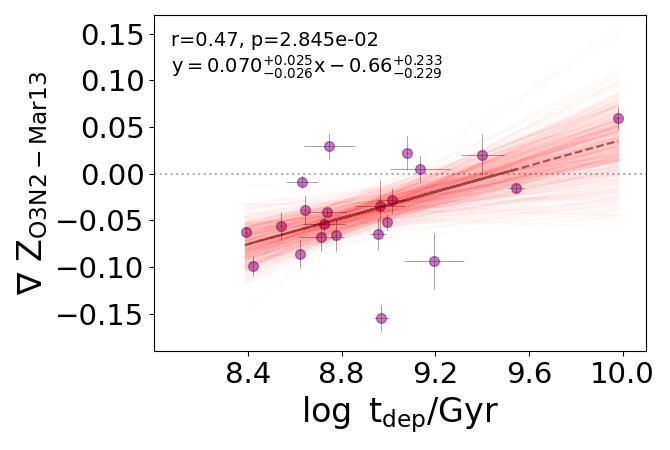}
    \end{minipage}
\end{minipage}
\end{adjustwidth}
\caption{Same as Fig~\ref{fig:main_plot}, but excluding objects which show potential evidence for a mis-alignement between their stellar and gas-phase kinematics.}
\label{fig:GradZ_vs_ltdep_Exclude_someVel}
\end{figure*}

\begin{figure*}
\centering
\includegraphics[width=0.49\textwidth, trim={0cm 0cm 0cm 0.35cm},clip]{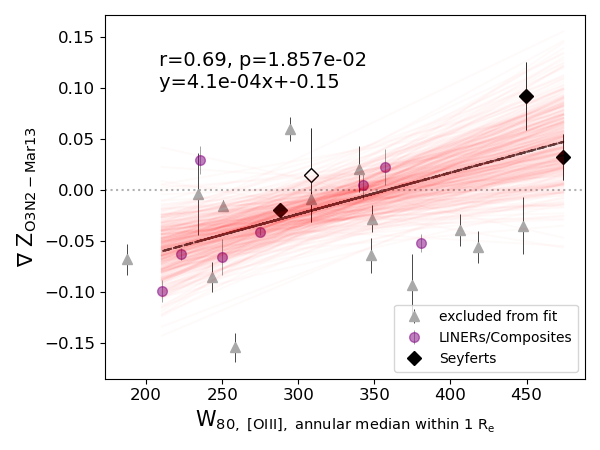}
\caption{Same as Fig~\ref{fig:Zgrad_vs_W80_0ptRe}, but excluding objects which show potential evidence for a mis-alignement between their stellar and gas-phase kinematics.}
\label{fig:Zgrad_vs_W80_0ptRe_Exclude_someVel}
\end{figure*}

In this Section, we briefly check that our results on AGN-like galaxies do not strongly depend on the inclusion of systems with potential post-merger signatures. As discussed in Section \ref{subsec:mergers}, we visually inspected the \OIII\ and stellar velocity maps to search for kinematic misalignements between the stellar and gas-phase components as a potential sign of post-merger status. The maps were produced with \texttt{Marvin} \citep{Cherinka2019} using hybrid binning with a $S/N$ cutoff of $3$. Fig \ref{fig:GradZ_vs_ltdep_Exclude_someVel} shows the $\nabla Z - \tdep$ relation that we obtain when excluding the galaxies with the following Plate-IFUs: 8588-6101, 8595-3703, 8084-6103. The relation remains robust. In Fig \ref{fig:Zgrad_vs_W80_0ptRe_Exclude_someVel}, we further show that the relation between $\nabla Z$ and the \oiii\ line width within $1 \ R_e$ remains significant when excluding those objects.

\end{document}

%% file: mycommands.tex
\newcommand{\todo}{\ifmmode {\color{red}\text{\Huge{\(\bullet\)}}} \else {\color{red} \Huge$\bullet$}\fi}
\newcommand{\tido}{\ifmmode {\bullet} \else $\bullet$\fi}

\newcommand{\E        }[1]{\ifmmode 10^{#1} \else $10^{#1}$\fi}
\newcommand{\tE        }[1]{\ifmmode \times10^{#1} \else $\times10^{#1}$\fi}
\newcommand{\til}{\ifmmode \sim \else $\sim$\fi}
\renewcommand{\~} {\ifmmode \sim \else $\sim$\fi}

\newcommand{\pc}	{\ifmmode {\rm pc} \else pc\fi}
\newcommand{\ld}	{\ifmmode {\rm l.d.} \else l.d.\fi}
\newcommand{\kms}	{\ifmmode {\rm km\,s}^{-1} \else km\,s$^{-1}$\fi}
\newcommand{\Jykms}	{\ifmmode {\rm Jy\,km\,s}^{-1} \else Jy\,km\,s$^{-1}$\fi}
\newcommand{\cc}	{\ifmmode {\rm cm}^{-3}    \else cm$^{-3}$\fi}
\newcommand{\cmii}	{\ifmmode {\rm cm}^{-2}    \else cm$^{-2}$\fi}
\newcommand{\ergs}	{\ifmmode {\rm erg\,s}^{-1} \else erg s$^{-1}$\fi}
\newcommand{\ergcms}	{\ifmmode {\rm erg\,cm}^{-2}\,{\rm s}^{-1} \else erg\,cm$^{-2}$\,s$^{-1}$\fi}
\newcommand{\ergcmsA}	{\ifmmode {\rm erg\,cm}^{-2}\,{\rm s}^{-1}\,{\rm\AA}^{-1}
\else erg\,cm$^{-2}$\,s$^{-1}$\,\AA$^{-1}$\fi}
\newcommand{  \ergcmsHz  }{\ifmmode{\rm erg\,cm}^{-2}\,{\rm s}^{-1}\,{\rm Hz}^{-1}
                       \else ergs\,cm$^{-2}$\,s$^{-1}$\,Hz$^{-1}$\fi}
\newcommand{\kev}	{\ifmmode {\rm keV} \else keV\fi}

\newcommand{\mic}	{\ifmmode {\rm \mu m} \else $\mu$m\fi}
\newcommand{\vFWHM}	{\ifmmode v_{\mbox{\tiny FWHM}} \else $v_{\mbox{\tiny FWHM}}$\fi}
\newcommand{\vBLR}	{\ifmmode v_{\mbox{\tiny BLR}} \else $v_{\mbox{\tiny BLR}}$\fi}
\newcommand{\sigBLR}	{\ifmmode \sigma_{\mbox{\tiny BLR}} \else $\sigma_{\mbox{\tiny BLR}}$\fi}
\newcommand{\vNLR}	{\ifmmode v_{\mbox{\tiny NLR}} \else $v_{\mbox{\tiny NLR}}$\fi}
\newcommand{\tauBLR}	{\ifmmode \tau_{\mbox{\tiny BLR}} \else $\tau_{\mbox{\tiny BLR}}$\fi}

\newcommand{\Hubble}	{\ifmmode {\rm km\,s}^{-1}\,{\rm Mpc}^{-1} \else km\,s$^{-1}$\,Mpc$^{-1}$\fi}
\newcommand{\NDunit}	{\ifmmode {\rm Mpc}^{-3} \else Mpc$^{-3}$\fi}
\newcommand{\LFunit}	{\ifmmode {\rm Mpc}^{-3}\,{\rm mag}^{-1} \else Mpc$^{-3}$\,mag$^{-1}$\fi}
\newcommand{\MFunit}	{\ifmmode {\rm Mpc}^{-3}\,{\rm dex}^{-1} \else Mpc$^{-3}$\,dex$^{-1}$\fi}

\newcommand{\Msun}{\ifmmode M_{\odot} \else $M_{\odot}$\fi}
\newcommand{\Lsun}{\ifmmode L_{\odot} \else $L_{\odot}$\fi}
\newcommand{\Zsun}{\ifmmode Z_{\odot} \else $Z_{\odot}$\fi}
\newcommand{\mpyr}{\ifmmode \Msun\,{\rm yr}^{-1} \else $\Msun\,{\rm yr}^{-1}$\fi}

\newcommand{\qnote}{\ifmmode q_{0} \else $q_{0}$\fi}
\newcommand{\Hnote}{\ifmmode H_{0} \else $H_{0}$\fi}
\newcommand{\hnote}{\ifmmode h_{0} \else $h_{0}$\fi}
\newcommand{\anote}{\ifmmode a_{0} \else $a_{0}$\fi}
\newcommand{\tnote}{\ifmmode t_{0} \else $t_{0}$\fi}

\def\gsim{\;\rlap{\lower 2.5pt \hbox{$\sim$}}\raise 1.5pt\hbox{$>$}\;}
\def\lsim{\;\rlap{\lower 2.5pt \hbox{$\sim$}}\raise 1.5pt\hbox{$<$}\;}

\newcommand{  \Halpha   }{\ifmmode {\rm H}\alpha \else H$\alpha$\fi}

\newcommand{  \ha       }{\Halpha}
\newcommand{  \Hbeta    }{\ifmmode {\rm H}\beta \else H$\beta$\fi}

\newcommand{  \hb       }{\Hbeta}
\newcommand{  \Hgamma   }{\ifmmode {\rm H}\gamma \else H$\gamma$\fi}
\newcommand{  \Hdelta   }{\ifmmode {\rm H}\delta \else H$\delta$\fi}
\newcommand{  \Lya      }{\ifmmode {\rm Ly}\alpha \else Ly$\alpha$\fi}
\newcommand{  \Lyb      }{\ifmmode {\rm Ly}\beta \else Ly$\beta$\fi}
\newcommand{  \Pa       }{\ifmmode {\rm P}\alpha \else P$\alpha$\fi}
\newcommand{  \Pb       }{\ifmmode {\rm P}\beta \else P$\beta$\fi}
\newcommand{  \Bra      }{\ifmmode {\rm Br}\alpha \else Br$\alpha$\fi}
\newcommand{  \Brg      }{\ifmmode {\rm Br}\gamma \else Br$\gamma$\fi}
\newcommand{  \hi       }{\ifmmode {\rm H}\,\textsc{i} \else H\,\textsc{i}\fi}
\newcommand{  \HI       }{\ifmmode {\rm H}\,\textsc{i} \else H\,\textsc{i}\fi}
\newcommand{  \hii      }{\ifmmode {\rm H}\,\textsc{ii} \else H\,\textsc{ii}\fi}
\newcommand{  \hei      }{\ifmmode {\rm He}\,\textsc{i} \else He\,\textsc{i}\fi}
\newcommand{  \heii     }{\ifmmode {\rm He}\,\textsc{ii} \else He\,\textsc{ii}\fi}
\newcommand{  \HeIIuv   }{\ifmmode {\rm He}\,\textsc{ii}\,\lambda1640 \else He\,\textsc{ii}\,$\lambda1640$\fi}
\newcommand{  \HeIIop   }{\ifmmode {\rm He}\,\textsc{ii}\,\lambda4686 \else He\,\textsc{ii}\,$\lambda4686$\fi}
\newcommand{  \CII	}{\ifmmode \left[{\rm C}\,\textsc{ii}\right]\,\lambda157.74\,\mu{\rm m} \else [C\,{\sc ii}]\ $\lambda157.74\,\mu{\rm m}$\fi}
\newcommand{  \cii	}{\ifmmode \left[{\rm C}\,\textsc{ii}\right] \else [C\,{\sc ii}]\fi}

\newcommand{  \ciii     }{\ifmmode {\rm C}\,\textsc{iii}\right] \else C\,\textsc{iii}]\fi}
\newcommand{  \CIII     }{\ifmmode {\rm C}\,\textsc{iii}\right]\,\lambda1909 \else C\,\textsc{iii}]\,$\lambda1909$\fi}
\newcommand{  \civ      }{\ifmmode {\rm C}\,\textsc{iv}  \else C\,\textsc{iv}\fi}
\newcommand{  \CIV      }{\ifmmode {\rm C}\,\textsc{iv}\,\lambda1549 \else C\,\textsc{iv}\,$\lambda1549$\fi}
\newcommand{  \nii      }{\ifmmode \left[{\rm N}\,\textsc{ii}\right]  \else [N\,\textsc{ii}]\fi}
\newcommand{\NII}{\ifmmode \left[{\rm N}\,\textsc{iii}\right]\,\lambda6583 \else [N\,{\sc iii}]\,$\lambda6583$\fi}
\newcommand{  \niii     }{\ifmmode {\rm N}\,\textsc{iii} \else N\,\textsc{iii}\fi}
\newcommand{  \niv      }{\ifmmode {\rm N}\,\textsc{iv}  \else N\,\textsc{iv}\fi}
\newcommand{  \NIVuv    }{\ifmmode {\rm N}\,\textsc{iv}\,\lambda1486 \else N\,\textsc{iv}\,$\lambda1486$\fi}
\newcommand{  \nv       }{\ifmmode {\rm N}\,\textsc{v}   \else N\,\textsc{v}\fi}
\newcommand{\oi}{\ifmmode \left[{\rm O}\,\textsc{i}\right] \else [O\,{\sc i}]\fi}
\newcommand{\OI}{\ifmmode \left[{\rm O}\,\textsc{i}\right]\,\lambda6300 \else [O\,{\sc i}]$\,\lambda6300$\fi}
\newcommand{\oii}{\ifmmode \left[{\rm O}\,\textsc{ii}\right] \else [O\,{\sc ii}]\fi}
\newcommand{\OII}{\ifmmode \left[{\rm O}\,\textsc{ii}\right]\,\lambda 3726, 3729 \else [O\,{\sc ii}]\,$\lambda 3726, 3729$\fi}
\newcommand{\oiii}{\ifmmode \left[{\rm O}\,\textsc{iii}\right] \else [O\,{\sc iii}]\fi}
\newcommand{\OIII}{\ifmmode \left[{\rm O}\,\textsc{iii}\right]\,\lambda5007 \else [O\,{\sc iii}]\,$\lambda5007$\fi}
\newcommand{  \OIIIuv   }{\ifmmode {\rm O}\,\textsc{iii}\,\lambda1663 \else O\,\textsc{iii}\,$\lambda1663$\fi}
\newcommand{  \oiv      }{\ifmmode {\rm O}\,\textsc{iv}  \else O\,\textsc{iv}\fi}
\newcommand{  \OIVuv    }{\ifmmode {\rm O}\,\textsc{iv}\,\lambda1402  \else O\,\textsc{iv}\,$\lambda1402$\fi}
\newcommand{  \OIVIR    }{\ifmmode {\rm O}\,\textsc{iv}\,25.9\,\mu {\rm m} \else O\,\textsc{iv}\,$25.9\,\mu$m\fi}
\newcommand{  \ovi      }{\ifmmode {\rm O}\,\textsc{vi}   \else O\,\textsc{vi}\fi}
\newcommand{  \Ovi      }{\ifmmode {\rm O}\,\textsc{vi}\,\lambda1035 \else O\,\textsc{vi}\,$\lambda1035$\fi}
\newcommand{  \nei      }{\ifmmode {\rm Ne}\,\textsc{i}   \else Ne\,\textsc{i}\fi}
\newcommand{  \neii     }{\ifmmode {\rm Ne}\,\textsc{ii}  \else Ne\,\textsc{ii}\fi}
\newcommand{  \NeiiIR   }{\ifmmode {\rm Ne}\,\textsc{ii}\,12.8\,\mu {\rm m} \else Ne\,\textsc{ii}\,$12.8\,\mu$m\fi}
\newcommand{  \neiii    }{\ifmmode {\rm Ne}\,\textsc{iii} \else Ne\,\textsc{iii}\fi}
\newcommand{  \neiv     }{\ifmmode {\rm Ne}\,\textsc{iv}  \else Ne\,\textsc{iv}\fi}
\newcommand{  \nev      }{\ifmmode {\rm Ne}\,\textsc{v}   \else Ne\,\textsc{v}\fi}
\newcommand{  \NevIR    }{\ifmmode {\rm Ne}\,\textsc{v}\,24.3\,\mu {\rm m} \else Ne\,\textsc{v}\,$24.3\,\mu$m\fi}
\newcommand{  \nevi     }{\ifmmode {\rm Ne}\,\textsc{vi}  \else Ne\,\textsc{vi}\fi}
\newcommand{  \mgi      }{\ifmmode {\rm Mg}\,\textsc{i} \else Mg\,\textsc{i}\fi}
\newcommand{  \mgii     }{\ifmmode {\rm Mg}\,\textsc{ii} \else Mg\,\textsc{ii}\fi}
\newcommand{  \MgII     }{\ifmmode {\rm Mg}\,\textsc{ii}\,\lambda2798 \else Mg\,\textsc{ii}\,$\lambda2798$\fi}
\newcommand{  \sii      }{\ifmmode {\rm S}\,\textsc{ii} \else S\,\textsc{ii}\fi}
\newcommand{\SII}{\ifmmode \left[{\rm S}\,\textsc{ii}\right]\,\lambda6716, 6731 \else [S\,{\sc ii}]\,$\lambda 6716, 6731$\fi}
\newcommand{  \siii     }{\ifmmode {\rm S}\,\textsc{iii} \else S\,\textsc{iii}\fi}
\newcommand{  \siv      }{\ifmmode {\rm S}\,\textsc{iv} \else S\,\textsc{iv}\fi}
\newcommand{  \sili     }{\ifmmode {\rm Si}\,\textsc{i}   \else Si\,\textsc{i}\fi}
\newcommand{  \silii    }{\ifmmode {\rm Si}\,\textsc{ii}  \else Si\,\textsc{ii}\fi}
\newcommand{  \Siliv    }{\ifmmode {\rm Si}\,\textsc{iv}  \else Si\,\textsc{iv}\fi}
\newcommand{  \SilIVuv  }{\ifmmode {\rm Si}\,\textsc{iv}\,\lambda1400  \else Si\,\textsc{iv}\,$\lambda1400$\fi}
\newcommand{  \AlIII   }{\ifmmode {\rm Al}\,\textsc{iii}\,\lambda1857 \else Al\,\textsc{iii}\,$\lambda1857$\fi}
\newcommand{  \Aliii   }{\ifmmode {\rm Al}\,\textsc{iii} \else Al\,\textsc{iii}\fi}
\newcommand{  \caii     }{\ifmmode {\rm Ca}\,\textsc{ii} \else Ca\,\textsc{ii}\fi}
\newcommand{  \feii     }{\ifmmode {\rm Fe}\,\textsc{ii} \else Fe\,\textsc{ii}\fi}
\newcommand{  \feiii    }{\ifmmode {\rm Fe}\,\textsc{iii} \else Fe\,\textsc{iii}\fi}
\newcommand{  \Kalpha   }{\ifmmode {\rm K}\alpha \else K$\alpha$\fi}

\newcommand{ \Lhb   }{\ifmmode L_{\hb} \else $L_{\hb}$\fi}
\newcommand{ \Lha   }{\ifmmode L_{\ha} \else $L_{\ha}$\fi}
\newcommand{ \fwhb  }{\ifmmode {\rm FWHM}\left(\hb\right) \else FWHM(\hb)\fi}
\newcommand{\sighb  }{\ifmmode \sigma\left(\hb\right) \else $\sigma\left(\hb\right)$\fi}
\newcommand{ \ewhb  }{\ifmmode {\rm EW}\left(\hb\right) \else EW(\hb)\fi}
\newcommand{ \fwha  }{\ifmmode {\rm FWHM}\left(\ha\right) \else FWHM(\ha)\fi}
\newcommand{ \ewha  }{\ifmmode {\rm EW}\left(\ha\right) \else EW(\ha)\fi}
\newcommand{ \Lmg   }{\ifmmode L\left(\mgii\right) \else $L\left(\mgii\right)$\fi}
\newcommand{ \fwmg  }{\ifmmode {\rm FWHM}\left(\mgii\right) \else FWHM(\mgii)\fi}
\newcommand{ \Lciv  }{\ifmmode L\left(\civ\right) \else $L\left(\civ\right)$\fi}
\newcommand{ \fwciv }{\ifmmode {\rm FWHM}\left(\civ\right) \else FWHM(\civ)\fi}
\newcommand{ \fwhm  }{\ifmmode {\rm FWHM} \else FWHM\fi} 
\newcommand{ \voff  }{\ifmmode v_{\rm off} \else $v_{\rm off}$\fi} 
\newcommand{ \vmax  }{\ifmmode v_{\rm max} \else $v_{\rm max}$\fi} 

\newcommand{ \mumg  }{\ifmmode \mu\left(\mgii\right) \else $\mu\left(\mgii\right)$\fi}
\newcommand{ \fmg   }{\ifmmode f\left(\mgii\right) \else $f\left(\mgii\right)$\fi}
\newcommand{ \muciv }{\ifmmode \mu\left(\civ\right) \else $\mu\left(\civ\right)$\fi}
\newcommand{ \fciv  }{\ifmmode f\left(\civ\right) \else $f\left(\civ\right)$\fi}

\newcommand{  \auvo     }{\ifmmode \alpha_{\nu,{\rm UVO}} \else $\alpha_{\nu,{\rm UVO}}$\fi}
\newcommand{  \Ledd     }{\ifmmode L_{\rm Edd} \else $L_{\rm Edd}$\fi}
\newcommand{  \lamLlam  }{\ifmmode \lambda L_{\lambda} \else $\lambda L_{\lambda}$\fi}
\newcommand{  \lLl      }{\ifmmode \lambda L_{\lambda} \else $\lambda L_{\lambda}$\fi}
\newcommand{  \nuLnu    }{\ifmmode \nu L_{\nu} \else $\nu L_{\nu}$\fi}
\newcommand{  \nLn      }{\ifmmode \nu L_{\nu} \else $\nu L_{\nu}$\fi}
\newcommand{  \Luv      }{\ifmmode L_{1450} \else $L_{1450}$\fi}
\newcommand{  \Lop      }{\ifmmode L_{5100} \else $L_{5100}$\fi}
\newcommand{  \lLop     }{\ifmmode \log\left(\Lop/\ergs\right) \else $\log\left(\Lop/\ergs\right)$\fi}
\newcommand{  \Lthree   }{\ifmmode L_{3000} \else $L_{3000}$\fi}
\newcommand{  \lLthree  }{\ifmmode \log\left(\Lthree/\ergs\right) \else $\log\left(\Lthree/\ergs\right)$\fi}
\newcommand{  \Lsix      }{\ifmmode L_{6200} \else $L_{6200}$\fi}
\newcommand{  \lLisx     }{\ifmmode \log\left(\Lop/\ergs\right) \else $\log\left(\Lop/\ergs\right)$\fi}
\newcommand{  \Lxray    }{\ifmmode L_{\rm X} \else $L_{\rm X}$\fi}
\newcommand{  \Lhard    }{\ifmmode L_{\rm 2-10} \else $L_{\rm 2-10}$\fi}
\newcommand{  \Lsoft    }{\ifmmode L_{\rm 0.5-2} \else $L_{\rm 0.5-2}$\fi}

\newcommand{\Fthree}{\ifmmode F_{3000} \else $F_{3000}$\fi}
\newcommand{\fuv}{\ifmmode f_{\lambda}\left(1450{\rm \AA}\right) \else $f_{\lambda}\left(1450 {\rm \AA}\right)$\fi}
\newcommand{\fthree}{\ifmmode f_{\lambda}\left(3000{\rm \AA}\right) \else $f_{\lambda}\left(3000{\rm \AA}\right)$\fi}
\newcommand{\fH}{\ifmmode f_{\lambda}\left(1.65\micron\right) \else
$f_{\lambda}\left(1.65\micron\right)$\fi}

\newcommand{\fbol}{\ifmmode f_{\rm bol} \else $f_{\rm bol}$\fi}
\newcommand{\fbolwv}{\ifmmode f_{\rm bol}\left(\lambda\right) \else $f_{\rm bol}\left(\lambda\right)$\fi}
\newcommand{\fbolopt}{\ifmmode f_{\rm bol}\left(5100{\rm \AA}\right) \else $f_{\rm bol}\left(5100{\rm \AA}\right)$\fi}
\newcommand{\fbolthree}{\ifmmode f_{\rm bol}\left(3000{\rm \AA}\right) \else $f_{\rm bol}\left(3000{\rm \AA}\right)$\fi}
\newcommand{\fboluv}{\ifmmode f_{\rm bol}\left(1450{\rm \AA}\right) \else $f_{\rm bol}\left(1450{\rm \AA}\right)$\fi}

\newcommand{\fobs}{\ifmmode f_{\rm obs} \else $f_{\rm obs}$\fi}

\newcommand{  \mbh      }{\ifmmode M_{\rm BH} \else $M_{\rm BH}$\fi}
\newcommand{  \lmbh     }{\ifmmode \log\left(\mbh/\Msun\right) \else $\log\left(\mbh/\Msun\right)$\fi} 
\newcommand{  \lledd    }{\ifmmode L/L_{\rm Edd} \else $L/L_{\rm Edd}$\fi}
\newcommand{  \mmedd    }{\ifmmode \dot{m}/\dot{m}_{\rm \,Edd} \else $\dot{m}/\dot{m}_{\rm \,Edd}$\fi}
\newcommand{  \Lbol     }{\ifmmode L_{\rm bol} \else $L_{\rm bol}$\fi}
\newcommand{  \lbol     }{\ifmmode L_{\rm bol} \else $L_{\rm bol}$\fi}
\newcommand{  \lLbol    }{\ifmmode \log\left(\Lbol/\ergs\right) \else $\log\left(\Lbol/\ergs\right)$\fi} 
\newcommand{  \Lagn     }{\ifmmode L_{\rm AGN} \else $L_{\rm AGN}$\fi}
\newcommand{  \lagn     }{\ifmmode L_{\rm AGN} \else $L_{\rm AGN}$\fi}

\newcommand{  \tgrow     }{\ifmmode t_{\rm growth} \else $t_{\rm growth}$\fi}
\newcommand{  \tUni      }{\ifmmode t_{\rm Universe} \else $t_{\rm Universe}$\fi}

\newcommand{  \Mindot	}{\ifmmode \dot{M}_{\rm infall} \else $\dot{M}_{\rm infall}$\fi}
\newcommand{  \Mbhdot	}{\ifmmode \dot{M}_{\rm BH} \else $\dot{M}_{\rm BH}$\fi}

\newcommand{  \Maddot	}{\ifmmode \dot{M}_{\rm AD} \else $\dot{M}_{\rm AD}$\fi}

\newcommand{  \Mdot	}{\ifmmode \dot{M} \else $\dot{M}$\fi}

\newcommand{  \as	}{\ifmmode a_{\rm *} \else $a_{\rm *}$\fi}
\newcommand{  \avec	}{\ifmmode \vec{a}_{\rm *} \else $\vec{a}_{\rm *}$\fi}
\newcommand{  \re	}{\ifmmode \eta      	 \else $\eta$\fi}
\newcommand{  \RISCO	}{\ifmmode R_{\rm ISCO}  \else $R_{\rm ISCO}$\fi}
\newcommand{  \rg	}{\ifmmode r_{\rm g}  \else $r_{\rm g}$\fi}
\newcommand{  \rS	}{\ifmmode r_{\rm S}  \else $r_{\rm S}$\fi}

\newcommand{  \mseed    }{\ifmmode M_{\rm seed} \else $M_{\rm seed}$\fi}
\newcommand{  \mbul     }{\ifmmode M_{\rm Bulge} \else $M_{\rm Bulge}$\fi} 
\newcommand{  \mstar    }{\ifmmode M_{\star} \else $M_{\star}$\fi} 
\newcommand{  \mgal     }{\ifmmode M_{*} \else $M_{*}$\fi} 
\newcommand{  \mhost    }{\ifmmode M_{\rm Host} \else $M_{\rm Host}$\fi}
\newcommand{  \mmsmall  }{\ifmmode M_{\rm BH}/M_{*} \else $M_{\rm BH}/M_{*}$\fi}
\newcommand{  \mmlarge  }{\ifmmode M_{*}/M_{\rm BH} \else $M_{*}/M_{\rm BH}$\fi}

\newcommand{  \mmwp     }{\ifmmode \left(M_{*}/M_{\rm BH}\right) \else $\left(M_{*}/M_{\rm BH}\right)$\fi}
\newcommand{  \ml       }{\ifmmode M_{*}/L_{*} \else $M_{*}/L_{*}$\fi}
\newcommand{  \mlwp     }{\ifmmode \left(M_{*}/L\right) \else $\left(M_{*}/L\right)$\fi}
\newcommand{  \mlk      }{\ifmmode \left(M_{*}/L_{K}\right) \else $\left(M_{*}/L_{K}\right)$\fi}
\newcommand{  \sigs     }{\ifmmode \sigma_{*} \else $\sigma_{*}$\fi}
\newcommand{  \Reff     }{\ifmmode R_{\rm e} \else $R_{\rm e}$\fi}
\newcommand{  \nser     }{\ifmmode n_{\rm s} \else $n_{\rm s}$\fi}
\newcommand{  \LFIR     }{\ifmmode L_{\rm FIR} \else $L_{\rm FIR}$\fi}
\newcommand{  \Lfir     }{\ifmmode L_{\rm FIR} \else $L_{\rm FIR}$\fi}

\newcommand{  \mdyn     }{\ifmmode M_{\rm dyn} \else $M_{\rm dyn}$\fi} 
\newcommand{  \mgas     }{\ifmmode M_{\rm gas} \else $M_{\rm gas}$\fi} 
\newcommand{  \mh       }{\ifmmode M_{\rm h} \else $M_{\rm h}$\fi}
\newcommand{  \mhalo    }{\ifmmode M_{\rm halo} \else $M_{\rm halo}$\fi}

\newcommand{  \Lcii     }{\ifmmode L_{\cii} \else $L_{\cii}$\fi}

\newcommand{\bj}{\ifmmode b_{\rm J} \else $b_{\rm J}$\fi}

\newcommand{\iab}{\ifmmode i_{\rm AB} \else $i_{\rm AB}$\fi}

\newcommand{\jab}{\ifmmode J_{\rm AB} \else $J_{\rm AB}$\fi}
\newcommand{\hab}{\ifmmode H_{\rm AB} \else $H_{\rm AB}$\fi}
\newcommand{\kab}{\ifmmode K_{\rm AB} \else $K_{\rm AB}$\fi}

\newcommand{\jveg}{\ifmmode J_{\rm Vega} \else $J_{\rm Vega}$\fi}
\newcommand{\hveg}{\ifmmode H_{\rm Vega} \else $H_{\rm Vega}$\fi}
\newcommand{\kveg}{\ifmmode K_{\rm Vega} \else $K_{\rm Vega}$\fi}

\newcommand{  \Chisq    }{\ifmmode \chi^{2} \else $\chi^{2}$}
\newcommand{  \nelec    }{\ifmmode n_{e} \else $n_{e}$\fi}     
\newcommand{  \nh       }{\ifmmode n_{H} \else $n_{H}$\fi}     
\newcommand{  \Ncol     }{\ifmmode N_{col} \else $N_{col}$\fi} 
\newcommand{  \NH       }{\ifmmode N_{H} \else $N_{H}$\fi}     

\newcommand{\MgasCO}{\ifmmode M_{\rm gas, \, CO} \else $M_{\rm gas, \, CO}$\fi}
\newcommand{\Mgasdust}{\ifmmode M_{\rm gas, \, dust} \else $M_{\rm gas, \, dust}$\fi}
\newcommand{\lMgasCO}{\relax\ifmmode \log \left( M_{\rm gas, \, CO} \right) \else $ \log \left( M_{\rm gas, \, CO}  \right)$\fi}
\newcommand{\lMgasdust}{\ifmmode  \log \left( M_{\rm gas, \, dust} \right) \else $ \log \left( M_{\rm gas, \, dust} \right)$\fi}
\newcommand{\Mdust}{\ifmmode M_{\rm dust} \else $M_{\rm dust}$\fi}
\newcommand{\lMdust}{\ifmmode  \log \left( M_{\rm dust} \right) \else $ \log \left( M_{\rm dust} \right)$\fi}

\newcommand{\Hmol }{\ifmmode {\rm H_2} \else ${\rm H_2}$ \fi}
\newcommand{\MHmol }{\ifmmode M_{\rm H_2} \else $M_{\rm H_2}$ \fi}
\newcommand{\fHmol }{\ifmmode f_{\rm H_2} \else $f_{\rm H_2}$ \fi}
\newcommand{\tdep }{\ifmmode t_{\rm dep} \else $t_{\rm dep}$ \fi}

\def\micron{\hbox{$\mu$m}}
\def\ion#1#2{#1$\;${\small\rm\@Roman{#2}}\relax}



%% file: MASCOT__Metallicity_gradients.bbl
\begin{thebibliography}{}
\makeatletter
\relax
\def\mn@urlcharsother{\let\do\@makeother \do\$\do\&\do\#\do\^\do\_\do\%\do\~}
\def\mn@doi{\begingroup\mn@urlcharsother \@ifnextchar [ {\mn@doi@}
  {\mn@doi@[]}}
\def\mn@doi@[#1]#2{\def\@tempa{#1}\ifx\@tempa\@empty \href
  {http://dx.doi.org/#2} {doi:#2}\else \href {http://dx.doi.org/#2} {#1}\fi
  \endgroup}
\def\mn@eprint#1#2{\mn@eprint@#1:#2::\@nil}
\def\mn@eprint@arXiv#1{\href {http://arxiv.org/abs/#1} {{\tt arXiv:#1}}}
\def\mn@eprint@dblp#1{\href {http://dblp.uni-trier.de/rec/bibtex/#1.xml}
  {dblp:#1}}
\def\mn@eprint@#1:#2:#3:#4\@nil{\def\@tempa {#1}\def\@tempb {#2}\def\@tempc
  {#3}\ifx \@tempc \@empty \let \@tempc \@tempb \let \@tempb \@tempa \fi \ifx
  \@tempb \@empty \def\@tempb {arXiv}\fi \@ifundefined
  {mn@eprint@\@tempb}{\@tempb:\@tempc}{\expandafter \expandafter \csname
  mn@eprint@\@tempb\endcsname \expandafter{\@tempc}}}

\bibitem[\protect\citeauthoryear{Accurso et~al.,}{Accurso
  et~al.}{2017}]{Accurso2017}
Accurso G.,  et~al., 2017, \mn@doi [Monthly Notices of the Royal Astronomical
  Society] {10.1093/mnras/stx1556}, 470, 4750

\bibitem[\protect\citeauthoryear{Albán, Wylezalek  \& et al.}{Albán
  et~al.}{in prep.}]{Alban2022}
Albán M.,  Wylezalek D.,   et al. in prep.

\bibitem[\protect\citeauthoryear{{Alvarez-Hurtado}, {Barrera-Ballesteros},
  {S{\'a}nchez}, {Colombo}, {L{\'o}pez-S{\'a}nchez}  \&
  {Aquino-Ort{\'\i}z}}{{Alvarez-Hurtado} et~al.}{2022}]{AlvarezHurtado2022}
{Alvarez-Hurtado} P.,  {Barrera-Ballesteros} J.~K.,  {S{\'a}nchez} S.~F.,
  {Colombo} D.,  {L{\'o}pez-S{\'a}nchez} A.~R.,   {Aquino-Ort{\'\i}z} E.,
  2022, \mn@doi [\apj] {10.3847/1538-4357/ac58fb}, \href
  {https://ui.adsabs.harvard.edu/abs/2022ApJ...929...47A} {929, 47}

\bibitem[\protect\citeauthoryear{{Avery} et~al.,}{{Avery}
  et~al.}{2021}]{Avery2021}
{Avery} C.~R.,  et~al., 2021, \mn@doi [\mnras] {10.1093/mnras/stab780}, \href
  {https://ui.adsabs.harvard.edu/abs/2021MNRAS.503.5134A} {503, 5134}

\bibitem[\protect\citeauthoryear{{Baker} et~al.,}{{Baker}
  et~al.}{2022a}]{Baker2022b}
{Baker} W.~M.,  et~al., 2022a, arXiv e-prints, \href
  {https://ui.adsabs.harvard.edu/abs/2022arXiv221003755B} {p. arXiv:2210.03755}

\bibitem[\protect\citeauthoryear{Baker, Maiolino, Bluck, Lin, Ellison,
  Belfiore, Pan  \& Thorp}{Baker et~al.}{2022b}]{Baker2022a}
Baker W.~M.,  Maiolino R.,  Bluck A. F.~L.,  Lin L.,  Ellison S.~L.,  Belfiore
  F.,  Pan H.-A.,   Thorp M.,  2022b, \mn@doi [\mnras]
  {10.1093/mnras/stab3672}, \href
  {https://ui.adsabs.harvard.edu/abs/2022MNRAS.510.3622B} {510, 3622}

\bibitem[\protect\citeauthoryear{Baldwin, Phillips  \& Terlevich}{Baldwin
  et~al.}{1981}]{BPT1981}
Baldwin J.~A.,  Phillips M.~M.,   Terlevich R.,  1981, \mn@doi [Publications of
  the Astronomical Society of the Pacific] {10.1086/130766}, 93, 5

\bibitem[\protect\citeauthoryear{{Barrera-Ballesteros}
  et~al.,}{{Barrera-Ballesteros} et~al.}{2016}]{BarreraBallesteros2016}
{Barrera-Ballesteros} J.~K.,  et~al., 2016, \mn@doi [\mnras]
  {10.1093/mnras/stw1984}, \href
  {https://ui.adsabs.harvard.edu/abs/2016MNRAS.463.2513B} {463, 2513}

\bibitem[\protect\citeauthoryear{{Belfiore} et~al.,}{{Belfiore}
  et~al.}{2017}]{Belfiore2017}
{Belfiore} F.,  et~al., 2017, \mn@doi [Monthly Notices of the Royal
  Astronomical Society] {10.1093/mnras/stx789}, \href
  {https://ui.adsabs.harvard.edu/abs/2017MNRAS.469..151B} {469, 151}

\bibitem[\protect\citeauthoryear{{Belfiore} et~al.,}{{Belfiore}
  et~al.}{2019}]{Belfiore2019}
{Belfiore} F.,  et~al., 2019, \mn@doi [\aj] {10.3847/1538-3881/ab3e4e}, \href
  {https://ui.adsabs.harvard.edu/abs/2019AJ....158..160B} {158, 160}

\bibitem[\protect\citeauthoryear{Bertemes \& Wuyts}{Bertemes \& Wuyts}{in
  prep.}]{Bertemes2022c}
Bertemes C.,  Wuyts S.,  in prep.

\bibitem[\protect\citeauthoryear{{Binette}, {Magris}, {Stasi{\'n}ska}  \&
  {Bruzual}}{{Binette} et~al.}{1994}]{Binette1994}
{Binette} L.,  {Magris} C.~G.,  {Stasi{\'n}ska} G.,   {Bruzual} A.~G.,  1994,
  \aap, \href {https://ui.adsabs.harvard.edu/abs/1994A&A...292...13B} {292, 13}

\bibitem[\protect\citeauthoryear{{Blanton}, {Kazin}, {Muna}, {Weaver}  \&
  {Price-Whelan}}{{Blanton} et~al.}{2011}]{Blanton2011}
{Blanton} M.~R.,  {Kazin} E.,  {Muna} D.,  {Weaver} B.~A.,   {Price-Whelan} A.,
   2011, \mn@doi [The Astrophysical Journal] {10.1088/0004-6256/142/1/31},
  \href {https://ui.adsabs.harvard.edu/abs/2011AJ....142...31B} {142, 31}

\bibitem[\protect\citeauthoryear{{Blanton} et~al.,}{{Blanton}
  et~al.}{2017}]{Blanton2017_SDSS-IV}
{Blanton} M.~R.,  et~al., 2017, \mn@doi [The Astrophysical Journal]
  {10.3847/1538-3881/aa7567}, \href
  {https://ui.adsabs.harvard.edu/abs/2017AJ....154...28B} {154, 28}

\bibitem[\protect\citeauthoryear{{Bluck} et~al.,}{{Bluck}
  et~al.}{2020}]{Bluck2020}
{Bluck} A. F.~L.,  et~al., 2020, \mn@doi [\mnras] {10.1093/mnras/staa2806},
  \href {https://ui.adsabs.harvard.edu/abs/2020MNRAS.499..230B} {499, 230}

\bibitem[\protect\citeauthoryear{{Boardman}, {Zasowski}, {Newman}, {Sanchez},
  {Schaefer}, {Lian}, {Bizyaev}  \& {Drory}}{{Boardman}
  et~al.}{2021}]{Boardman2021}
{Boardman} N.~F.,  {Zasowski} G.,  {Newman} J.~A.,  {Sanchez} S.~F.,
  {Schaefer} A.,  {Lian} J.,  {Bizyaev} D.,   {Drory} N.,  2021, \mn@doi
  [\mnras] {10.1093/mnras/staa3785}, \href
  {https://ui.adsabs.harvard.edu/abs/2021MNRAS.501..948B} {501, 948}

\bibitem[\protect\citeauthoryear{{Boardman} et~al.,}{{Boardman}
  et~al.}{2022}]{Boardman2022}
{Boardman} N.,  et~al., 2022, \mn@doi [\mnras] {10.1093/mnras/stac1475}, \href
  {https://ui.adsabs.harvard.edu/abs/2022MNRAS.514.2298B} {514, 2298}

\bibitem[\protect\citeauthoryear{{Bolatto} et~al.,}{{Bolatto}
  et~al.}{2017}]{Bolatto2017}
{Bolatto} A.~D.,  et~al., 2017, \mn@doi [\apj] {10.3847/1538-4357/aa86aa},
  \href {https://ui.adsabs.harvard.edu/abs/2017ApJ...846..159B} {846, 159}

\bibitem[\protect\citeauthoryear{{Brennan}, {Choi}, {Somerville}, {Hirschmann},
  {Naab}  \& {Ostriker}}{{Brennan} et~al.}{2018}]{Brennan2018}
{Brennan} R.,  {Choi} E.,  {Somerville} R.~S.,  {Hirschmann} M.,  {Naab} T.,
  {Ostriker} J.~P.,  2018, \mn@doi [\apj] {10.3847/1538-4357/aac2c4}, \href
  {https://ui.adsabs.harvard.edu/abs/2018ApJ...860...14B} {860, 14}

\bibitem[\protect\citeauthoryear{Brinchmann, Charlot, White, Tremonti,
  Kauffmann, Heckman  \& Brinkmann}{Brinchmann et~al.}{2004}]{Brinchmann2004}
Brinchmann J.,  Charlot S.,  White S.,  Tremonti C.,  Kauffmann G.,  Heckman
  T.,   Brinkmann J.,  2004, \mn@doi [Monthly Notices of the Royal Astronomical
  Society] {10.1111/j.1365-2966.2004.07881.x}, 351, 1151

\bibitem[\protect\citeauthoryear{Bruzual \& Charlot}{Bruzual \&
  Charlot}{2003}]{Bruzual2003}
Bruzual G.,  Charlot S.,  2003, Monthly Notices of the Royal Astronomical
  Society, 344, 1000

\bibitem[\protect\citeauthoryear{{Bundy} et~al.,}{{Bundy}
  et~al.}{2015}]{Bundy2015}
{Bundy} K.,  et~al., 2015, \mn@doi [The Astrophysical Journal]
  {10.1088/0004-637X/798/1/7}, \href
  {https://ui.adsabs.harvard.edu/abs/2015ApJ...798....7B} {798, 7}

\bibitem[\protect\citeauthoryear{{Calzetti}, {Armus}, {Bohlin}, {Kinney},
  {Koornneef}  \& {Storchi-Bergmann}}{{Calzetti} et~al.}{2000}]{Calzetti2000}
{Calzetti} D.,  {Armus} L.,  {Bohlin} R.~C.,  {Kinney} A.~L.,  {Koornneef} J.,
   {Storchi-Bergmann} T.,  2000, \mn@doi [The Astrophysical Journal]
  {10.1086/308692}, \href
  {https://ui.adsabs.harvard.edu/abs/2000ApJ...533..682C} {533, 682}

\bibitem[\protect\citeauthoryear{Carnall, McLure, Dunlop  \&
  Dav{\'{e}}}{Carnall et~al.}{2018}]{Carnall2018}
Carnall A.~C.,  McLure R.~J.,  Dunlop J.~S.,   Dav{\'{e}} R.,  2018, \mn@doi
  [Monthly Notices of the Royal Astronomical Society] {10.1093/MNRAS/STY2169},
  480, 4379

\bibitem[\protect\citeauthoryear{Carnall et~al.,}{Carnall
  et~al.}{2019}]{Carnall2019b}
Carnall A.~C.,  et~al., 2019, \mn@doi [Monthly Notices of the Royal
  Astronomical Society] {10.1093/mnras/stz2544}, 490, 417

\bibitem[\protect\citeauthoryear{{Cazzoli} et~al.,}{{Cazzoli}
  et~al.}{2018}]{Cazzoli2018}
{Cazzoli} S.,  et~al., 2018, \mn@doi [\mnras] {10.1093/mnras/sty1811}, \href
  {https://ui.adsabs.harvard.edu/abs/2018MNRAS.480.1106C} {480, 1106}

\bibitem[\protect\citeauthoryear{{Chauke} et~al.,}{{Chauke}
  et~al.}{2019}]{Chauke2019}
{Chauke} P.,  et~al., 2019, \mn@doi [\apj] {10.3847/1538-4357/ab164d}, \href
  {https://ui.adsabs.harvard.edu/abs/2019ApJ...877...48C} {877, 48}

\bibitem[\protect\citeauthoryear{{Cherinka} et~al.,}{{Cherinka}
  et~al.}{2019}]{Cherinka2019}
{Cherinka} B.,  et~al., 2019, \mn@doi [\aj] {10.3847/1538-3881/ab2634}, \href
  {https://ui.adsabs.harvard.edu/abs/2019AJ....158...74C} {158, 74}

\bibitem[\protect\citeauthoryear{{Chisholm}, {Tremonti}  \&
  {Leitherer}}{{Chisholm} et~al.}{2018}]{Chisholm2018}
{Chisholm} J.,  {Tremonti} C.,   {Leitherer} C.,  2018, \mn@doi [\mnras]
  {10.1093/mnras/sty2380}, \href
  {https://ui.adsabs.harvard.edu/abs/2018MNRAS.481.1690C} {481, 1690}

\bibitem[\protect\citeauthoryear{{Colombo} et~al.,}{{Colombo}
  et~al.}{2020}]{Colombo2020}
{Colombo} D.,  et~al., 2020, \mn@doi [\aap] {10.1051/0004-6361/202039005},
  \href {https://ui.adsabs.harvard.edu/abs/2020A&A...644A..97C} {644, A97}

\bibitem[\protect\citeauthoryear{{Comerford} et~al.,}{{Comerford}
  et~al.}{2020}]{Comerford2020}
{Comerford} J.~M.,  et~al., 2020, \mn@doi [\apj] {10.3847/1538-4357/abb2ae},
  \href {https://ui.adsabs.harvard.edu/abs/2020ApJ...901..159C} {901, 159}

\bibitem[\protect\citeauthoryear{{Concas} \& {Popesso}}{{Concas} \&
  {Popesso}}{2019}]{Concas2019}
{Concas} A.,  {Popesso} P.,  2019, \mn@doi [\mnras] {10.1093/mnrasl/slz065},
  \href {https://ui.adsabs.harvard.edu/abs/2019MNRAS.486L..91C} {486, L91}

\bibitem[\protect\citeauthoryear{{Dom{\'\i}nguez S{\'a}nchez}, {Margalef},
  {Bernardi}  \& {Huertas-Company}}{{Dom{\'\i}nguez S{\'a}nchez}
  et~al.}{2022}]{DominguezSanchez2022}
{Dom{\'\i}nguez S{\'a}nchez} H.,  {Margalef} B.,  {Bernardi} M.,
  {Huertas-Company} M.,  2022, \mn@doi [\mnras] {10.1093/mnras/stab3089}, \href
  {https://ui.adsabs.harvard.edu/abs/2022MNRAS.509.4024D} {509, 4024}

\bibitem[\protect\citeauthoryear{{Ellison}, {Patton}, {Simard}  \&
  {McConnachie}}{{Ellison} et~al.}{2008}]{Ellison2008}
{Ellison} S.~L.,  {Patton} D.~R.,  {Simard} L.,   {McConnachie} A.~W.,  2008,
  \mn@doi [\apjl] {10.1086/527296}, \href
  {https://ui.adsabs.harvard.edu/abs/2008ApJ...672L.107E} {672, L107}

\bibitem[\protect\citeauthoryear{{Ellison}, {Thorp}, {Pan}, {Lin}, {Scudder},
  {Bluck}, {S{\'a}nchez}  \& {Sargent}}{{Ellison} et~al.}{2020}]{Ellison2020}
{Ellison} S.~L.,  {Thorp} M.~D.,  {Pan} H.-A.,  {Lin} L.,  {Scudder} J.~M.,
  {Bluck} A. F.~L.,  {S{\'a}nchez} S.~F.,   {Sargent} M.,  2020, \mn@doi
  [\mnras] {10.1093/mnras/staa001}, \href
  {https://ui.adsabs.harvard.edu/abs/2020MNRAS.492.6027E} {492, 6027}

\bibitem[\protect\citeauthoryear{{Falc{\'o}n-Barroso},
  {S{\'a}nchez-Bl{\'a}zquez}, {Vazdekis}, {Ricciardelli}, {Cardiel}, {Cenarro},
  {Gorgas}  \& {Peletier}}{{Falc{\'o}n-Barroso}
  et~al.}{2011}]{FalconBarroso2011}
{Falc{\'o}n-Barroso} J.,  {S{\'a}nchez-Bl{\'a}zquez} P.,  {Vazdekis} A.,
  {Ricciardelli} E.,  {Cardiel} N.,  {Cenarro} A.~J.,  {Gorgas} J.,
  {Peletier} R.~F.,  2011, \mn@doi [\aap] {10.1051/0004-6361/201116842}, \href
  {https://ui.adsabs.harvard.edu/abs/2011A&A...532A..95F} {532, A95}

\bibitem[\protect\citeauthoryear{{Ferland}, {Korista}, {Verner}, {Ferguson},
  {Kingdon}  \& {Verner}}{{Ferland} et~al.}{1998}]{Ferland1998}
{Ferland} G.~J.,  {Korista} K.~T.,  {Verner} D.~A.,  {Ferguson} J.~W.,
  {Kingdon} J.~B.,   {Verner} E.~M.,  1998, \mn@doi [\pasp] {10.1086/316190},
  \href {https://ui.adsabs.harvard.edu/abs/1998PASP..110..761F} {110, 761}

\bibitem[\protect\citeauthoryear{{Fischer}, {Dom{\'\i}nguez S{\'a}nchez}  \&
  {Bernardi}}{{Fischer} et~al.}{2019}]{Fischer2019}
{Fischer} J.~L.,  {Dom{\'\i}nguez S{\'a}nchez} H.,   {Bernardi} M.,  2019,
  \mn@doi [\mnras] {10.1093/mnras/sty3135}, \href
  {https://ui.adsabs.harvard.edu/abs/2019MNRAS.483.2057F} {483, 2057}

\bibitem[\protect\citeauthoryear{{Franchetto} et~al.,}{{Franchetto}
  et~al.}{2021}]{Franchetto2021}
{Franchetto} A.,  et~al., 2021, \mn@doi [\apj] {10.3847/1538-4357/ac2510},
  \href {https://ui.adsabs.harvard.edu/abs/2021ApJ...923...28F} {923, 28}

\bibitem[\protect\citeauthoryear{{Gibson}, {Pilkington}, {Brook}, {Stinson}  \&
  {Bailin}}{{Gibson} et~al.}{2013}]{Gibson2013}
{Gibson} B.~K.,  {Pilkington} K.,  {Brook} C.~B.,  {Stinson} G.~S.,   {Bailin}
  J.,  2013, \mn@doi [\aap] {10.1051/0004-6361/201321239}, \href
  {https://ui.adsabs.harvard.edu/abs/2013A&A...554A..47G} {554, A47}

\bibitem[\protect\citeauthoryear{{Hemler} et~al.,}{{Hemler}
  et~al.}{2021}]{Hemler2021}
{Hemler} Z.~S.,  et~al., 2021, \mn@doi [\mnras] {10.1093/mnras/stab1803}, \href
  {https://ui.adsabs.harvard.edu/abs/2021MNRAS.506.3024H} {506, 3024}

\bibitem[\protect\citeauthoryear{Hermosa~Mu$\rm\tilde{n}$oz, M{{\'a}}rquez,
  Cazzoli, Masegosa  \& Ag{{\'\i}}s-Gonz{{\'a}}lez}{Hermosa~Mu$\rm\tilde{n}$oz
  et~al.}{2022}]{HermosaMunoz2022}
Hermosa~Mu$\rm\tilde{n}$oz L.,  M{{\'a}}rquez I.,  Cazzoli S.,  Masegosa J.,
  Ag{{\'\i}}s-Gonz{{\'a}}lez B.,  2022, \mn@doi [\aap]
  {10.1051/0004-6361/202142629}, \href
  {https://ui.adsabs.harvard.edu/abs/2022A&A...660A.133H} {660, A133}

\bibitem[\protect\citeauthoryear{{Janowiecki}, {Catinella}, {Cortese},
  {Saintonge}  \& {Wang}}{{Janowiecki} et~al.}{2020}]{Janowiecki2020}
{Janowiecki} S.,  {Catinella} B.,  {Cortese} L.,  {Saintonge} A.,   {Wang} J.,
  2020, \mn@doi [\mnras] {10.1093/mnras/staa178}, \href
  {https://ui.adsabs.harvard.edu/abs/2020MNRAS.493.1982J} {493, 1982}

\bibitem[\protect\citeauthoryear{{Kauffmann} et~al.,}{{Kauffmann}
  et~al.}{2003}]{Kauffmann2003c}
{Kauffmann} G.,  et~al., 2003, \mn@doi [Monthly Notices of the Royal
  Astronomical Society] {10.1111/j.1365-2966.2003.07154.x}, \href
  {https://ui.adsabs.harvard.edu/abs/2003MNRAS.346.1055K} {346, 1055}

\bibitem[\protect\citeauthoryear{Kennicutt}{Kennicutt}{1998}]{Kennicutt1998a}
Kennicutt R.~C.,  1998, \mn@doi [The Astrophysical Journal] {10.1086/305588},
  498, 541

\bibitem[\protect\citeauthoryear{{Kennicutt} \& {Evans}}{{Kennicutt} \&
  {Evans}}{2012}]{Kennicutt2012}
{Kennicutt} R.~C.,  {Evans} N.~J.,  2012, \mn@doi [Annual Review of Astronomy
  and Astrophysics] {10.1146/annurev-astro-081811-125610}, \href
  {https://ui.adsabs.harvard.edu/abs/2012ARA&A..50..531K} {50, 531}

\bibitem[\protect\citeauthoryear{{Kewley}, {Dopita}, {Sutherland}, {Heisler}
  \& {Trevena}}{{Kewley} et~al.}{2001}]{Kewley2001}
{Kewley} L.~J.,  {Dopita} M.~A.,  {Sutherland} R.~S.,  {Heisler} C.~A.,
  {Trevena} J.,  2001, \mn@doi [\apj] {10.1086/321545}, \href
  {https://ui.adsabs.harvard.edu/abs/2001ApJ...556..121K} {556, 121}

\bibitem[\protect\citeauthoryear{{Kewley}, {Nicholls}  \&
  {Sutherland}}{{Kewley} et~al.}{2019}]{Kewley2019}
{Kewley} L.~J.,  {Nicholls} D.~C.,   {Sutherland} R.~S.,  2019, \mn@doi [\araa]
  {10.1146/annurev-astro-081817-051832}, \href
  {https://ui.adsabs.harvard.edu/abs/2019ARA&A..57..511K} {57, 511}

\bibitem[\protect\citeauthoryear{{Kroupa} \& {Boily}}{{Kroupa} \&
  {Boily}}{2002}]{Kroupa2002}
{Kroupa} P.,  {Boily} C.~M.,  2002, \mn@doi [Monthly Notices of the Royal
  Astronomical Society] {10.1046/j.1365-8711.2002.05848.x}, \href
  {https://ui.adsabs.harvard.edu/abs/2002MNRAS.336.1188K} {336, 1188}

\bibitem[\protect\citeauthoryear{{Kumari}, {Maiolino}, {Belfiore}  \&
  {Curti}}{{Kumari} et~al.}{2019}]{Kumari2019}
{Kumari} N.,  {Maiolino} R.,  {Belfiore} F.,   {Curti} M.,  2019, \mn@doi
  [\mnras] {10.1093/mnras/stz366}, \href
  {https://ui.adsabs.harvard.edu/abs/2019MNRAS.485..367K} {485, 367}

\bibitem[\protect\citeauthoryear{{Lacerda} et~al.,}{{Lacerda}
  et~al.}{2018}]{Lacerda2018}
{Lacerda} E.~A.~D.,  et~al., 2018, \mn@doi [\mnras] {10.1093/mnras/stx3022},
  \href {https://ui.adsabs.harvard.edu/abs/2018MNRAS.474.3727L} {474, 3727}

\bibitem[\protect\citeauthoryear{{Lacerda}, {S{\'a}nchez},
  {Mej{\'\i}a-Narv{\'a}ez}, {Camps-Fari{\~n}a}, {Espinosa-Ponce},
  {Barrera-Ballesteros}, {Ibarra-Medel}  \& {Lugo-Aranda}}{{Lacerda}
  et~al.}{2022}]{Lacerda2022}
{Lacerda} E. A.~D.,  {S{\'a}nchez} S.~F.,  {Mej{\'\i}a-Narv{\'a}ez} A.,
  {Camps-Fari{\~n}a} A.,  {Espinosa-Ponce} C.,  {Barrera-Ballesteros} J.~K.,
  {Ibarra-Medel} H.,   {Lugo-Aranda} A.~Z.,  2022, \mn@doi [\na]
  {10.1016/j.newast.2022.101895}, \href
  {https://ui.adsabs.harvard.edu/abs/2022NewA...9701895L} {97, 101895}

\bibitem[\protect\citeauthoryear{{Law} et~al.,}{{Law} et~al.}{2015}]{Law2015}
{Law} D.~R.,  et~al., 2015, \mn@doi [The Astrophysical Journal]
  {10.1088/0004-6256/150/1/19}, \href
  {https://ui.adsabs.harvard.edu/abs/2015AJ....150...19L} {150, 19}

\bibitem[\protect\citeauthoryear{Leja, Carnall, Johnson, Conroy  \&
  Speagle}{Leja et~al.}{2019}]{Leja2019a}
Leja J.,  Carnall A.~C.,  Johnson B.~D.,  Conroy C.,   Speagle J.~S.,  2019,
  \mn@doi [The Astrophysical Journal] {10.3847/1538-4357/ab133c}, 876, 3

\bibitem[\protect\citeauthoryear{{Lian}, {Thomas}, {Maraston}, {Goddard},
  {Comparat}, {Gonzalez-Perez}  \& {Ventura}}{{Lian} et~al.}{2018a}]{Lian2018}
{Lian} J.,  {Thomas} D.,  {Maraston} C.,  {Goddard} D.,  {Comparat} J.,
  {Gonzalez-Perez} V.,   {Ventura} P.,  2018a, \mn@doi [\mnras]
  {10.1093/mnras/stx2829}, \href
  {https://ui.adsabs.harvard.edu/abs/2018MNRAS.474.1143L} {474, 1143}

\bibitem[\protect\citeauthoryear{{Lian} et~al.,}{{Lian}
  et~al.}{2018b}]{Lian2018b}
{Lian} J.,  et~al., 2018b, \mn@doi [Monthly Notices of the Royal Astronomical
  Society] {10.1093/mnras/sty425}, \href
  {https://ui.adsabs.harvard.edu/abs/2018MNRAS.476.3883L} {476, 3883}

\bibitem[\protect\citeauthoryear{{Lin} et~al.,}{{Lin} et~al.}{2019}]{Lin2019}
{Lin} L.,  et~al., 2019, \mn@doi [\apjl] {10.3847/2041-8213/ab4815}, \href
  {https://ui.adsabs.harvard.edu/abs/2019ApJ...884L..33L} {884, L33}

\bibitem[\protect\citeauthoryear{{Lin} et~al.,}{{Lin} et~al.}{2020}]{Lin2020}
{Lin} L.,  et~al., 2020, \mn@doi [\apj] {10.3847/1538-4357/abba3a}, \href
  {https://ui.adsabs.harvard.edu/abs/2020ApJ...903..145L} {903, 145}

\bibitem[\protect\citeauthoryear{{Lin} et~al.,}{{Lin} et~al.}{2022}]{Lin2022}
{Lin} L.,  et~al., 2022, arXiv e-prints, \href
  {https://ui.adsabs.harvard.edu/abs/2022arXiv220105318L} {p. arXiv:2201.05318}

\bibitem[\protect\citeauthoryear{{L{\'o}pez-Cob{\'a}}, {S{\'a}nchez},
  {Bland-Hawthorn}, {Moiseev}, {Cruz-Gonz{\'a}lez}, {Garc{\'\i}a-Benito},
  {Barrera-Ballesteros}  \& {Galbany}}{{L{\'o}pez-Cob{\'a}}
  et~al.}{2019}]{LopezCoba2019}
{L{\'o}pez-Cob{\'a}} C.,  {S{\'a}nchez} S.~F.,  {Bland-Hawthorn} J.,  {Moiseev}
  A.~V.,  {Cruz-Gonz{\'a}lez} I.,  {Garc{\'\i}a-Benito} R.,
  {Barrera-Ballesteros} J.~K.,   {Galbany} L.,  2019, \mn@doi [\mnras]
  {10.1093/mnras/sty2960}, \href
  {https://ui.adsabs.harvard.edu/abs/2019MNRAS.482.4032L} {482, 4032}

\bibitem[\protect\citeauthoryear{{L{\'o}pez-Cob{\'a}}
  et~al.,}{{L{\'o}pez-Cob{\'a}} et~al.}{2020}]{LopezCoba2020}
{L{\'o}pez-Cob{\'a}} C.,  et~al., 2020, \mn@doi [\aj]
  {10.3847/1538-3881/ab7848}, \href
  {https://ui.adsabs.harvard.edu/abs/2020AJ....159..167L} {159, 167}

\bibitem[\protect\citeauthoryear{{Lotz}, {Jonsson}, {Cox}  \& {Primack}}{{Lotz}
  et~al.}{2008}]{Lotz2008}
{Lotz} J.~M.,  {Jonsson} P.,  {Cox} T.~J.,   {Primack} J.~R.,  2008, \mn@doi
  [\mnras] {10.1111/j.1365-2966.2008.14004.x}, \href
  {https://ui.adsabs.harvard.edu/abs/2008MNRAS.391.1137L} {391, 1137}

\bibitem[\protect\citeauthoryear{{Lutz} et~al.,}{{Lutz}
  et~al.}{2021}]{Lutz2021}
{Lutz} K.~A.,  et~al., 2021, \mn@doi [\aap] {10.1051/0004-6361/202038961},
  \href {https://ui.adsabs.harvard.edu/abs/2021A&A...649A..39L} {649, A39}

\bibitem[\protect\citeauthoryear{{Ma}, {Hopkins}, {Feldmann}, {Torrey},
  {Faucher-Gigu{\`e}re}  \& {Kere{\v{s}}}}{{Ma} et~al.}{2017}]{Ma2017}
{Ma} X.,  {Hopkins} P.~F.,  {Feldmann} R.,  {Torrey} P.,  {Faucher-Gigu{\`e}re}
  C.-A.,   {Kere{\v{s}}} D.,  2017, \mn@doi [\mnras] {10.1093/mnras/stx034},
  \href {https://ui.adsabs.harvard.edu/abs/2017MNRAS.466.4780M} {466, 4780}

\bibitem[\protect\citeauthoryear{{Maiolino} \& {Mannucci}}{{Maiolino} \&
  {Mannucci}}{2019}]{Maiolino2019}
{Maiolino} R.,  {Mannucci} F.,  2019, \mn@doi [\aapr]
  {10.1007/s00159-018-0112-2}, \href
  {https://ui.adsabs.harvard.edu/abs/2019A&ARv..27....3M} {27, 3}

\bibitem[\protect\citeauthoryear{{Maiolino} et~al.,}{{Maiolino}
  et~al.}{2008}]{Maiolino2008}
{Maiolino} R.,  et~al., 2008, \mn@doi [\aap] {10.1051/0004-6361:200809678},
  \href {https://ui.adsabs.harvard.edu/abs/2008A&A...488..463M} {488, 463}

\bibitem[\protect\citeauthoryear{{Mannucci}, {Cresci}, {Maiolino}, {Marconi}
  \& {Gnerucci}}{{Mannucci} et~al.}{2010}]{Mannucci2010}
{Mannucci} F.,  {Cresci} G.,  {Maiolino} R.,  {Marconi} A.,   {Gnerucci} A.,
  2010, \mn@doi [\mnras] {10.1111/j.1365-2966.2010.17291.x}, \href
  {https://ui.adsabs.harvard.edu/abs/2010MNRAS.408.2115M} {408, 2115}

\bibitem[\protect\citeauthoryear{{Marino} et~al.,}{{Marino}
  et~al.}{2013}]{Marino2013}
{Marino} R.~A.,  et~al., 2013, \mn@doi [\aap] {10.1051/0004-6361/201321956},
  \href {https://ui.adsabs.harvard.edu/abs/2013A&A...559A.114M} {559, A114}

\bibitem[\protect\citeauthoryear{Martig, Bournaud, Teyssier  \& Dekel}{Martig
  et~al.}{2009}]{Martig2009}
Martig M.,  Bournaud F.,  Teyssier R.,   Dekel A.,  2009, \mn@doi
  [Astrophysical Journal] {10.1088/0004-637X/707/1/250}, 707, 250

\bibitem[\protect\citeauthoryear{{Martins}, {Gonz{\'a}lez Delgado},
  {Leitherer}, {Cervi{\~n}o}  \& {Hauschildt}}{{Martins}
  et~al.}{2005}]{Martins2005}
{Martins} L.~P.,  {Gonz{\'a}lez Delgado} R.~M.,  {Leitherer} C.,  {Cervi{\~n}o}
  M.,   {Hauschildt} P.,  2005, \mn@doi [\mnras]
  {10.1111/j.1365-2966.2005.08703.x}, \href
  {https://ui.adsabs.harvard.edu/abs/2005MNRAS.358...49M} {358, 49}

\bibitem[\protect\citeauthoryear{{Mast} et~al.,}{{Mast}
  et~al.}{2014}]{Mast2014}
{Mast} D.,  et~al., 2014, \mn@doi [\aap] {10.1051/0004-6361/201321789}, \href
  {https://ui.adsabs.harvard.edu/abs/2014A&A...561A.129M} {561, A129}

\bibitem[\protect\citeauthoryear{{Nevin} et~al.,}{{Nevin}
  et~al.}{2021}]{Nevin2021}
{Nevin} R.,  et~al., 2021, \mn@doi [\apj] {10.3847/1538-4357/abe2a9}, \href
  {https://ui.adsabs.harvard.edu/abs/2021ApJ...912...45N} {912, 45}

\bibitem[\protect\citeauthoryear{Noeske et~al.,}{Noeske
  et~al.}{2007}]{Noeske2007_MS}
Noeske K.,  et~al., 2007, \mn@doi [The Astrophysical Journall]
  {10.1086/517926}, 660, L43

\bibitem[\protect\citeauthoryear{{Oliveira Jr.}, {Krabbe}, {Hernandez-Jimenez},
  {Dors Jr.}, {Zinchenko}, {H{\"a}gele}, {Cardaci}  \& {Monteiro}}{{Oliveira
  Jr.} et~al.}{2022}]{OliveiraJr2022}
{Oliveira Jr.} C.~B.,  {Krabbe} A.~C.,  {Hernandez-Jimenez} J.~A.,  {Dors Jr.}
  O.~L.,  {Zinchenko} I.~A.,  {H{\"a}gele} G.~F.,  {Cardaci} M.~V.,
  {Monteiro} A.~F.,  2022, arXiv e-prints, \href
  {https://ui.adsabs.harvard.edu/abs/2022arXiv220710260O} {p. arXiv:2207.10260}

\bibitem[\protect\citeauthoryear{Pettini \& Pagel}{Pettini \&
  Pagel}{2004}]{PP04_metal}
Pettini M.,  Pagel B.,  2004, \mn@doi [Monthly Notices of the Royal
  Astronomical Society] {10.1111/j.1365-2966.2004.07591.x}, 348, L59

\bibitem[\protect\citeauthoryear{{Piotrowska}, {Bluck}, {Maiolino}  \&
  {Peng}}{{Piotrowska} et~al.}{2021}]{Piotrowska2021}
{Piotrowska} J.~M.,  {Bluck} A. F.~L.,  {Maiolino} R.,   {Peng} Y.,  2021,
  \mn@doi [\mnras] {10.1093/mnras/stab3673}, \href
  {https://ui.adsabs.harvard.edu/abs/2021MNRAS.tmp.3414P} {}

\bibitem[\protect\citeauthoryear{{Poetrodjojo} et~al.,}{{Poetrodjojo}
  et~al.}{2018}]{Poetrodjojo2018}
{Poetrodjojo} H.,  et~al., 2018, \mn@doi [\mnras] {10.1093/mnras/sty1782},
  \href {https://ui.adsabs.harvard.edu/abs/2018MNRAS.479.5235P} {479, 5235}

\bibitem[\protect\citeauthoryear{{Renzini} \& {Peng}}{{Renzini} \&
  {Peng}}{2015}]{Renzini2015}
{Renzini} A.,  {Peng} Y.-j.,  2015, \mn@doi [The Astrophysical Journall]
  {10.1088/2041-8205/801/2/L29}, \href
  {https://ui.adsabs.harvard.edu/abs/2015ApJ...801L..29R} {801, L29}

\bibitem[\protect\citeauthoryear{{Rich}, {Torrey}, {Kewley}, {Dopita}  \&
  {Rupke}}{{Rich} et~al.}{2012}]{Rich2012}
{Rich} J.~A.,  {Torrey} P.,  {Kewley} L.~J.,  {Dopita} M.~A.,   {Rupke}
  D.~S.~N.,  2012, \mn@doi [\apj] {10.1088/0004-637X/753/1/5}, \href
  {https://ui.adsabs.harvard.edu/abs/2012ApJ...753....5R} {753, 5}

\bibitem[\protect\citeauthoryear{{Rosales-Ortega}, {S{\'a}nchez},
  {Iglesias-P{\'a}ramo}, {D{\'\i}az}, {V{\'\i}lchez}, {Bland-Hawthorn},
  {Husemann}  \& {Mast}}{{Rosales-Ortega} et~al.}{2012}]{RosalesOrtega2012}
{Rosales-Ortega} F.~F.,  {S{\'a}nchez} S.~F.,  {Iglesias-P{\'a}ramo} J.,
  {D{\'\i}az} A.~I.,  {V{\'\i}lchez} J.~M.,  {Bland-Hawthorn} J.,  {Husemann}
  B.,   {Mast} D.,  2012, \mn@doi [\apjl] {10.1088/2041-8205/756/2/L31}, \href
  {https://ui.adsabs.harvard.edu/abs/2012ApJ...756L..31R} {756, L31}

\bibitem[\protect\citeauthoryear{{Rupke}, {Kewley}  \& {Chien}}{{Rupke}
  et~al.}{2010}]{Rupke2010}
{Rupke} D. S.~N.,  {Kewley} L.~J.,   {Chien} L.~H.,  2010, \mn@doi [\apj]
  {10.1088/0004-637X/723/2/1255}, \href
  {https://ui.adsabs.harvard.edu/abs/2010ApJ...723.1255R} {723, 1255}

\bibitem[\protect\citeauthoryear{{Saintonge} \& {Catinella}}{{Saintonge} \&
  {Catinella}}{2022}]{Saintonge2022}
{Saintonge} A.,  {Catinella} B.,  2022, arXiv e-prints, \href
  {https://ui.adsabs.harvard.edu/abs/2022arXiv220200690S} {p. arXiv:2202.00690}

\bibitem[\protect\citeauthoryear{{Saintonge} et~al.,}{{Saintonge}
  et~al.}{2017}]{Saintonge2017}
{Saintonge} A.,  et~al., 2017, \mn@doi [The Astrophysical Journals]
  {10.3847/1538-4365/aa97e0}, \href
  {https://ui.adsabs.harvard.edu/abs/2017ApJS..233...22S} {233, 22}

\bibitem[\protect\citeauthoryear{Salim}{Salim}{2014}]{Salim2014}
Salim S.,  2014, \mn@doi [Serbian Astronomical Journal] {10.2298/SAJ1489001S},
  1, 1

\bibitem[\protect\citeauthoryear{{S{\'a}nchez}}{{S{\'a}nchez}}{2020}]{Sanchez2020}
{S{\'a}nchez} S.~F.,  2020, \mn@doi [\araa]
  {10.1146/annurev-astro-012120-013326}, \href
  {https://ui.adsabs.harvard.edu/abs/2020ARA&A..58...99S} {58, 99}

\bibitem[\protect\citeauthoryear{{S{\'a}nchez Almeida} \&
  {S{\'a}nchez-Menguiano}}{{S{\'a}nchez Almeida} \&
  {S{\'a}nchez-Menguiano}}{2019}]{SanchezAlmeida2019}
{S{\'a}nchez Almeida} J.,  {S{\'a}nchez-Menguiano} L.,  2019, \mn@doi [\apjl]
  {10.3847/2041-8213/ab218d}, \href
  {https://ui.adsabs.harvard.edu/abs/2019ApJ...878L...6S} {878, L6}

\bibitem[\protect\citeauthoryear{{S{\'a}nchez-Bl{\'a}zquez}
  et~al.,}{{S{\'a}nchez-Bl{\'a}zquez} et~al.}{2006}]{SanchezBlazquez2006}
{S{\'a}nchez-Bl{\'a}zquez} P.,  et~al., 2006, \mn@doi [\mnras]
  {10.1111/j.1365-2966.2006.10699.x}, \href
  {https://ui.adsabs.harvard.edu/abs/2006MNRAS.371..703S} {371, 703}

\bibitem[\protect\citeauthoryear{{S{\'a}nchez} et~al.,}{{S{\'a}nchez}
  et~al.}{2013}]{Sanchez2013}
{S{\'a}nchez} S.~F.,  et~al., 2013, \mn@doi [\aap]
  {10.1051/0004-6361/201220669}, \href
  {https://ui.adsabs.harvard.edu/abs/2013A&A...554A..58S} {554, A58}

\bibitem[\protect\citeauthoryear{{S{\'a}nchez} et~al.,}{{S{\'a}nchez}
  et~al.}{2014}]{Sanchez2014}
{S{\'a}nchez} S.~F.,  et~al., 2014, \mn@doi [\aap]
  {10.1051/0004-6361/201322343}, \href
  {https://ui.adsabs.harvard.edu/abs/2014A&A...563A..49S} {563, A49}

\bibitem[\protect\citeauthoryear{{S{\'a}nchez} et~al.,}{{S{\'a}nchez}
  et~al.}{2016}]{Sanchez2016b}
{S{\'a}nchez} S.~F.,  et~al., 2016, \rmxaa, \href
  {https://ui.adsabs.harvard.edu/abs/2016RMxAA..52..171S} {52, 171}

\bibitem[\protect\citeauthoryear{{S{\'a}nchez} et~al.,}{{S{\'a}nchez}
  et~al.}{2017}]{Sanchez2017}
{S{\'a}nchez} S.~F.,  et~al., 2017, \mn@doi [\mnras] {10.1093/mnras/stx808},
  \href {https://ui.adsabs.harvard.edu/abs/2017MNRAS.469.2121S} {469, 2121}

\bibitem[\protect\citeauthoryear{{S{\'a}nchez} et~al.,}{{S{\'a}nchez}
  et~al.}{2018}]{Sanchez2018}
{S{\'a}nchez} S.~F.,  et~al., 2018, \rmxaa, \href
  {https://ui.adsabs.harvard.edu/abs/2018RMxAA..54..217S} {54, 217}

\bibitem[\protect\citeauthoryear{{S{\'a}nchez}, {Walcher}, {Lopez-Cob{\'a}},
  {Barrera-Ballesteros}, {Mej{\'\i}a-Narv{\'a}ez}, {Espinosa-Ponce}  \&
  {Camps-Fari{\~n}a}}{{S{\'a}nchez} et~al.}{2021a}]{Sanchez2021a}
{S{\'a}nchez} S.~F.,  {Walcher} C.~J.,  {Lopez-Cob{\'a}} C.,
  {Barrera-Ballesteros} J.~K.,  {Mej{\'\i}a-Narv{\'a}ez} A.,  {Espinosa-Ponce}
  C.,   {Camps-Fari{\~n}a} A.,  2021a, \mn@doi [\rmxaa]
  {10.22201/ia.01851101p.2021.57.01.01}, \href
  {https://ui.adsabs.harvard.edu/abs/2021RMxAA..57....3S} {57, 3}

\bibitem[\protect\citeauthoryear{{S{\'a}nchez} et~al.,}{{S{\'a}nchez}
  et~al.}{2021b}]{Sanchez2021b}
{S{\'a}nchez} S.~F.,  et~al., 2021b, \mn@doi [\mnras] {10.1093/mnras/stab442},
  \href {https://ui.adsabs.harvard.edu/abs/2021MNRAS.503.1615S} {503, 1615}

\bibitem[\protect\citeauthoryear{{Schaefer} et~al.,}{{Schaefer}
  et~al.}{2019}]{Schaefer2019}
{Schaefer} A.~L.,  et~al., 2019, \mn@doi [\apj] {10.3847/1538-4357/ab43ca},
  \href {https://ui.adsabs.harvard.edu/abs/2019ApJ...884..156S} {884, 156}

\bibitem[\protect\citeauthoryear{{Schawinski}, {Koss}, {Berney}  \&
  {Sartori}}{{Schawinski} et~al.}{2015}]{Schawinski2015}
{Schawinski} K.,  {Koss} M.,  {Berney} S.,   {Sartori} L.~F.,  2015, \mn@doi
  [\mnras] {10.1093/mnras/stv1136}, \href
  {https://ui.adsabs.harvard.edu/abs/2015MNRAS.451.2517S} {451, 2517}

\bibitem[\protect\citeauthoryear{{Schmidt}}{{Schmidt}}{1959}]{Schmidt1959}
{Schmidt} M.,  1959, \mn@doi [The Astrophysical Journal] {10.1086/146614},
  \href {https://ui.adsabs.harvard.edu/abs/1959ApJ...129..243S} {129, 243}

\bibitem[\protect\citeauthoryear{{Sharda}, {Wisnioski}, {Krumholz}  \&
  {Federrath}}{{Sharda} et~al.}{2021}]{Sharda2021}
{Sharda} P.,  {Wisnioski} E.,  {Krumholz} M.~R.,   {Federrath} C.,  2021,
  \mn@doi [\mnras] {10.1093/mnras/stab1836}, \href
  {https://ui.adsabs.harvard.edu/abs/2021MNRAS.506.1295S} {506, 1295}

\bibitem[\protect\citeauthoryear{{Simons} et~al.,}{{Simons}
  et~al.}{2021}]{Simons2021}
{Simons} R.~C.,  et~al., 2021, \mn@doi [\apj] {10.3847/1538-4357/ac28f4}, \href
  {https://ui.adsabs.harvard.edu/abs/2021ApJ...923..203S} {923, 203}

\bibitem[\protect\citeauthoryear{{Singh} et~al.,}{{Singh}
  et~al.}{2013}]{Singh2013}
{Singh} R.,  et~al., 2013, \mn@doi [\aap] {10.1051/0004-6361/201322062}, \href
  {https://ui.adsabs.harvard.edu/abs/2013A&A...558A..43S} {558, A43}

\bibitem[\protect\citeauthoryear{{Smee} et~al.,}{{Smee}
  et~al.}{2013}]{Smee2013}
{Smee} S.~A.,  et~al., 2013, \mn@doi [The Astrophysical Journal]
  {10.1088/0004-6256/146/2/32}, \href
  {https://ui.adsabs.harvard.edu/abs/2013AJ....146...32S} {146, 32}

\bibitem[\protect\citeauthoryear{{Stinson}, {Bailin}, {Couchman}, {Wadsley},
  {Shen}, {Nickerson}, {Brook}  \& {Quinn}}{{Stinson}
  et~al.}{2010}]{Stinson2010}
{Stinson} G.~S.,  {Bailin} J.,  {Couchman} H.,  {Wadsley} J.,  {Shen} S.,
  {Nickerson} S.,  {Brook} C.,   {Quinn} T.,  2010, \mn@doi [\mnras]
  {10.1111/j.1365-2966.2010.17187.x}, \href
  {https://ui.adsabs.harvard.edu/abs/2010MNRAS.408..812S} {408, 812}

\bibitem[\protect\citeauthoryear{{Thorp}, {Ellison}, {Simard}, {S{\'a}nchez}
  \& {Antonio}}{{Thorp} et~al.}{2019}]{Thorp2019}
{Thorp} M.~D.,  {Ellison} S.~L.,  {Simard} L.,  {S{\'a}nchez} S.~F.,
  {Antonio} B.,  2019, \mn@doi [\mnras] {10.1093/mnrasl/sly185}, \href
  {https://ui.adsabs.harvard.edu/abs/2019MNRAS.482L..55T} {482, L55}

\bibitem[\protect\citeauthoryear{{Trayford} \& {Schaye}}{{Trayford} \&
  {Schaye}}{2019}]{Trayford2019}
{Trayford} J.~W.,  {Schaye} J.,  2019, \mn@doi [\mnras] {10.1093/mnras/stz757},
  \href {https://ui.adsabs.harvard.edu/abs/2019MNRAS.485.5715T} {485, 5715}

\bibitem[\protect\citeauthoryear{{Tremonti} et~al.,}{{Tremonti}
  et~al.}{2004}]{Tremonti2004}
{Tremonti} C.~A.,  et~al., 2004, \mn@doi [\apj] {10.1086/423264}, \href
  {https://ui.adsabs.harvard.edu/abs/2004ApJ...613..898T} {613, 898}

\bibitem[\protect\citeauthoryear{{Tumlinson}, {Peeples}  \& {Werk}}{{Tumlinson}
  et~al.}{2017}]{Tumlinson2017}
{Tumlinson} J.,  {Peeples} M.~S.,   {Werk} J.~K.,  2017, \mn@doi [\araa]
  {10.1146/annurev-astro-091916-055240}, \href
  {https://ui.adsabs.harvard.edu/abs/2017ARA&A..55..389T} {55, 389}

\bibitem[\protect\citeauthoryear{{Vale Asari}, {Couto}, {Cid Fernandes},
  {Stasi{\'n}ska}, {de Amorim}, {Ruschel-Dutra}, {Werle}  \& {Florido}}{{Vale
  Asari} et~al.}{2019}]{ValeAsari2019}
{Vale Asari} N.,  {Couto} G.~S.,  {Cid Fernandes} R.,  {Stasi{\'n}ska} G.,  {de
  Amorim} A.~L.,  {Ruschel-Dutra} D.,  {Werle} A.,   {Florido} T.~Z.,  2019,
  \mn@doi [\mnras] {10.1093/mnras/stz2470}, \href
  {https://ui.adsabs.harvard.edu/abs/2019MNRAS.489.4721V} {489, 4721}

\bibitem[\protect\citeauthoryear{{Vazdekis}, {S{\'a}nchez-Bl{\'a}zquez},
  {Falc{\'o}n-Barroso}, {Cenarro}, {Beasley}, {Cardiel}, {Gorgas}  \&
  {Peletier}}{{Vazdekis} et~al.}{2010}]{Vazdekis2010}
{Vazdekis} A.,  {S{\'a}nchez-Bl{\'a}zquez} P.,  {Falc{\'o}n-Barroso} J.,
  {Cenarro} A.~J.,  {Beasley} M.~A.,  {Cardiel} N.,  {Gorgas} J.,   {Peletier}
  R.~F.,  2010, \mn@doi [\mnras] {10.1111/j.1365-2966.2010.16407.x}, \href
  {https://ui.adsabs.harvard.edu/abs/2010MNRAS.404.1639V} {404, 1639}

\bibitem[\protect\citeauthoryear{{Wake} et~al.,}{{Wake}
  et~al.}{2017}]{Wake2017}
{Wake} D.~A.,  et~al., 2017, \mn@doi [The Astrophysical Journal]
  {10.3847/1538-3881/aa7ecc}, \href
  {https://ui.adsabs.harvard.edu/abs/2017AJ....154...86W} {154, 86}

\bibitem[\protect\citeauthoryear{{Wylezalek}, {Flores}, {Zakamska}, {Greene}
  \& {Riffel}}{{Wylezalek} et~al.}{2020}]{Wylezalek2020}
{Wylezalek} D.,  {Flores} A.~M.,  {Zakamska} N.~L.,  {Greene} J.~E.,   {Riffel}
  R.~A.,  2020, \mn@doi [\mnras] {10.1093/mnras/staa062}, \href
  {https://ui.adsabs.harvard.edu/abs/2020MNRAS.492.4680W} {492, 4680}

\bibitem[\protect\citeauthoryear{{Wylezalek} et~al.,}{{Wylezalek}
  et~al.}{2022}]{Wylezalek2022}
{Wylezalek} D.,  et~al., 2022, \mn@doi [\mnras] {10.1093/mnras/stab3356}, \href
  {https://ui.adsabs.harvard.edu/abs/2022MNRAS.510.3119W} {510, 3119}

\bibitem[\protect\citeauthoryear{{Yan} et~al.,}{{Yan} et~al.}{2016a}]{Yan2016a}
{Yan} R.,  et~al., 2016a, \mn@doi [The Astrophysical Journal]
  {10.3847/0004-6256/151/1/8}, \href
  {https://ui.adsabs.harvard.edu/abs/2016AJ....151....8Y} {151, 8}

\bibitem[\protect\citeauthoryear{{Yan} et~al.,}{{Yan} et~al.}{2016b}]{Yan2016b}
{Yan} R.,  et~al., 2016b, \mn@doi [The Astrophysical Journal]
  {10.3847/0004-6256/152/6/197}, \href
  {https://ui.adsabs.harvard.edu/abs/2016AJ....152..197Y} {152, 197}

\bibitem[\protect\citeauthoryear{{Zhang} et~al.,}{{Zhang}
  et~al.}{2017}]{Zhang2017}
{Zhang} K.,  et~al., 2017, \mn@doi [\mnras] {10.1093/mnras/stw3308}, \href
  {https://ui.adsabs.harvard.edu/abs/2017MNRAS.466.3217Z} {466, 3217}

\bibitem[\protect\citeauthoryear{{Zheng} et~al.,}{{Zheng}
  et~al.}{2017}]{Zheng2017}
{Zheng} Z.,  et~al., 2017, \mn@doi [Monthly Notices of the Royal Astronomical
  Society] {10.1093/mnras/stw3030}, \href
  {https://ui.adsabs.harvard.edu/abs/2017MNRAS.465.4572Z} {465, 4572}

\makeatother
\end{thebibliography}
